\documentclass[twocolumn]{aastex62}
\pdfoutput=1 
\usepackage{amsmath,amstext}
\usepackage[T1]{fontenc}
\usepackage{apjfonts} 
\usepackage[figure,figure*]{hypcap}
\usepackage{url}
\usepackage{hyperref}
\usepackage{natbib}

\shorttitle{Planet Occurrence Rates}
\shortauthors{Hsu et al.}

\begin{document}

\title{Occurrence Rates of Planets orbiting FGK Stars: Combining Kepler DR25, Gaia DR2 and Bayesian Inference}
\author{Danley C. Hsu}
\affiliation{Department of Astronomy \& Astrophysics, 525 Davey Laboratory, The Pennsylvania State University, University Park, PA, 16802, USA}
\affiliation{Center for Exoplanets and Habitable Worlds, 525 Davey Laboratory, The Pennsylvania State University, University Park, PA, 16802, USA}
\affiliation{Center for Astrostatistics, 525 Davey Laboratory, The Pennsylvania State University, University Park, PA, 16802, USA}
\affiliation{Institute for CyberScience, The Pennsylvania State University}
\author{Eric B. Ford}
\affiliation{Department of Astronomy \& Astrophysics, 525 Davey Laboratory, The Pennsylvania State University, University Park, PA, 16802, USA}
\affiliation{Center for Exoplanets and Habitable Worlds, 525 Davey Laboratory, The Pennsylvania State University, University Park, PA, 16802, USA}
\affiliation{Center for Astrostatistics, 525 Davey Laboratory, The Pennsylvania State University, University Park, PA, 16802, USA}
\affiliation{Institute for CyberScience, The Pennsylvania State University}
\author{Darin Ragozzine}
\affiliation{Department of Physics \& Astronomy, N283 ESC, Brigham Young University, Provo, UT 84602, USA}
\author{Keir Ashby}
\affiliation{Department of Physics \& Astronomy, N283 ESC, Brigham Young University, Provo, UT 84602, USA}

\begin{abstract}
We characterize the occurrence rate of planets, ranging in size from 0.5-16 R$_\oplus$, orbiting FGK stars with orbital periods from 0.5-500 days.  
Our analysis is based on results from the ``DR25'' catalog of planet candidates produced by NASA's Kepler mission and stellar radii from Gaia ``DR2''. 
We incorporate additional Kepler data products to accurately characterize the efficiency of planets being recognized as a ``threshold crossing events'' (TCE) by Kepler's Transiting Planet Search pipeline {\em and} labeled as a planet candidate by the robovetter.  
Using a hierarchical Bayesian model, we derive planet occurrence rates for a wide range of planet sizes and orbital periods.  
For planets with sizes $0.75-1.5$ R$_\oplus$ and orbital periods of 237-500 days, we find a rate of planets per FGK star of $<0.27$ ($84.13$th percentile).  
While the true rate of such planets could be lower by a factor of $\sim~2$
(primarily due to potential contamination of planet candidates by false alarms), 
the upper limits on the occurrence rate of such planets are robust to $\sim~10\%$.  
We recommend that mission concepts aiming to characterize potentially rocky planets in or near the habitable zone of sun-like stars prepare compelling science programs that would be robust for a true rate in the range $f_{R,P} = $ $0.03-0.40$ for $0.75-1.5$ R$_\oplus$ planets with orbital periods in 237-500 days, or a differential rate of $\Gamma_\oplus \equiv (d^2 f)/[d(\ln P)~d(\ln R_{p})] = $ $0.06-0.76$.

\end{abstract}


\section{Introduction}
\label{secIntro}
In 2009, NASA's Kepler mission began its mission to characterize the abundance and diversity of exoplanets. 
Four years of precise photometry of 189,516 target stars (as part of its exoplanet survey) allowed the Kepler project to identify 4,034 strong planet candidates via transit, over half of which have been confirmed by another method or validated based on detailed analyses of light curves and statistical considerations \citep{TCH+2018}.  
The Kepler mission was enormously successful, with an impressive array of discoveries, both anticipated and surprising.  
However, one of the primary goals of the mission, characterizing the frequency of Earth-size planets in or near the habitable zone of solar-like stars has remained elusive.  
A combination of factors, from intrinsic stellar variability to failure of Kepler's reaction wheels, resulted in the Kepler data being sensitive to such planets around a small fraction of the stars surveyed and the detection efficiency of such planets depending sensitively on their host star properties.  
As a result, considerable care is necessary when inferring the rate of small planets near the habitable zone from Kepler data.  

Several previous studies have estimated the rate of planets as a function of planet size and orbital period based on Kepler data.  Early studies had less data, but just as importantly, the accuracy of early studies were constrained by limited knowledge of the Kepler pipeline's detection efficiency and the host star properties.  
Further complicating matters, \citet{HFR+2018} showed that one of the most common algorithms for estimating the planet occurrence rate was systematically biased for planets near Kepler's detection threshold.  
While this bias is modest for much of the parameter space explored by Kepler, it can be of order unity for small planets near the habitable zone.  
Only a few recent studies analyzed the planet occurrence rates using a hierarchical Bayesian model, so as to avoid this bias \citep{FHM2014,BCM+2015,HFR+2018}.    
\citet{FHM2014} inferred planet occurrence rates using a hierarchical Bayesian model with non-parametric model and a prior that favored the planet occurrence rate changing smoothly.  This study analyzed a list of planet candidates from a non-standard pipeline that was restricted to finding a single planet around each star.  
\citet{BCM+2015} inferred planet occurrence rates using a parametric model (i.e., power-law in planet size and orbital period) and a hierarchical Bayesian model.  
\citet{HFR+2018} inferred planet occurrence rates using a non-parametric model, a hierarchical Bayesian model and Approximate Bayesian Computing (ABC), and performed extensive tests on simulated catalogs to validate the algorithm.  
However, none of these studies made use of Kepler's final planet catalog \citep[known as DR25;][]{TCH+2018}, additional DR25 data products characterizing the detection pipeline's efficiency \citep{C2017Pixel,BC2017Wf,BC2017Flux,BC2017DetCon,C2017Robo}, or recent improvements in the knowledge of host star properties from the ESA's Gaia mission \citep{GBV+2018}.  
Recent studies indicate that the updates in stellar properties thanks to Gaia can have a significant effect on the inferred planet occurrence rate \citep{SBT+2019}.

In this manuscript, we report occurrence rates inferred using a hierarchical Bayesian model and Approximate Bayesian Computing (ABC), and incorporating several improvements relative to previous studies, including the DR25 catalog of planet candidates, updates to the host star properties from Gaia, and an improved model for the Kepler efficiency for detecting and vetting of planet candidates.  
We review the statistical methodology in \S\ref{secABC} and 
describe improvements to our model since \citet{HFR+2018} in \S\ref{secModelImprov}.
In \S\ref{secCat}, we describe the selection of target stars, and present the resulting planet occurrence rates in \S\ref{secRates}.
We conclude with a summary of key findings, implications for the rate of Earth-size planets in or near the habitable zone of FGK stars, and future of occurrence rate studies in \S\ref{secDiscussion}.

\section{Methodology}
\label{secModel}

\subsection{Approximate Bayesian Computation}
\label{secABC}
Approximate Bayesian Computation (ABC) is a tool for performing Bayesian inference when the likelihood is not available.
For example, ABC can be applied to perform inference using a hierarchical Bayesian model to infer the properties of a population based on observations from a survey.  
Real-world surveys can have many complexities (e.g., measurement uncertainties, target selection, instrumental effects, data reduction pipeline, detection probability that depends on unobserved properties) that make it impractical to write down the correct likelihood.
Instead, ABC relies on: 1) the ability to generate samples from a forward-model for the intrinsic population and survey, and 2) a physically-motivated distance function that quantifies the distance between the actual survey results and the results of a simulated survey.    
For a sufficiently large number of simulated catalogs and a well-chosen distance function, the ABC posterior estimate converges around the true posterior. 
Thus, ABC is ideally suited to inferring planet occurrence rates based on transit surveys such as Kepler.  

Here we provide a qualitative description of naive ABC and the sequential importance sampling method applied in this study.  
The occurrence rate of planets within a given range of sizes and orbital periods is characterized by model and a set of hyperparameters.  
We apply a model that is a piece-wise constant rate per logarithmic bin in orbital period and planet size.  
The hyperparameters are simply the occurrence rates for each bin in the 2-d grid.
For this study, we chose a period-radius grid with similar period and radius bins: $P = $ \{0.5, 1, 2, 4, 8, 16, 32, 64, 128, 256, 500\} days \& $R_p = $ \{0.5, 0.75, 1, 1.25, 1.5, 1.75, 2, 2.5, 3, 4, 6, 8, 12, 16\} R$_{\oplus}$.  
We repeatedly draw planets from this model, simulate observations by the Kepler mission, compute a set of summary statistics (in this case the number of detected planets in each bin) and compare the number of planets (within a certain range of planet sizes and/or orbital periods) that would have been detected in the simulated dataset to the actual number of such planets detected by the Kepler mission.  
A naive application of ABC would involve drawing hyperparameters from their prior distribution and keeping parameters that resulted in matching the observed number of planets in each bin exactly.  
While straightforward, this approach is computationally unfeasible for our application.  

The implementation of ABC used in this study is Population Monte Carlo, wherein multiple generations of simulated data are created.  
In each generation, a population of model parameters are used to generate and evaluate multiple simulated catalogs. 
Simulated catalogs with a distance less than some threshold (``$\epsilon$'') are kept and the values of their hyperparameters inform the values of hyperparameters proposed for the following generation.
Each set of hyperparameter values is assigned a weight, and sequential importance sampling is used to ensure that the weights properly account for the prior probability distribution at each generation.  
This sequential importance sampling continues until one of several stopping criteria are met which indicate sufficient convergence has been achieved.  
For a full description of ABC and the particulars of the ABC-PMC algorithm we use, see \citet{HFR+2018}.

While we generally apply the ABC procedure detailed in \citet{HFR+2018}, there were a few modifications to account for this study.  
In this study, each generation of ABC-PMC consists of 100 to 500 ``particles'' (increased from 50) where each particle is a simulated catalog produced from a draw of occurrence rates from importance sampling. 
We found an increased number of particles to be helpful when applying the algorithm to infer several occurrence rates simultaneously, particularly when some of the rates were weakly constrained by the data (see \S\ref{secMultiRates}).
Instead of always using the same number of target stars and value of $\tau$ (which controls the width of the importance sampler's proposal distribution), we split the ABC simulation into three steps.  
For the first 15 generations of each ABC simulation, we create simulated catalogs with only $N_{*} = 1000$ targets and set $\tau = 1.5$.   
This accelerates the ABC-PMC algorithm as it ``zooms in'' on the relevant region of parameter space.  
For future generations, we set $\tau=2$ and generate larger simulated catalogs of $N_{*} = 10,000$ targets, so as to provide a good importance sampling density for future generations.
The ABC-PMC algorithm continues for more generations until satisfying the stopping criteria described in \citet{HFR+2018}.
In the third step, if the simulation was using less than 200 particles, we run one more generation of ABC using 200 particles drawn using the distance goal ($\epsilon$) of the last generation to provide a larger posterior sample for plotting and computing percentiles.
To prevent to selection of an outlier as the reported result, we perform five ABC-PMC simulations for each set of parameters and report the median value for each quantile.

\subsection{Model Improvements}
\label{secModelImprov}
The majority of the physical model in the Exoplanet System Simulator (known as ``SysSim'') used in this study has remained the same as in our previous application of ABC to estimate the occurrence rates of planet candidates around FGK stars based on the Q1-Q16 Kepler catalogs in \citet{HFR+2018} (see Section 3).
SysSim \citep{ExoplanetsSysSim2019} is available publicly on Github.  The version used in this study which implements the older v0.6 release of the programming language Julia \citep{Julia} is available at \url{https://github.com/dch216/ExoplanetsSysSim.jl/tree/hsu_etal_2019-v1.0} while a version updated to be compatible with the current release of Julia is available at \url{https://github.com/ExoJulia/ExoplanetsSysSim.jl/releases/tag/v1.0.0}, with both repositories under the MIT Expat License.
Below we summarize several improvements made for this study. 

\subsubsection{Stellar Properties}
\label{secImproveStars}
The second Gaia data release (DR2) \citep{GBV+2018} provides significantly improved stellar parameters for the overwhelming majority of the Kepler target stars.  
Improvements in stellar radii determinations result in more accurate planet radii (and more accurate assigning of planet candidates to their appropriate radii bins).
For this study, we use the stellar radius from Gaia DR2 \citep{AFC+2018}. For comparing observations to simulations, we use the median transit depths computed from the MCMC posterior chains provided with the Kepler DR25 catalog \citep{H2017MCMC}  (see \S\ref{secDR25Prod}: MCMC Transit Depths).  Note that this differs from the values listed in the DR25 catalog which are based on a maximum likelihood fit.  For assigning each planet to a radius bin, we also use planet-star radius ratios which we derived from the MCMC posterior chains provided along with the DR 25, accounting for limb-darkening, plus the transit depth and impact parameter of each state in the Markov chain.
An initial cross-match was performed using \url{https://github.com/megbedell/gaia-kepler.fun}.
We filter this target list further to remove poor planet search targets, as described in \S\ref{secCat}.  

Note that we do not report results which use the stellar radii derived by \citet{BHG+2018}, as that catalog makes use of additional observations (e.g., spectroscopy, asteroseismology) that are not available for all stars.
Comparing the stellar radii values for the same targets reveals that the Berger catalog generally reports larger values than Gaia DR2 for stars with radii $R_* < 0.75 R_{\odot}$  and vice versa at larger radii.
Additionally, the Berger catalog reports a much narrower spread of radii uncertainties than the Gaia catalog, although the Gaia catalog has smaller uncertainties on average.
Following-up on a suggestion from the anonymous referee, we repeated our ABC simulations making use of stellar properties from the Berger catalog instead of Gaia DR2.  Typically, we found very similar results, but there were some modest differences in regions of the  period-radius grid which have the most planet candidates and thus best measurement precision.
While incorporating additional observational constraints (as done by the Berger catalog) likely improves the precision for individual stars, using such an inhomogeneous set of stellar parameters risks biasing our results since high-resolution spectroscopic observations are available for nearly all stars with known planet candidates, but only a small fraction of stars without any planet candidates.

Additionally, Gaia's precise parallax measurements allow for more accurate identification of main-sequence stars based on their position in the color-luminosity diagram.
We use this information (in place of estimates based on $\log g$) to define a clean sample of FGK main-sequence target stars.  
Finally, we make use of Gaia's astrometric information to identify targets likely to be multiple star systems with stars of similar masses/luminosities.  
The details of our target selection are described in \S\ref{secCat}.
This results in a total of $79,935$ Kepler targets for which the Kepler DR25 pipeline and robovetter identified $2,525$ planet candidates with $P=0.5-500$d and $R_p = 0.5-16 R_{\oplus}$.  

\subsubsection{Incorporating Kepler DR 25 data products}
\label{secDR25Prod}
{\em Planet Catalog:}  
This study makes use of the Kepler DR 25 planet catalog and its associated data products  \citep{TCH+2018}.  
This represents a substantial improvement over the Q1-16 catalog used in \citet{HFR+2018}.
The DR25 catalog makes use of the light curves from the full Kepler prime mission and incorporates several improvements to the data reduction, planet search, and vetting algorithms.  Even more importantly, DR25 was the first Kepler catalog to be generated using a uniform data reduction, planet search and vetting process.   For planet occurrence studies, uniformity of processing is particularly important, as it enables statistical modeling of the detection process.  

{\em MCMC Transit Depths:}
We replace the catalog maximum-likelihood estimator value of transit depth with the median value of the transit depth posterior, which was determined by taking the DR25 MCMC posterior chains \citep{H2017MCMC} for each planet candidate and applying the quadratic limb-darkening law from \citet{MA2002}.  The median depth value is then used to determine the planet-star radius ratio and therefore planet radius in a consistent manner with the simulated planets.

{\em Planet Detection Efficiency:}  
When computing occurrence rates, it is important to use an accurate detection efficiency model for the probability that a planet with known properties is included in the result catalog of planet candidates.  
\citet{HFR+2018} used a detection efficiency model from \cite{CCB+2015} that was calibrated to transit injection tests performed using a previous pipeline.  
In this study, we update the planet detection efficiency model based on transit injection tests based on the DR 25 pipeline \cite{C2017Pixel}.  
Initially, we adopted the detection probability for a transiting planet as a function of the expected effective signal-to-noise ratio (SNR), 
\begin{equation}
p_{\mathrm{det,DR25}}(SNR) = c \times \gamma(\alpha,\beta\times SNR)/\Gamma(\alpha),
\label{eqnDetC2017}
\end{equation}
where $\gamma$ is the incomplete gamma function, $\Gamma$ is the gamma function, $\alpha=30.87$, $\beta=0.271$, and $c=0.94$, based on \S5 of \citet{C2017Pixel}.
This expression models the probability that a transiting planet is detected by the pipeline as a threshold crossing event, but does not attempt to model the probability that a threshold crossing event is promoted to a Kepler Object of Interest or a planet candidate.  
The Kepler pipeline identifies threshold crossing events for further investigation based on a ``multiple event statistic'' or MES exceeding a threshold of 7.1.  The MES is defined to be what the pipeline returns and can not be computed exactly without applying the Kepler pipeline to a lightcurve.  As described later in this section, our definition of expected SNR represents the best available approximation of the expected MES in the absence of applying the Kepler pipeline to a light curve.  
\citet{C2017Pixel} suggest that occurrence rate studies fit custom detection efficiency models that are optimized for the specific stellar sample and range of orbital periods being studied.  While most previously published occurrence rate studies have ignored this advice, we fit a custom detection efficiency model to the \citet{C2017Pixel} simulations for this study's baseline model (described below).  We explicitly give Eqn. \ref{eqnDetC2017}, because we also report results using Eqn. \ref{eqnDetC2017} so as to facilitate comparisons to previous studies and our baseline model.

Most previous studies have simply assumed that all true planets that are identified as threshold crossing events survive the vetting process.  
\citet{MPA+2018} presented a first attempt to estimate the impact of the vetting process. 
In this study, we consider three models for the combined planet detection process (i.e., being identified as a threshold crossing event by the pipeline and labeled as a planet candidate by the robovetter).  
First, we adopt $p_{\mathrm{det,DR25}}$ from Eqn. \ref{eqnDetC2017} and assume perfect vetting (i.e., $p_{\mathrm{vet}}=1$).
Second, we adopt the product of $p_{\mathrm{det,DR25}}$ and the \citet{MPA+2018} model for vetting efficiency,
\begin{equation}
p_{\mathrm{vet,Mulders}}(R_{p},P) = cR_{p}^{a_{R}}
\left\{
\begin{array}{ll}
\left(P/P_\mathrm{break}\right)^{a_P},~\mathrm{if~}P < P_{\mathrm{break}}\\
\left(P/P_\mathrm{break}\right)^{b_P},~\mathrm{otherwise}
\end{array}
\right.,
\label{eqnVetMulders}
\end{equation}
where $R_p$ is planet radius (in R$_\oplus$), $P$ is the orbital period (in days), 
$c = 0.93, a_R = -0.03, P_\mathrm{break} = 205, a_P = 0.00$ and $b_P = -0.24$\footnote{These parameter values differ from the values reported in \citet{MPA+2018} which were based on fitting the same functional form to planet candidates with a vetting score of $\ge~0.9$ and using the DR25 stellar parameters.  This study uses updated parameters provided by G. Mulders that are updated to use Gaia DR2 stellar parameters and make use of all Kepler DR 25 KOIs that are labeled as a planet candidate by the robovetter, regardless of the vetting score.}.
Note that this model implies a vetting efficiency that does not explicitly depend on the effective transit SNR, but rather has only an indirect dependence via the planet size and orbital period.  
Since the power of the vetting process depends primarily on the ability of the robovetter to features in the shape of the light curve, it is likely that the vetting efficiency indeed depends more directly on the effective measurement noise for the target star.  
Another concern with this model is that it assumes the probability of a planet being detected by the pipeline and a planet passing vetting are uncorrelated.  
Since both the pipeline detection efficiency and vetting efficiency depend on the transit light curve (and thus key properties such as the transit signal to noise and the number of transits observed), assuming that these are independent seems unwise.  
These (as well as results described in \S\ref{secVV}) motivated us to create a new model of the planet detection and vetting process.

For the third model, we derive a combined model for the probability of a transiting planet being detected and labeled as a planet candidate by the robovetter using the pixel-level transit injection tests \citep{C2017Pixel} and the associated robovetter results \citep{C2017Robo}.  
\begin{equation}
p_{\mathrm{det\&vet}}(SNR,N_{tr}) = c_{N_{tr}} \times \gamma(\alpha_{N_{tr}},\beta_{N_{tr}} \times SNR)/\Gamma(\alpha_{N_{tr}}),
\label{eqnDetAndVet}
\end{equation}
where $N_{tr}$ is the number of ``valid'' transits observed by Kepler and values for $\alpha$, $\beta$, and $c$ are given in Table \ref{tab:detvetmodel}.  
(The Kepler photometric reduction pipeline assigned weights to each flux measurement, so as to deweight observations that may be spurious for a variety of reasons (e.g., measurement near a data gap or anomaly). Following \citet{C2017Pixel}, a transit is labeled as ``valid'' if the central flux measurement during the transit has a weight greater than 0.5.)  

We follow the general procedure recommended in \cite{C2017Pixel} for fitting a probability as a function of the effective SNR.  However, we only consider an injected transit to have been detected if it also is labeled as a planet candidate by the robovetter.
The choice of $N_{tr}$ as the second model parameter was motivated by the observation that the vetting efficiency decreases for planets with longer orbital periods and thus smaller $N_{tr}$.  Theoretically, we expect that the dispersion in the measured SNR will depend directly on the number of transits that are contributing to the candidate and only indirectly on orbital period.  For example, if two planets have a similar orbital period, but fewer transits of one planet were observed (e.g., due to falling on the CCD module that died early in the Kepler mission), then the distribution of effective SNR would be broader for the planet with fewer observed transits. 
We adopt this as our baseline model and compare the results using these three models for the combined planet detection process in \S\ref{secVet}.  

\begin{deluxetable*}{rrrr}
\tablewidth{0pt}
\tablecaption{Parameters for $p_{\mathrm{det \& vet}}$
\label{tab:detvetmodel}}
\tablehead{
\colhead{$N_{tr}$}&
\colhead{$\alpha$}&
\colhead{$\beta$}&
\colhead{$c$}
}
\startdata
 3       & 33.3884& 0.264472& 0.699093\\
 4       & 32.8860& 0.269577& 0.768366\\
 5       & 31.5196& 0.282741& 0.833673\\
 6       & 30.9919& 0.286979& 0.859865\\
 7-9     & 30.1906& 0.294688& 0.875042\\
 10-18   & 31.6342& 0.279425& 0.886144\\
 19-36   & 32.6448& 0.268898& 0.889724\\
 $\ge37$ & 27.8185& 0.32432 & 0.945075\\
\enddata
\tablecomments{We model the probability that a transiting planet is detected by the Kepler DR25 pipeline and labeled as a planet candidate by the robovetter with Eqn. \ref{eqnDetAndVet} and these parameters.  
For large SNR the detection/vetting probability asymptotes at $c$.  The detection probability increases most rapidly at an expected effective SNR of $(\alpha-1)\beta$.  The rate of increase in the detection probability is characterized by $\alpha/\beta^2$. 
For comparison, \citep{C2017Pixel} reported a fit to the DR25 detection efficiency (not including vetting) of $\alpha=30.87$, $\beta=0.271$, and $c=0.94$.  
$p_{\mathrm{det \& vet}}$ and Eqn.\ \ref{eqnDetC2017} are similar for large $N_{tr}$, but deviate significantly for small $N_{tr}$.  
}

\end{deluxetable*}

{\em $1-\sigma$ Depth Function:}  Each of the above planet detection models depends on computing the expected effective SNR.  
In \citet{HFR+2018}, we used:
\begin{equation}
    SNR = d*\sqrt{N_{tr}*D} / \mathrm{CDPP}
\end{equation} where the numerator represents the ``signal'', $N_{tr}$ is the number of transits, $D$ is the transit duration, and $d$ is the fractional transit depth.  For ``noise'', we took the mission average of the 4.5 hour duration combined differential photometric precision (CDPP) for each target star and assumed that the fractional noise scaled with one over the square root of the number of measurements.

In this study, we make use of a new data product provided by DR25, the ``1-$\sigma$ depth function'' (OSDF), which describes the true transit depth that would be expected to result in a MES of unity for a given target star, orbital period, and transit duration, after averaging over the epoch of transit \citep{BC2017Wf}.  
Thus, the expected MES and effective SNR become simply the transit depth divided by the OSDF for the target star interpolated to the orbital period and appropriate transit duration. The CDPP is effectively an averaged version of the OSDF that assumes a specific transit duration and averages over orbital period and phase. Thus, we expect the OSDF to be a more accurate measure of the effective noise as it accounts for how the photometric noise deviates from white noise.  For planets with a modest number of transits, the SNR can be affected by whether they occur at times with more or less noise than typical.   This is particularly significant for planets with orbital periods greater than 90 days, since there is a smaller number of transits and the number of transits observed with one CCD module may differ significantly from the number of transits observed with another CCD module.  

The original Kepler DR25 OSDFs are defined for every star, all 15 pulse durations that were searched by the pipeline, and for each of a a large ($\sim10^4$) number of orbital periods \citep{BC2017Wf}.  In order to reduce the memory requirements, we downsample these through linear interpolation to a common period grid for all stars of 1000 logarithmically-spaced periods from 0.49 to 700 days, which captures the vast majority of the information in these OSDFs.\footnote{The OSDFs for each star are available from the Kepler mission at \url{https://exoplanetarchive.ipac.caltech.edu/docs/Kepler_completeness_reliability.html}. Our downsampled versions are in a more convenient format and available on GitHub at \url{https://github.com/ExoJulia/SysSimData}.}  Though the amount of memory required for SysSim to use OSDFs is much higher than before, it's run-time performance did not increase significantly. 

When evaluating the noise of a particular planet, we typically use bi-linear interpolation between the period and duration. However, OSDFs are not defined for all periods and durations because the Kepler Transit Search pipeline did not search every combination of period and duration. When the simulated planet's duration was shorter than the shortest duration searched (and available in OSDF), we assume that the SNR was diluted by the ratio of the searched  duration to the actual duration since the search pipeline effectively adds noise to such short duration transits. 

When the simulation planet's duration was longer than the longest duration searched at that period, we adopt the noise of the longest duration searched.  The SNR is reduced due to the signal only accumulating for the duration searched rather than the true duration.  

The ``timeout'' corrections for which stars and durations were searched in DR25 have not been incorporated to our model. 
This affects only a small fraction of our target stars, as discussed in \S\ref{secPipelineTimeouts}.

{\em Window Function:}  
For the Kepler pipeline to detect a planet candidate, there must be good data for at least three transits. 
This becomes non-trivial for long period planets and/or stars not observed for the full mission duration.
Following \citep{BCM+2015}, \citet{HFR+2018} adopted a binomial model for the probability that at least three transits were observed, based the expected number of transits if there were no data gaps and the target-specific duty cycle.

In this study, we make use of the DR25 Window Function data products which tabulate the probability of detecting at least three transits as a function of orbital period \citep{BC2017Wf}. 
As with OSDFs, the provided Kepler DR25 Window Functions are a function of star, duration, and period. 
These were downsampled to 1000 linearly-spaced periods from 0.5 to 700 days using linear interpolation.\footnote{The original versions are also at \url{https://exoplanetarchive.ipac.caltech.edu/docs/Kepler_completeness_reliability.html} and our downsampled versions are available on GitHub at \url{https://github.com/ExoJulia/SysSimData}.}
For a simulated planet, SysSim uses bilinear interpolation on these window functions to determine the value of the window function for each simulated planet. 
As in \cite{HFR+2018}, the SNR-detection probability is multiplied by this value to return the final probability of detecting a planet. 

Since window functions are identical for targets that were observed in the same sequence of quarters, we identify the 100 most common window functions for  targets that were observed for at least 4 quarters as part of the Exoplanet target list (determined by requiring "EX" to be in the Investigation ID as recorded on MAST (\url{archive.stsci.edu/kepler}). 
These 100 window functions provide an exact match for practically all of our targets.  
Any targets for which the exact window function was not available were assigned a randomly drawn window function. 

{\em Minor improvements affecting detection efficiency:}  
We also incorporate several minor model improvements when calculating the effective signal to noise.  
For example, \citet{HFR+2018} ignored limb-darkening.  
In this study, we account for limb-darkening when computing the transit depth, using the limb-darkening parameters from the Kepler DR 25 stellar catalog, the same values as used for the Kepler pipeline run that produced the DR25 planet candidate catalog. 
(Note: These limb-darkening parameters are based off of Kepler DR25 stellar parameters, so as to be consistent with DR25 data products such as the OSDF.)
Additionally, we account for dilution of transit depth due to contamination as tabulated in the Kepler Input Catalog. Finally, we have improved the calculation of the transit duration, by using  Eqn.\ 15 of \citet{K2010}, so as to avoid making the small angle approximation.

\subsection{Sky-averaged detection probabilities using CORBITS}
\label{secCORBITS}
In order to improve the wall-time efficiency of our ABC calculation, we have implemented a probabilistic simulated catalog approach.  
Previously, when simulating observations to determine the detection probability of each simulated planet, we assigned a geometric transit probability of either zero or unity depending on whether the planet would appear to transit for a single observer orientation.  
In this study, we assign each potentially detectable planet a weight proportional to its transit detection probability marginalized over all possible observer orientations.
The geometric transit probability for each planet is multiplied by a transit detection efficiency that is marginalized over all impact parameters (that result in a transit) to give the total detection probability.  
For a single planet, the sky-averaged geometric transit probability for each planet is $R_\star/[a(1-e^2)]$, where $R_\star$ is the stellar radius, $a$ is the semi-major axis and $e$ is the orbital eccentricity.  

For stars with multiple planets, SysSim calculates the sky-averaged geometric transit probability for each planet, as well as each pair of planets and each number of transiting planets in the system.
The CORBITS package \citep{BR2016} makes this computationally practical by using semi-analytic methods to determine the transit probability of each combination of planets associated with a given planetary system.
\footnote{CORBITS requires that orbits not ``cross'', i.e., if $a_1<a_2$, then it requires that $a_1\times(1+e_1) < a_2\times(1-e_2)$.
Physically, such configurations are unlikely due to long-term orbital stability.  
Therefore, when multiple planets are assigned to a single star, we reject configurations that result in crossing orbits and redraw until a configuration without crossing orbits is created (up to a maximum of 20 times).
}
Using the sky-averaged detection probability instead of the detection probability for a single viewing geometry reduces the number of simulated planetary systems needed to accurately infer planet occurrence rates.  
As part of our model and algorithm verification, we determined that the number of planetary systems drawn could be reduced by a factor of $\simeq~8$ relative to the number of target stars without significantly affecting the accuracy of the inferred occurrence rates.  
This reduction offsets the overhead from CORBITS, which increases the calculation time from single observer by a factor of $\simeq~6$. 
Therefore, the overall reduction is about one-third off the previous calculation time.
While this provides only a modest performance improvement for this study, we anticipate that the sky-averaged observing geometry mode will be particularly valuable for future studies that investigate the occurrence rate of systems with multiple transiting planets.  

In \citet{HFR+2018}, the distance function used by ABC was simply the absolute value of the difference in the ratios of detected planets to stars surveyed for the simulated and observed planet catalogs.  
Since this study averages over viewing geometries, simply summing the detection probabilities would result in the expected number of planet detections which is not directly comparable with the observed number of planets of the actual Kepler data.  
The expected number of planet detections has a smaller variance than the actual number of planet detections.  
Therefore, after calculating the detection probabilities for each planet in a simulated catalog, we label each planet as either detected or not-detected based on a Bernoulli draw with success probability equal to each planet's total detection probability. 

\subsection{Inferring multiple occurrence rates simultaneously}
\label{secMultiRates}
In \citet{HFR+2018}, we inferred the occurrence rate for each ``bin'' in period and radius independently of other bins.  
Orbital periods are measured so precisely that there is effectively no ambiguity about which period bin a planet should be assigned to.  
In principle, three effects could cause a planet to be assigned to a different bin than it would be if its properties were known perfectly.  
First, measurement errors cause the observed planet-star radius ratio to differ from it's true value.  
Fortunately, this happens for only a small fraction of planets.
Indeed, \citet{HFR+2018} found that this effect did not result in significant correlation between neighboring planet radii bins.  
Second, uncertainties in the stellar properties cause the measured planet radius to differ from the true planet radius, even if the planet-star radius ratio were known precisely.  
Nearly all previously published studies, including \citet{HFR+2018}, have neglected this effect.
\citet{SBT+2019} showed that uncertainties in host star radii could have a significant effect on planet occurrence rates.  
In this study, we infer multiple occurrence rates simultaneously, so as to account for this effect.
A third source of uncertainty (unrecognized contamination) is  discussed in \ref{secFutureContam}, but is not modeled in this study.  

Once Gaia provided accurate parallaxes for most Kepler target stars \citep{GBV+2018}, accurate stellar radii and uncertainties were derived for those targets \citep{AFC+2018}. 
This makes it feasible to accurately model the effects of uncertainty in stellar radii on planet occurrence rates.  
Since planet radii are proportional to the host star radius,  uncertainty in host star radius directly translate into uncertainty in the planet radius.
Even with Gaia DR2 parallaxes, the uncertainty in stellar radius can result in a planet being assigned to the wrong radius bin.  
For each planet we simulate, the true transit depth is based on the true planet radius and the stellar radius from Gaia \citep{AFC+2018}. 
We draw an assumed stellar radius from an equal mixture of two half-normal distributions with median equal to the Gaia best-fit radius and use Gaia's upper and lower uncertainties for the standard deviations of the half-normals \citep{AFC+2018}.  
If the uncertainty in stellar radius (in either direction) is less than $6\%$ the value of the radius, then we increase that uncertainty to $6\%$ the stellar radius (based on the distribution of stellar radius uncertainties in \citet{BHG+2018}).  
Then, we draw the observed transit depth centered on the true transit depth with a width based on its SNR and the diagonal noise model of \citet{PR2014} that accounts for finite integration time.  
Next, we compute the ``observed'' planet radius from the observed plant-star radius ratio and the assumed stellar radius.
In most cases, the observed planet radius results in the planet being assigned to the same bin as it would be if its radius were known precisely.
However, sometimes, the observed planet radius results in the planet being assigned to a neighboring bin containing slightly larger or smaller planets.  
If one were to include the uncertainty in stellar properties, but only modeled planets drawn from a single bin, then the inferred planet occurrence rate would be biased to be larger than the true rate, since simulated planets can ``leak'' beyond of the boundaries of the radius bin.
The magnitude of the effect depends on the size of the bin in radius relative to the size of uncertainties in stellar properties and on the differences between occurrence rates in neighboring bins. 
In tests, we found that this effect could lead to occurrence rates being inflated by $\sim~20\%$ of the measured rate (for bins focusing on small planets with a width of 0.25 R$_\oplus$), if each bin were analyzed individually, rather than in a group.

\subsubsection{Model Parameterization}
\label{secImproveParam}

Fortunately, our ABC approach naturally accounts for this effect, as long as we simultaneously model the occurrence rate of planets in each size bin.  
When performing inference with occurrence rates for multiple radius bins, we performed calculations using two sets of priors, each with their own model parameterization.  

First, we assign each bin in period and radius for the entire grid its own occurrence rate, with a uniform prior.  
We adopt a uniform prior for $f_{i,j}$ over $[0,f_{\mathrm{\max,i,j}})$.
The upper limit was set to $f_{\mathrm{\max,i,j}}=C\times \log(P_{\max,j}/P_{\min,j})/\log(2)\times \log(R_{\max,i}/R_{\min,i})/\log(2)$, with $C = 2$.
This upper limit is small enough that proposals with more than 3 planets per factor of 2 in period are extremely rare, consistent with expectations based on long-term orbital stability.
This has the advantage that the prior for any individual bin is easy to understand.  
However, it has the disadvantage that the total number of planets within a range of orbital periods could be larger than plausible if one considers long-term orbital stability.

When deriving our best estimate of planet occurrence rates in the $\eta_{\oplus}$ regime, we use a different parameterization.
For each bin in orbital period (indexed by $j$), we infer: 1) $f_{\mathrm{tot},j}$, the total occurrence rate for all planets within the range of orbital periods and the full range of planet sizes being considered, and 2) a vector $f_{\mathrm{rel},i,j}$: the fraction of planets in the $j$th orbital period bin that have sizes falling within the range of the $i$th radius bin.
Thus, the occurrence rate for planets in the $i$th radius bin and $j$th period bin is given by $f_{i,j} = f_{\mathrm{tot,j}}f_{\mathrm{rel},i,j}$.
We adopt a uniform prior for $f_{\mathrm{tot,j}}$ over $[0,f_{\mathrm{\max,tot,j}})$.
The upper limit $f_{\mathrm{\max,tot,j}}=3\times \log(P_{\max,j}/P_{\min,j})/\log(2)$ is motivated by long-term orbital stability, as it is very rare for more than 3 planets to have orbital periods within a factor of 2 of each other (though a few cases are known to exist for very small planets). 
For each $f_{\mathrm{rel},i,j}$ (a vector with $j$ fixed), we adopt a Dirichlet prior with concentration parameters proportional to $\log(R_{\max,i}/R_{\min,i})$.  The smallest concentration parameter is set to unity.  The Dirichlet prior ensures that $\sum_i f_{\mathrm{rel},i,j} = 1$ to within numerical precision.
This prior has the advantage that the rate of planets summed over all sizes within one period bin is constrained.  
Since the Dirichlet prior over the fractions in each radius bin makes it more challenging to visualize the prior, we report results and show posterior distributions for both sets of priors in \S\ref{secMultiRates}.

For most periods and sizes, the choice of prior did not have a significant impact on the posterior for occurrence rates.  
The differences were more noticeable for small, long-period planets.  
Therefore, we will report results for both choices of prior in the $\eta_\oplus$ regime.

\subsubsection{Distance Function}
\label{secImproveDistance}
Now that we allow for uncertainty in stellar properties, we revisit the choice of distance function.  
We performed tests on simulated data to verify our algorithms and to compare the performance of multiple distance functions.  
Simply summing the absolute value of the difference in the ratios of detected planets to stars surveyed for the simulated and observed planet catalogs often resulted in an ABC posterior with highly disparate widths for different bins.  
Since bins with a greater number of observed planets have a greater contribution to the total distance, ABC would result in precise occurrence rates for well-populated bins, but much less precise estimates of occurrence rates for bins with low rates of planet candidates.  
For many scientific purposes, one would prefer a similar fractional error in occurrence rates for all bins, rather than a similar absolute error in occurrence rates for all bins.
Therefore, using the Canberra distance \citep{LW1967} as a foundation
we choose a new distance function that weights the absolute value of the difference in the ratio of detected planets to targets for each bin by the square root of the sum of the ratio of detected planets to targets for the simulated and observed planet catalogs for that bin.  
Our distance is given by
\begin{equation}
    \rho (s_{\mathrm{obs},k},s^*_k) = \sum_k \frac{|s_{\mathrm{obs},k} - s^*_k|}{\sqrt{s_{\mathrm{obs},k} + s^*_k}},
\end{equation}
where, for each $k$th period-radius bin, $s_{\mathrm{obs}, k}$ is the ratio of number of planet candidates detected by Kepler to the number of target stars searched and $s^*_k$ is ratio of the number of planets detected to target stars in the simulated catalog.  
We found that this distance function allows ABC to converge more rapidly when inferring occurrence rates for multiple bins simultaneously. 
We tested that this algorithm accurately simulates both the number of planets detected and the dispersion in the number of planets detected for the simulated catalogs.
We also tested that this distance function accurately estimates the true planet occurrence rate and its uncertainty (by comparing to simulations that do not make use of sky-averaged detection probabilities; see \S\ref{secVV} and Figure \ref{figSkySingle}).

\subsubsection{Importance Sampler Proposal Distribution}
\label{secImproveProposal}
Initially, we attempted to apply the sequential importance sampler in the ABC-PMC algorithm using a multivariate Gaussian for the proposal distribution, as described in \citet{BCM+2008} and \citet{HFR+2018}.
However, this proved computationally prohibitive.  
For example, a proposal of a negative value for any one bin would result in rejecting the entire proposal; 
when considering several bins that have small occurrence rates, such proposals are common.  
In order to make the ABC-PMC algorithm practical, we experimented with a variety of replacement proposal distributions.  
Eventually, we settled on using a Beta distribution for the proposal for both parameterizations.
When using independent priors for each $f_{i,j}$, the ABC-PMC algorithm uses a Beta proposal distribution for each $f_{i,j}$.
When using a Dirichlet prior for the relative occurrence rates within a given period bin, the ABC-PMC algorithm using a Beta proposal for $f_{\mathrm{tot},j}$ and each $f_{\mathrm{rel},i,j}/f_{\mathrm{\max,tot,j}}$.  
After each proposal, the initially proposed values for the relative rates were rescaled according to $f_{\mathrm{rel},i,j} \leftarrow f_{\mathrm{rel},i,j} / \sum_i f_{\mathrm{rel},i,j}$, so that the rescaled rates sum to unity.  
We validated the algorithm on simulated data sets and found that this dramatically improved the computational efficiency.

\subsection{Verification and Validation}
\label{secVV}

Before applying our improved model and ABC algorithm to actual Kepler catalogs, we performed extensive tests to validate and and verify the model, the algorithms, and our implementations.  

\begin{figure}
\centering
\includegraphics[scale=0.28]{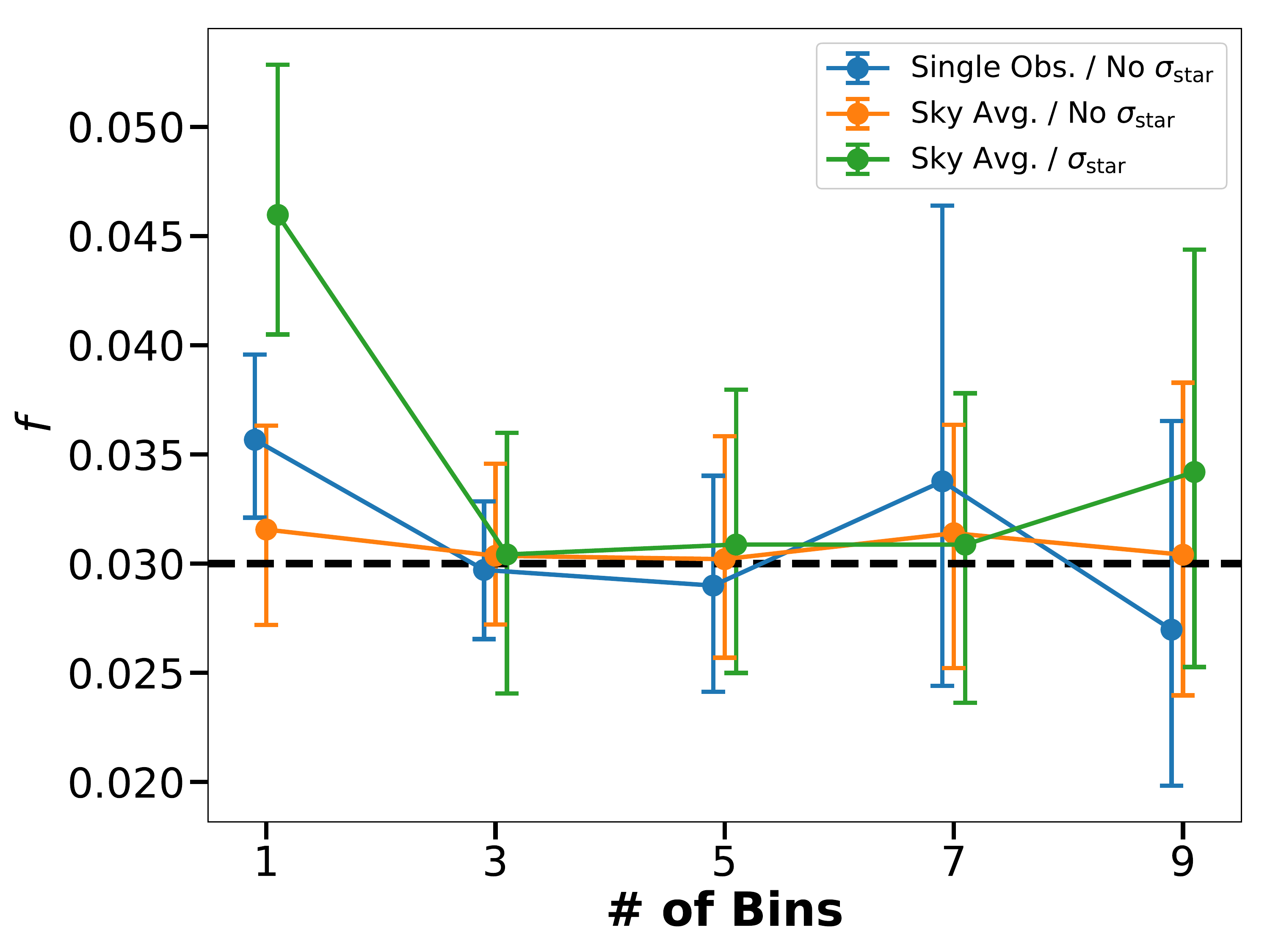}
\caption{Occurrence rate estimates for the 8-16d period, 1.25-1.5 R$_{\oplus}$ planet radius bin using a simulated ''true`` catalog.
This catalog was created with the rates $f = $ \{$0.07$, $0.06$, $0.05$, $0.04$, $0.03$, $0.02$, $0.01$, $0.005$, $0.0025$\} for the bins $R_p = $ \{$0.25-0.5$, $0.5-0.75$, $0.75-1$, $1-1.25$, $1.25-1.5$, $1.5-1.75$, $1.75-2$, $2-3$, $3-4$\} R$_\oplus$.  
For each run, 0, 1, 2, 3, and 4 neighboring radii bins on each side of the $8-16$d period, $1.25-1.5$ R$_{\oplus}$ planet radius bin are simultaneously fit for a total number of bins of 1, 3, 5, 7, and 9 bins.
In the case of 9 bins, the full range of bins used in the creation of the catalog are fit, while for runs with fewer bins the outer two bins were consecutively removed until the appropriate number of bins is reached.
The dashed black line indicates the true rate for the $1.25-1.5$ R$_{\oplus}$ bin ($f = 3\%$).
Estimates based on a single observing geometry (blue) and sky-averaged viewing geometries (orange) are consistent.
Excluding (orange) or including (green) stellar uncertainties does not affect the median estimated rate significantly, except when only fitting one bin (i.e., left-most points).
However, the uncertainty in the estimate of the occurrence rate increases when uncertainties in the stellar radius are included in the model.  
The median estimated rate does not significantly vary with increasing number of bins, but the width of the ABC-posterior does increase with the number of bins inferred simultaneously.
}
\label{figSkySingle}
\end{figure}

Figure \ref{figSkySingle} shows a comparison of results using either the sky-averaging mode (described in \S\ref{secCORBITS}) or the single observing geometry mode for one particular bin (orbital periods of 8-16 days and planet sizes of 1.25-1.5R$_\oplus$) where the ''true`` catalog is set to $f = 3\%$.  
The blue (orange) points show the median planet occurrence rate inferred for this bin using a single observing geometry (sky-averaged viewing geometry) and the vertical error bars indicate the 68.3\% credible region.
The horizontal axis indicates the number of occurrence rates inferred simultaneously.  
For any single simulated catalog, the number of observed planets is likely greater (or less) than the expected number of observed planets, leading to the posterior median rate being larger (or smaller) than the true rate ($f = 3\%$).  
We performed several simulations to verify that the median posterior rate from multiple runs was distributed about the true value.  
Figure \ref{figSkySingle} shows the results of multiple analyses for a single catalog demonstrating that we recover a posterior median close to the true value.

First, we will focus on results for three bins.  
Note that SysSim yields very similar results, regardless of whether using a single observing geometry (blue) or when using averaging over all viewing geometries (orange).  
As expected, including stellar radii uncertainties (green) results in an increased width of posterior distribution relative to an analysis which assumes stellar radii are known perfectly (orange).  

Next, we compare results as a function of the number of occurrence rates inferred simultaneously.  
Each method (single observing geometry or sky-averaged viewing geometries, excluding or including uncertainties in stellar properties) gives statistically indistinguishable results for the median occurrence rate, as long as the number of model parameters being inferred simultaneously does not exceed seven.  
When analyzing three or five radius bins simultaneously, the posterior widths are very similar.  
In either case, the posterior width is observed to increase modestly when the model accounts for uncertainties in stellar radii.  
When analyzing nine radius bins simultaneously, the posterior width has increased noticeably.  
This behavior is typical of the sequential importance sampler used by the ABC-PMC algorithm.  
When performing inference on nine parameters simultaneously, it becomes less likely that the proposed values for all nine parameters will result in a precise match for the number of observed planets in all nine bins.  
As a result, the computational time required for ABC-PMC to achieve a given distance increases significantly.
In practice, the final distance achieved by the ABC-PMC algorithm begins to increase substantially as the number of dimensions increases beyond seven, resulting in an increased width of the ABC-posterior.  
Thus, we must balance the desire to minimize bias with the desire to minimize unnecessary increase in posterior width due to computational limitations.  
Minimizing bias leads us to perform inference on at least three bins simultaneously.  
Avoiding an unnecessary increase in posterior width limits us to performing inference on $\le~7$ bins simultaneously.

Next, we performed several tests to further fine tune the choice of how many occurrence rates to infer simultaneously.
When inferring occurrence rates for multiple radius bins simultaneously, we find that only the top and bottom bins display noticeable bias.  
This is expected since the model effectively assumes that there are no detectable planets with radius greater than the upper limit of the top bin or less than the lower limit of the bottom bin. 
Since the bias is proportional to the difference between the true and assumed rate of such planets, it is maximized when the assumed rate is zero.  
When inferring occurrence rates for just three bins simultaneously, one could worry that the estimated rates for both edge bins will be overestimated and this could have an indirect effect on the estimated rate for the central bin. 
In practice, the bias for the central bin is a second-order effect and significantly smaller than the other uncertainties.
A related effect occurs when two bins (neighboring in radii) have a large difference in occurrence rate.
If the ``leaking'' of planets out-of and in-to the radius bin in question is not symmetric, then the impact of stellar uncertainties is amplified compared to other sets of bins where the occurrence rate is similar above and below the bin in question.  
This demonstrates that it is important to model the uncertainties in stellar radii and to simultaneously infer the planet occurrence rates when estimating planet occurrence rates for a wide range of planet radii and orbital periods.  

Based on the above tests, we settled on reporting results using simulations that simultaneously infer planet occurrence rates for 5 (or 7) radius bins, reporting the results for the interior 3 (or 5) radius bins.  
(One exception is that we also report the occurrence rate for our largest bin, 12-16 R$_\oplus$, as explained below.)  
In order to span the full range of planet radii and to avoid significantly overestimating the ABC-posterior width, we perform multiple simulations, each using 5 to 7 radius bins.  
The specific bin boundaries chosen for the subsets were: 
$R_p =$ \{0.25, 0.5, 0.75, 1, 1.25, 1.5\}, \{0.5, 0.75, 1, 1.25, 1.5, 1.75, 2, 2.5\}, \{1, 1.25, 1.5, 1.75, 2, 2.5\}, \{1, 2, 2.5, 3, 4, 6\}, \{3, 4, 6, 8, 12, 16\} R$_{\oplus}$ for each period range.

Formally, one should be cautious in using results from an ``edge bin''.  
Therefore, for each subset we throw out the two edge bins and take estimates for only interior bins from each subset.  
In the case of bins which are interior to multiple subsets, we combine the final populations produced by the separate runs on each subset to produce the occurrence rate estimates.
The one edge bin which we do not throw out is the 12-16 R$_\oplus$ bin.
Even though the 12-16 R$_\oplus$ bin is an edge, the rate of planets larger than 16 R$_\oplus$ is so small that there is no practical impact on the measured occurrence rates for 12-16 R$_\oplus$ planets.  
While we include occurrence rates for $0.25-0.5$ R$_\oplus$ planets in our model calculations, we do so primarily so as to ensure that our estimates for 0.5-0.75 R$_\oplus$ planets are accurate.  
Thus, we do not include the inferred rates for $0.25-0.5$ R$_\oplus$ planets in Figure \ref{figRates} out of an abundance of caution.  

\section{Application to DR25}
\label{secResults}
\subsection{Catalog Selection}
\label{secCat}
The second Gaia data release (DR2) \citep{GBV+2018} provides significantly improved stellar parameters for the overwhelming majority of the Kepler target stars.  
Improvements in stellar radii determinations result in more accurate planet radii and more accurate assigning of planet candidates to their appropriate radii bins.
Additionally, Gaia's precise parallax measurements allow for more accurate identification of main-sequence stars based on their position in the color-luminosity diagram.
We use this information (in place of estimates of $\log g$) to define a clean sample of FGK main-sequence FGK target stars.  
Finally, we make use of Gaia's astrometric information to identify targets likely to be multiple star systems with stars of similar masses/luminosities.  

We construct a catalog of Kepler target stars starting from the Kepler DR25 stellar properties catalog and augmented with data from Gaia DR2 \citep{GBV+2018}.
The selection criteria used to select Kepler targets likely to be main-sequence FGK stars are as follows (applied in order):
\begin{enumerate}

\item \textbf{Require that the Kepler magnitude and the Gaia G magnitude are consistent.}
We compute the median and standard deviation of the difference in the Kepler magnitude and Gaia G magnitude based on the initial cross-matched catalog based on position.  If a target's Kepler magnitude minus Gaia G magnitude differs from the median by more than 1.5 times the standard deviation, then it is excluded from our cleaned catalog.  
This filter makes sure that a Kepler target is not erroneously matched to a Gaia target corresponding to a background star.
\item \textbf{Require Gaia GOF\_AL $\leq 20$ and Gaia astrometric excess noise $\leq 5$ cut.}  These criteria indicate a poor astrometric fit.  This is often caused by an unresolved astrometric binary star \citep{E2018}.  
\item \textbf{Require Gaia Priam processing flags that indicate the parallax value is strictly positive and both colors are close to the standard locus for main-sequence stars}.  
This rejects Kepler targets which are unlikely FGK main sequence stars or are so distant that Gaia does not have a good parallax measurement and the stellar radius will be highly uncertain.  
\item \textbf{Require Gaia parallax error is less than $10\%$ the parallax value}.
This excludes some faint/distant Kepler targets for which accurate stellar radii would not be available.  
Of all our target selection criteria based on data quality this results in the largest decrease in the number of target stars and planet candidates.  
In principle, future observations (e.g., improved Gaia data releases) could increase the number of stars with accurate stellar properties, perhaps allowing for more precise planet occurrence rate measurements.   
However, for FGK main-sequence stars, this criterion only excludes distant and hence faint targets for which Kepler does not have significant sensitivity to small planets in the habitable zone.  
Therefore, we expect that even future Gaia data releases and other follow-up observations to improve stellar characterization will not lead to significant improvements in the constraints on the occurrence rate of small planets in or near the habitable zone of main-sequence FGK stars.
\item \textbf{Require that Kepler DR 25 provide a valid Kepler stellar mass, data span, duty cycle, and limb-darkening coefficients, and that Gaia's Apsis module provide a valid stellar radius based on Gaia data} \citep{AFC+2018}.
\item \textbf{Require that the Kepler target was observed for $>4$ quarters and must have been on the Exoplanet target list for at least one quarter}.
These criteria exclude targets that were observed by Kepler for purposes other than the exoplanet search, as these are unlikely to be FGK main sequence stars.  
A small number of stars that were part of the exoplanet search will also be excluded if they were only observed for $\le~4$ quarters.  Kepler does not have significant sensitivity to small planets in the habitable zone of such FGK stars.
\item \textbf{Require the target to have a color $0.5 \leq B_p - R_p \leq 1.7$ from Gaia photometry}.  
This color cut results in selecting FGK stars and is more precise than using the temperature from the Kepler Input Catalog and more uniform than using temperatures from the DR25 stellar catalog.
\item \textbf{Require the target star to have a luminosity, $L \leq 1.75 L_{MS}(B_p - R_p)$, where $L_{MS}(x) = 10^{2.62-3.74x+0.962x^2}$ represents the luminosity of a main-sequence star for a given $B_p - R_p$ color}.
To compute $L_{MS}(B_p - R_p)$ we fit a quadratic function to the observed values of $\log_{10} L$ as a function of $B_p - R_p$ for our Kepler targets.  We reject those targets whose observed $L$ deviates from $L_{MS}$ by more than 75\% and refit the model to the remaining targets.  We iterate this process six times before converging on our final model for the main-sequence luminosity $L_{MS}$.  
This rejects targets that are significantly above the main sequence, either due to stellar evolution (primarily for F stars) or due to a multiple star system where stars other than the primary contribute >75\% of the total flux.  
While this final criteria likely removes some viable targets for planet hunting (e.g., F stars starting to evolve off the main sequence), it is a small fraction of the total stars searched and Kepler does not have significant sensitivity to small planets near their habitable zone.  
\end{enumerate}
\begin{figure*}
\centering
\includegraphics[scale=0.5]{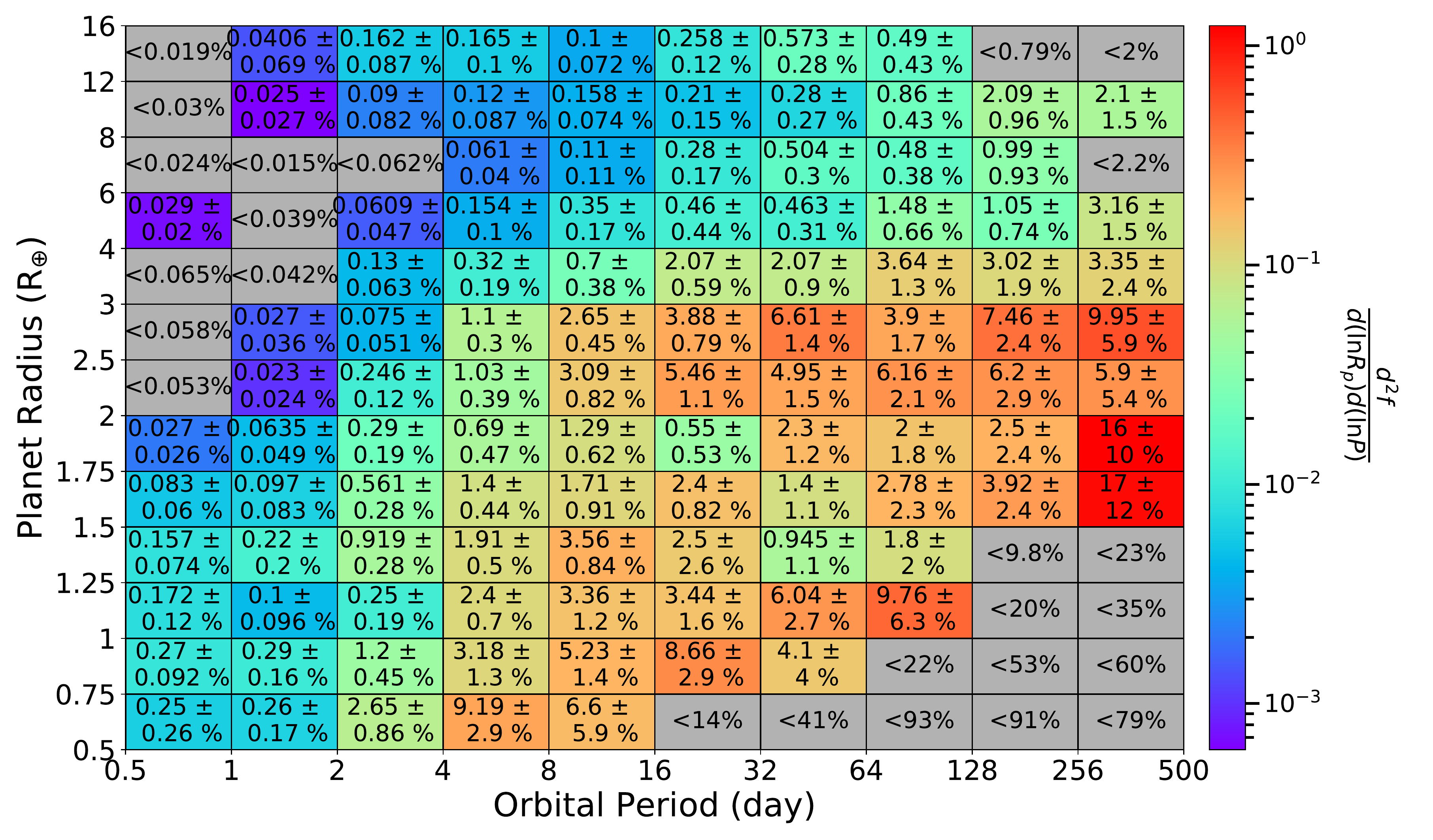}
\caption{Inferred occurrence rates for Kepler's DR25 planet candidates associated with high-quality FGK target stars.  
These rares are based on a combined detection and vetting efficiency model that was fit to flux-level planet injection tests.  
The numerical values of the occurrence rates are stated as percentage (i.e. $10^{-2}$). 
The color coding of each cell is based on $(d^{2}f)/[d (\ln{R_{p}})~d(\ln{P})]$, which provides an occurrence rate normalized to the width of the bin and therefore is not dependent on choice of grid density.
The uncertainties shown are the differences between the median and either the 15.87th or 84.13th percentile (whichever has the larger absolute difference).
Cells colored gray have estimated upper limits for the occurrence rate.
Note that the bin sizes are not constant.
}
\label{figRates}
\end{figure*}

The total number of FGK Kepler target stars remaining after these cuts is $79,935$.  
While the number of targets is significantly less than the total number observed by Kepler, a clean sample of target stars is preferable for performing planet occurrence rate studies.  
In principle, a less restrictive set of cuts might provide a larger stellar sample and provide more precise estimates for occurrence rates of large planets.  
However, such a strategy is unlikely to be useful for measuring the occurrence rate of small planets near the habitable zone, since the vast majority of stars excluded were not FGK main-sequence stars for which Kepler had significant sensitivity to small planets in or near the habitable zone.  
Our selection criterion intentionally exclude M stars, which will be the subject of a future study.

The Kepler DR25 pipeline and robovetter identified $2,524$ planet candidates associated with these targets with $P=0.5-500$d and $R_p = 0.5-16 R_{\oplus}$.  
One noticeable change from our previous study is the relatively few planet candidates in the $\eta_{\oplus}$ regime.  
The primary cause for this change is the usage of updated stellar radii from Gaia, which are often larger than the previously determined Kepler stellar radii.  
This trend tends to boost the inferred planet radii.
As a result, the majority of long period, small radii planet candidates shifted to larger radii bins within the period-radius grid once we incorporated Gaia stellar radii.
For example, the inferred radius KOI 7016.01\citep[also known as Kepler-452 b]{JTB+2015,MTC+2018} reported in DR25 is  $1.06^{+0.2}_{-0.1}$ R$_\oplus$, but incorporating the updated stellar radii from Gaia DR results in the estimated radius increasing to 1.51 R$_\oplus$ (with uncertainties increasing proportionally).  While the best-estimate for the planetary radius no longer falls within the $1-1.25$ R$_\oplus$ bin, the uncertainty in the radius of such planets results in it to contributing to the estimated occurrence rate.  

In order to explore which $\eta_{\oplus}$ regime planet candidates were removed due to our stellar cuts (as opposed to updated stellar radii), we created a separate target list where we relax the two stellar cuts which most significantly reduced the number of stars in our sample:  the FGK luminosity cut and the cuts designed to remove suspected binaries.  
This catalog has a total number of $139,232$ target stars, with $3,170$ planet candidates within the limits of the period-radius grid.  
Most significantly, we recover two long-period, small radius planet candidates (associated with targets KIC 5097856 and 5098334) in the $P=256-500$d, $R_p = 1.25-1.5 R_{\oplus}$ bin.  
After investigating the properties of these two planet candidates, we determined that the candidates were not in our final catalog because their associated target stars had poor astrometric GOF which suggests that their host star is likely part of an unresolved binary.  
If true, then the unmodeled flux from the binary companion would be diluting the transit depth, causing the true planet radius to be larger than currently estimated. 
We conclude that our process for identifying a clean sample of target stars worked as intended.  
\begin{figure}
\centering
\includegraphics[scale=0.28]{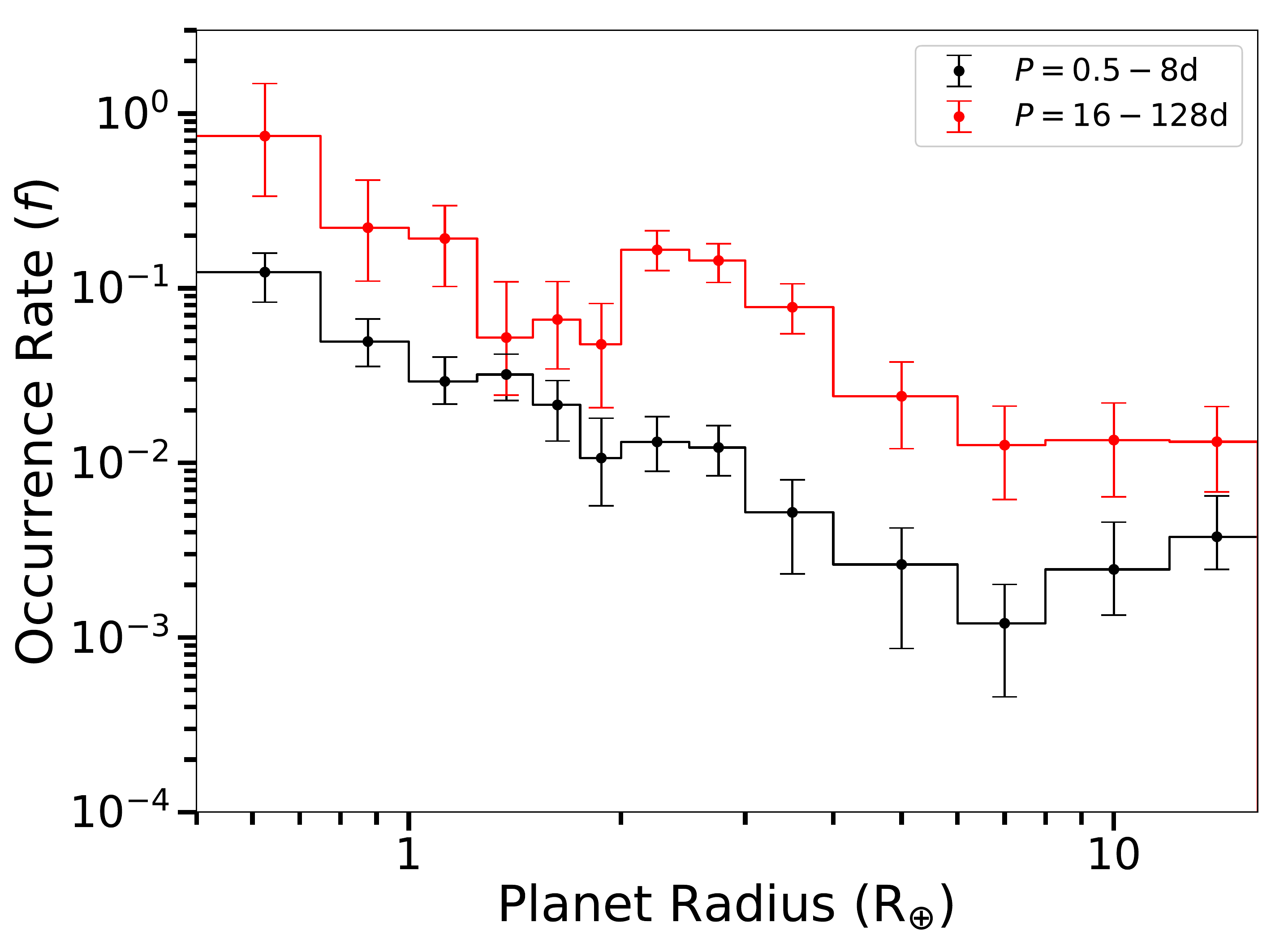}
\caption{Occurrence rate estimates marginalized over two period ranges: $0.5-8$d (black) and $16-128$d (red).  
}
\label{figInteg}
\end{figure}

\subsection{Baseline Planet Occurrence Rates}
\label{secRates}

The values in each bin of Figure \ref{figRates} state the median of the ABC-posterior for the occurrence rates over the full period-radius grid.  
Results are tabulated in Table \ref{tab:occ_rates}.
The uncertainties shown are the differences between the median and either the 15.87th or 84.13th percentile (whichever has the larger absolute difference) of the ABC-posterior for each occurrence rate.
These rates make use of the combined detection and vetting efficiency (Eqn. \ref{eqnDetAndVet}).
For period-radius bins with zero or one observed planet candidates, we report the 84.13th percentile as an upper limit instead of a median rate and uncertainty.  
For these bins, the posterior distribution is asymmetrical and summarizing the posterior with a point estimate runs a greater risk of misinterpretation.  
Figure \ref{figPost} depicts the ABC posterior for the occurrence rate for several long period, small radius bins to demonstrate this asymmetry.  

The reported upper limits are conservative in the sense that a full calculation (i.e., inferring rate for all radius bins simultaneously within a period range, rather than just 5 at a time) would likely cause a modest reduction in the upper limit. 
This is because we set a maximum total rate of 3 per factor of 2 in period (as a stability criterion), so a full calculation which adds well-constrained bins at larger radii would reduce the remaining total rate that can be allotted to the bins with only upper limits.
Given the occurrence rates inferred, the effect is negligible for most of the parameter space considered. While a few effects (e.g., unmodeled contaminating flux, detection efficiency decreasing for systems with multiple transiting planets, pipeline timeouts) could cause our model to overestimate the detection efficiency for individual planetary systems, each of these affect only a small fraction of stars (see \S\ref{secLimitations}).  
Therefore, we expect that any revisions to our upper limits due to these effects would be limited to a few percent.  
However, for bins with so little constraining data the estimated posterior may be significantly affected by the choice of prior (see \S\ref{secEtaEarth}). 

In Figure \ref{figInteg}, we show the occurrence rate as a function of planet size, marginalized over two separate period ranges, 0.5-8d and 16-128d.  
The planet radius valley \citet[e.g.,][]{FPH+2017,VAL+2017} is apparent for periods of 16-128d, but not for periods less than 8 days.  
Our use of a cleaned target star sample is designed to reduce effects due to
contamination from binary stars, similar to \citet{TeCH+2018}.  
The depth and width of the valley are expected to depend on orbital period on theoretical grounds \citep{LF2014,OW2017,GS2018}.  
A narrow and deep valley whose location depends on period would appear to be broader and shallower when one marginalizes over a broad range of periods, as in this figure.   
Therefore, the depth of the valley shown in Figure \ref{figInteg} should be interpreted as a minimum depth when comparing to predictions of model predictions over a narrower range of periods.   
Comparing the rate of 1.25-1.5 R$_\oplus$ planets as a function of orbital periods, we find no evidence that the occurrence rate is decreasing beyond 64 days, as predicted by some models.  

\startlongtable
\begin{deluxetable*}{rrrr}
\tablewidth{0pt}
\tablecaption{Inferred Planet Occurrence Rates
\label{tab:occ_rates}}
\tablehead{
\colhead{Period}&
\colhead{Radius}&
\colhead{Combined Detection \&}&
\colhead{No Vetting}\\
\colhead{(days)}&
\colhead{(R$_{\oplus}$)}&
\colhead{Vetting Efficiency}&
\colhead{Efficiency}}

\startdata
$\phn\phn 0.50-\phn\phn 1.00$&$\phn0.50-\phn0.75$&$2.5^{+2.6}_{-1.7}\times10^{-3}$&$1.71^{+1.34}_{-0.58}\times10^{-3}$\\
$\phn\phn 0.50-\phn\phn 1.00$&$\phn0.75-\phn1.00$&$2.70^{+0.80}_{-0.92}\times10^{-3}$&$2.5^{+1.2}_{-1.0}\times10^{-3}$\\
$\phn\phn 0.50-\phn\phn 1.00$&$\phn1.00-\phn1.25$&$1.7^{+1.2}_{-1.1}\times10^{-3}$&$1.93^{+1.15}_{-0.74}\times10^{-3}$\\
$\phn\phn 0.50-\phn\phn 1.00$&$\phn1.25-\phn1.50$&$1.57^{+0.74}_{-0.69}\times10^{-3}$&$1.73^{+0.87}_{-1.07}\times10^{-3}$\\
$\phn\phn 0.50-\phn\phn 1.00$&$\phn1.50-\phn1.75$&$8.3^{+6.0}_{-5.3}\times10^{-4}$&$7.3^{+5.3}_{-4.0}\times10^{-4}$\\
$\phn\phn 0.50-\phn\phn 1.00$&$\phn1.75-\phn2.00$&$2.7^{+2.6}_{-1.9}\times10^{-4}$&$4.1^{+4.3}_{-2.1}\times10^{-4}$\\
$\phn\phn 0.50-\phn\phn 1.00$&$\phn2.00-\phn2.50$&$<5.3\times10^{-4}$&$<3.9\times10^{-4}$\\
$\phn\phn 0.50-\phn\phn 1.00$&$\phn2.50-\phn3.00$&$<5.8\times10^{-4}$&$<5.8\times10^{-4}$\\
$\phn\phn 0.50-\phn\phn 1.00$&$\phn3.00-\phn4.00$&$<6.5\times10^{-4}$&$<4.9\times10^{-4}$\\
$\phn\phn 0.50-\phn\phn 1.00$&$\phn4.00-\phn6.00$&$2.9^{+1.4}_{-2.0}\times10^{-4}$&$3.2^{+1.4}_{-1.3}\times10^{-4}$\\
$\phn\phn 0.50-\phn\phn 1.00$&$\phn6.00-\phn8.00$&$<2.4\times10^{-4}$&$<1.6\times10^{-4}$\\
$\phn\phn 0.50-\phn\phn 1.00$&$\phn8.00-12.00$&$<3.0\times10^{-4}$&$<1.8\times10^{-4}$\\
$\phn\phn 0.50-\phn\phn 1.00$&$12.00-16.00$&$<1.9\times10^{-4}$&$<2.4\times10^{-4}$\\
\hline
$\phn\phn 1.00-\phn\phn 2.00$&$\phn0.50-\phn0.75$&$2.6^{+1.4}_{-1.7}\times10^{-3}$&$2.7^{+2.2}_{-1.5}\times10^{-3}$\\
$\phn\phn 1.00-\phn\phn 2.00$&$\phn0.75-\phn1.00$&$2.9^{+1.6}_{-1.1}\times10^{-3}$&$2.49^{+0.93}_{-1.77}\times10^{-3}$\\
$\phn\phn 1.00-\phn\phn 2.00$&$\phn1.00-\phn1.25$&$1.03^{+0.96}_{-0.68}\times10^{-3}$&$1.02^{+0.65}_{-0.65}\times10^{-3}$\\
$\phn\phn 1.00-\phn\phn 2.00$&$\phn1.25-\phn1.50$&$2.2^{+2.0}_{-1.3}\times10^{-3}$&$2.7^{+3.1}_{-1.2}\times10^{-3}$\\
$\phn\phn 1.00-\phn\phn 2.00$&$\phn1.50-\phn1.75$&$9.7^{+6.7}_{-8.3}\times10^{-4}$&$6.4^{+8.4}_{-3.4}\times10^{-4}$\\
$\phn\phn 1.00-\phn\phn 2.00$&$\phn1.75-\phn2.00$&$6.4^{+4.9}_{-3.3}\times10^{-4}$&$6.1^{+3.3}_{-3.7}\times10^{-4}$\\
$\phn\phn 1.00-\phn\phn 2.00$&$\phn2.00-\phn2.50$&$2.3^{+2.4}_{-1.5}\times10^{-4}$&$2.9^{+2.5}_{-1.4}\times10^{-4}$\\
$\phn\phn 1.00-\phn\phn 2.00$&$\phn2.50-\phn3.00$&$2.7^{+3.6}_{-2.1}\times10^{-4}$&$2.2^{+4.1}_{-2.1}\times10^{-4}$\\
$\phn\phn 1.00-\phn\phn 2.00$&$\phn3.00-\phn4.00$&$<4.2\times10^{-4}$&$<3.9\times10^{-4}$\\
$\phn\phn 1.00-\phn\phn 2.00$&$\phn4.00-\phn6.00$&$<3.9\times10^{-4}$&$<3.8\times10^{-4}$\\
$\phn\phn 1.00-\phn\phn 2.00$&$\phn6.00-\phn8.00$&$<1.5\times10^{-4}$&$<2.6\times10^{-4}$\\
$\phn\phn 1.00-\phn\phn 2.00$&$\phn8.00-12.00$&$2.5^{+2.7}_{-1.0}\times10^{-4}$&$2.9^{+1.5}_{-1.8}\times10^{-4}$\\
$\phn\phn 1.00-\phn\phn 2.00$&$12.00-16.00$&$4.1^{+6.9}_{-1.2}\times10^{-4}$&$4.4^{+2.6}_{-3.1}\times10^{-4}$\\
\hline
$\phn\phn 2.00-\phn\phn 4.00$&$\phn0.50-\phn0.75$&$2.65^{+0.86}_{-0.77}\times10^{-2}$&$2.49^{+0.62}_{-0.75}\times10^{-2}$\\
$\phn\phn 2.00-\phn\phn 4.00$&$\phn0.75-\phn1.00$&$1.20^{+0.15}_{-0.45}\times10^{-2}$&$10.0^{+3.7}_{-1.7}\times10^{-3}$\\
$\phn\phn 2.00-\phn\phn 4.00$&$\phn1.00-\phn1.25$&$2.5^{+1.9}_{-1.3}\times10^{-3}$&$2.3^{+1.0}_{-1.4}\times10^{-3}$\\
$\phn\phn 2.00-\phn\phn 4.00$&$\phn1.25-\phn1.50$&$9.2^{+2.8}_{-2.3}\times10^{-3}$&$9.4^{+2.8}_{-2.3}\times10^{-3}$\\
$\phn\phn 2.00-\phn\phn 4.00$&$\phn1.50-\phn1.75$&$5.6^{+2.5}_{-2.8}\times10^{-3}$&$5.4^{+3.1}_{-2.4}\times10^{-3}$\\
$\phn\phn 2.00-\phn\phn 4.00$&$\phn1.75-\phn2.00$&$2.9^{+1.9}_{-1.2}\times10^{-3}$&$2.9^{+5.8}_{-1.5}\times10^{-3}$\\
$\phn\phn 2.00-\phn\phn 4.00$&$\phn2.00-\phn2.50$&$2.46^{+0.74}_{-1.19}\times10^{-3}$&$2.5^{+1.1}_{-1.6}\times10^{-3}$\\
$\phn\phn 2.00-\phn\phn 4.00$&$\phn2.50-\phn3.00$&$7.5^{+5.1}_{-4.7}\times10^{-4}$&$9.8^{+8.7}_{-5.8}\times10^{-4}$\\
$\phn\phn 2.00-\phn\phn 4.00$&$\phn3.00-\phn4.00$&$1.28^{+0.53}_{-0.63}\times10^{-3}$&$1.21^{+1.04}_{-0.61}\times10^{-3}$\\
$\phn\phn 2.00-\phn\phn 4.00$&$\phn4.00-\phn6.00$&$6.1^{+4.7}_{-4.1}\times10^{-4}$&$7.1^{+4.6}_{-4.1}\times10^{-4}$\\
$\phn\phn 2.00-\phn\phn 4.00$&$\phn6.00-\phn8.00$&$<6.2\times10^{-4}$&$<5.5\times10^{-4}$\\
$\phn\phn 2.00-\phn\phn 4.00$&$\phn8.00-12.00$&$9.0^{+8.2}_{-3.6}\times10^{-4}$&$8.4^{+7.6}_{-5.1}\times10^{-4}$\\
$\phn\phn 2.00-\phn\phn 4.00$&$12.00-16.00$&$1.62^{+0.87}_{-0.54}\times10^{-3}$&$1.84^{+0.65}_{-0.82}\times10^{-3}$\\
\hline
$\phn\phn 4.00-\phn\phn 8.00$&$\phn0.50-\phn0.75$&$9.2^{+2.2}_{-2.9}\times10^{-2}$&$7.6^{+3.1}_{-2.6}\times10^{-2}$\\
$\phn\phn 4.00-\phn\phn 8.00$&$\phn0.75-\phn1.00$&$3.18^{+1.32}_{-0.73}\times10^{-2}$&$2.90^{+0.77}_{-0.97}\times10^{-2}$\\
$\phn\phn 4.00-\phn\phn 8.00$&$\phn1.00-\phn1.25$&$2.40^{+0.70}_{-0.45}\times10^{-2}$&$2.26^{+0.66}_{-0.52}\times10^{-2}$\\
$\phn\phn 4.00-\phn\phn 8.00$&$\phn1.25-\phn1.50$&$1.91^{+0.43}_{-0.50}\times10^{-2}$&$1.93^{+0.70}_{-0.48}\times10^{-2}$\\
$\phn\phn 4.00-\phn\phn 8.00$&$\phn1.50-\phn1.75$&$1.40^{+0.44}_{-0.40}\times10^{-2}$&$1.38^{+0.40}_{-0.41}\times10^{-2}$\\
$\phn\phn 4.00-\phn\phn 8.00$&$\phn1.75-\phn2.00$&$6.9^{+4.7}_{-3.3}\times10^{-3}$&$7.0^{+5.2}_{-3.2}\times10^{-3}$\\
$\phn\phn 4.00-\phn\phn 8.00$&$\phn2.00-\phn2.50$&$1.03^{+0.39}_{-0.28}\times10^{-2}$&$1.07^{+0.30}_{-0.32}\times10^{-2}$\\
$\phn\phn 4.00-\phn\phn 8.00$&$\phn2.50-\phn3.00$&$1.10^{+0.28}_{-0.30}\times10^{-2}$&$1.10^{+0.27}_{-0.30}\times10^{-2}$\\
$\phn\phn 4.00-\phn\phn 8.00$&$\phn3.00-\phn4.00$&$3.2^{+1.9}_{-1.8}\times10^{-3}$&$3.6^{+2.2}_{-1.8}\times10^{-3}$\\
$\phn\phn 4.00-\phn\phn 8.00$&$\phn4.00-\phn6.00$&$1.54^{+0.80}_{-1.02}\times10^{-3}$&$1.49^{+0.76}_{-0.74}\times10^{-3}$\\
$\phn\phn 4.00-\phn\phn 8.00$&$\phn6.00-\phn8.00$&$6.1^{+4.0}_{-3.7}\times10^{-4}$&$5.7^{+5.6}_{-4.2}\times10^{-4}$\\
$\phn\phn 4.00-\phn\phn 8.00$&$\phn8.00-12.00$&$1.17^{+0.87}_{-0.54}\times10^{-3}$&$1.19^{+0.84}_{-0.60}\times10^{-3}$\\
$\phn\phn 4.00-\phn\phn 8.00$&$12.00-16.00$&$1.65^{+1.05}_{-0.59}\times10^{-3}$&$1.75^{+0.84}_{-0.71}\times10^{-3}$\\
\hline
$\phn\phn 8.00-\phn16.00$&$\phn0.50-\phn0.75$&$6.6^{+5.9}_{-3.3}\times10^{-2}$&$5.4^{+3.5}_{-2.9}\times10^{-2}$\\
$\phn\phn 8.00-\phn16.00$&$\phn0.75-\phn1.00$&$5.2^{+1.4}_{-1.3}\times10^{-2}$&$4.81^{+1.33}_{-0.77}\times10^{-2}$\\
$\phn\phn 8.00-\phn16.00$&$\phn1.00-\phn1.25$&$3.36^{+1.18}_{-0.94}\times10^{-2}$&$2.90^{+0.79}_{-0.48}\times10^{-2}$\\
$\phn\phn 8.00-\phn16.00$&$\phn1.25-\phn1.50$&$3.56^{+0.84}_{-0.74}\times10^{-2}$&$3.25^{+0.78}_{-0.92}\times10^{-2}$\\
$\phn\phn 8.00-\phn16.00$&$\phn1.50-\phn1.75$&$1.71^{+0.91}_{-0.62}\times10^{-2}$&$1.64^{+0.78}_{-0.55}\times10^{-2}$\\
$\phn\phn 8.00-\phn16.00$&$\phn1.75-\phn2.00$&$1.29^{+0.53}_{-0.62}\times10^{-2}$&$1.29^{+0.72}_{-0.59}\times10^{-2}$\\
$\phn\phn 8.00-\phn16.00$&$\phn2.00-\phn2.50$&$3.09^{+0.82}_{-0.80}\times10^{-2}$&$3.18^{+0.53}_{-0.56}\times10^{-2}$\\
$\phn\phn 8.00-\phn16.00$&$\phn2.50-\phn3.00$&$2.65^{+0.43}_{-0.45}\times10^{-2}$&$2.63^{+0.48}_{-0.53}\times10^{-2}$\\
$\phn\phn 8.00-\phn16.00$&$\phn3.00-\phn4.00$&$7.0^{+3.8}_{-3.2}\times10^{-3}$&$7.2^{+3.5}_{-3.8}\times10^{-3}$\\
$\phn\phn 8.00-\phn16.00$&$\phn4.00-\phn6.00$&$3.5^{+1.5}_{-1.7}\times10^{-3}$&$3.9^{+2.5}_{-1.7}\times10^{-3}$\\
$\phn\phn 8.00-\phn16.00$&$\phn6.00-\phn8.00$&$1.07^{+1.12}_{-0.70}\times10^{-3}$&$1.096^{+0.095}_{-0.450}\times10^{-3}$\\
$\phn\phn 8.00-\phn16.00$&$\phn8.00-12.00$&$1.58^{+0.74}_{-0.73}\times10^{-3}$&$1.45^{+1.59}_{-0.60}\times10^{-3}$\\
$\phn\phn 8.00-\phn16.00$&$12.00-16.00$&$1.02^{+0.72}_{-0.66}\times10^{-3}$&$9.7^{+6.4}_{-6.2}\times10^{-4}$\\
\hline
$\phn16.00-\phn32.00$&$\phn0.50-\phn0.75$&$<1.4\times10^{-1}$&$<1.1\times10^{-1}$\\
$\phn16.00-\phn32.00$&$\phn0.75-\phn1.00$&$8.7^{+2.5}_{-2.9}\times10^{-2}$&$7.2^{+2.0}_{-1.4}\times10^{-2}$\\
$\phn16.00-\phn32.00$&$\phn1.00-\phn1.25$&$3.4^{+1.5}_{-1.6}\times10^{-2}$&$3.0^{+1.7}_{-1.4}\times10^{-2}$\\
$\phn16.00-\phn32.00$&$\phn1.25-\phn1.50$&$2.47^{+2.57}_{-0.83}\times10^{-2}$&$2.52^{+0.74}_{-1.00}\times10^{-2}$\\
$\phn16.00-\phn32.00$&$\phn1.50-\phn1.75$&$2.40^{+0.82}_{-0.68}\times10^{-2}$&$2.28^{+0.30}_{-0.26}\times10^{-2}$\\
$\phn16.00-\phn32.00$&$\phn1.75-\phn2.00$&$5.5^{+5.3}_{-3.3}\times10^{-3}$&$5.1^{+5.4}_{-3.2}\times10^{-3}$\\
$\phn16.00-\phn32.00$&$\phn2.00-\phn2.50$&$5.46^{+1.11}_{-0.89}\times10^{-2}$&$5.4^{+1.0}_{-1.2}\times10^{-2}$\\
$\phn16.00-\phn32.00$&$\phn2.50-\phn3.00$&$3.88^{+0.79}_{-0.68}\times10^{-2}$&$3.80^{+1.06}_{-0.70}\times10^{-2}$\\
$\phn16.00-\phn32.00$&$\phn3.00-\phn4.00$&$2.07^{+0.59}_{-0.39}\times10^{-2}$&$2.14^{+0.60}_{-0.62}\times10^{-2}$\\
$\phn16.00-\phn32.00$&$\phn4.00-\phn6.00$&$4.6^{+4.4}_{-2.8}\times10^{-3}$&$3.4^{+2.8}_{-1.6}\times10^{-3}$\\
$\phn16.00-\phn32.00$&$\phn6.00-\phn8.00$&$2.8^{+1.7}_{-1.1}\times10^{-3}$&$2.6^{+1.7}_{-1.8}\times10^{-3}$\\
$\phn16.00-\phn32.00$&$\phn8.00-12.00$&$2.1^{+1.5}_{-1.1}\times10^{-3}$&$1.9^{+1.4}_{-1.0}\times10^{-3}$\\
$\phn16.00-\phn32.00$&$12.00-16.00$&$2.58^{+0.67}_{-1.23}\times10^{-3}$&$2.6^{+2.1}_{-1.6}\times10^{-3}$\\
\hline
$\phn32.00-\phn64.00$&$\phn0.50-\phn0.75$&$<4.1\times10^{-1}$&$<2.5\times10^{-1}$\\
$\phn32.00-\phn64.00$&$\phn0.75-\phn1.00$&$4.1^{+4.0}_{-2.3}\times10^{-2}$&$3.7^{+4.9}_{-2.5}\times10^{-2}$\\
$\phn32.00-\phn64.00$&$\phn1.00-\phn1.25$&$6.0^{+2.7}_{-2.5}\times10^{-2}$&$5.1^{+2.6}_{-1.9}\times10^{-2}$\\
$\phn32.00-\phn64.00$&$\phn1.25-\phn1.50$&$9.5^{+11.3}_{-6.3}\times10^{-3}$&$8.4^{+9.6}_{-5.5}\times10^{-3}$\\
$\phn32.00-\phn64.00$&$\phn1.50-\phn1.75$&$1.44^{+1.11}_{-0.70}\times10^{-2}$&$1.33^{+0.85}_{-0.90}\times10^{-2}$\\
$\phn32.00-\phn64.00$&$\phn1.75-\phn2.00$&$2.3^{+1.1}_{-1.2}\times10^{-2}$&$2.1^{+1.1}_{-1.2}\times10^{-2}$\\
$\phn32.00-\phn64.00$&$\phn2.00-\phn2.50$&$5.0^{+1.5}_{-1.2}\times10^{-2}$&$4.8^{+1.5}_{-1.2}\times10^{-2}$\\
$\phn32.00-\phn64.00$&$\phn2.50-\phn3.00$&$6.6^{+1.4}_{-1.3}\times10^{-2}$&$6.2^{+1.5}_{-1.3}\times10^{-2}$\\
$\phn32.00-\phn64.00$&$\phn3.00-\phn4.00$&$2.07^{+0.90}_{-0.76}\times10^{-2}$&$2.09^{+1.01}_{-0.79}\times10^{-2}$\\
$\phn32.00-\phn64.00$&$\phn4.00-\phn6.00$&$4.6^{+3.1}_{-2.6}\times10^{-3}$&$4.0^{+2.8}_{-2.4}\times10^{-3}$\\
$\phn32.00-\phn64.00$&$\phn6.00-\phn8.00$&$5.0^{+3.0}_{-2.5}\times10^{-3}$&$5.0^{+2.8}_{-2.3}\times10^{-3}$\\
$\phn32.00-\phn64.00$&$\phn8.00-12.00$&$2.8^{+2.7}_{-1.7}\times10^{-3}$&$2.7^{+2.4}_{-1.9}\times10^{-3}$\\
$\phn32.00-\phn64.00$&$12.00-16.00$&$5.7^{+2.8}_{-1.9}\times10^{-3}$&$5.2^{+3.1}_{-2.0}\times10^{-3}$\\
\hline
$\phn64.00-128.00$&$\phn0.50-\phn0.75$&$<9.3\times10^{-1}$&$<9.5\times10^{-1}$\\
$\phn64.00-128.00$&$\phn0.75-\phn1.00$&$<2.2\times10^{-1}$&$<1.7\times10^{-1}$\\
$\phn64.00-128.00$&$\phn1.00-\phn1.25$&$9.8^{+6.3}_{-4.9}\times10^{-2}$&$7.6^{+5.6}_{-3.5}\times10^{-2}$\\
$\phn64.00-128.00$&$\phn1.25-\phn1.50$&$1.8^{+2.0}_{-1.3}\times10^{-2}$&$1.24^{+1.75}_{-0.95}\times10^{-2}$\\
$\phn64.00-128.00$&$\phn1.50-\phn1.75$&$2.8^{+2.3}_{-1.8}\times10^{-2}$&$2.5^{+2.1}_{-1.4}\times10^{-2}$\\
$\phn64.00-128.00$&$\phn1.75-\phn2.00$&$2.0^{+1.8}_{-1.2}\times10^{-2}$&$1.8^{+1.4}_{-1.1}\times10^{-2}$\\
$\phn64.00-128.00$&$\phn2.00-\phn2.50$&$6.2^{+2.1}_{-1.9}\times10^{-2}$&$5.5^{+1.9}_{-1.6}\times10^{-2}$\\
$\phn64.00-128.00$&$\phn2.50-\phn3.00$&$3.9^{+1.4}_{-1.7}\times10^{-2}$&$3.7^{+2.1}_{-1.4}\times10^{-2}$\\
$\phn64.00-128.00$&$\phn3.00-\phn4.00$&$3.6^{+1.3}_{-1.1}\times10^{-2}$&$3.35^{+1.21}_{-0.89}\times10^{-2}$\\
$\phn64.00-128.00$&$\phn4.00-\phn6.00$&$1.48^{+0.62}_{-0.66}\times10^{-2}$&$1.40^{+0.68}_{-0.52}\times10^{-2}$\\
$\phn64.00-128.00$&$\phn6.00-\phn8.00$&$4.8^{+3.8}_{-2.9}\times10^{-3}$&$4.8^{+4.7}_{-2.5}\times10^{-3}$\\
$\phn64.00-128.00$&$\phn8.00-12.00$&$8.6^{+4.3}_{-4.3}\times10^{-3}$&$8.6^{+4.3}_{-3.5}\times10^{-3}$\\
$\phn64.00-128.00$&$12.00-16.00$&$4.9^{+4.3}_{-3.3}\times10^{-3}$&$4.2^{+4.3}_{-2.3}\times10^{-3}$\\
\hline
$128.00-256.00$&$\phn0.50-\phn0.75$&$<9.1\times10^{-1}$&$<9.4\times10^{-1}$\\
$128.00-256.00$&$\phn0.75-\phn1.00$&$<5.3\times10^{-1}$&$<4.5\times10^{-1}$\\
$128.00-256.00$&$\phn1.00-\phn1.25$&$<2.0\times10^{-1}$&$<1.9\times10^{-1}$\\
$128.00-256.00$&$\phn1.25-\phn1.50$&$<9.8\times10^{-2}$&$<6.4\times10^{-2}$\\
$128.00-256.00$&$\phn1.50-\phn1.75$&$3.9^{+2.0}_{-2.4}\times10^{-2}$&$3.2^{+2.4}_{-2.0}\times10^{-2}$\\
$128.00-256.00$&$\phn1.75-\phn2.00$&$2.5^{+2.4}_{-2.0}\times10^{-2}$&$1.9^{+2.1}_{-1.3}\times10^{-2}$\\
$128.00-256.00$&$\phn2.00-\phn2.50$&$6.2^{+2.6}_{-2.9}\times10^{-2}$&$5.5^{+2.8}_{-2.0}\times10^{-2}$\\
$128.00-256.00$&$\phn2.50-\phn3.00$&$7.5^{+2.4}_{-2.1}\times10^{-2}$&$6.5^{+2.1}_{-2.3}\times10^{-2}$\\
$128.00-256.00$&$\phn3.00-\phn4.00$&$3.0^{+1.9}_{-1.7}\times10^{-2}$&$2.8^{+1.6}_{-1.1}\times10^{-2}$\\
$128.00-256.00$&$\phn4.00-\phn6.00$&$1.05^{+0.74}_{-0.54}\times10^{-2}$&$9.6^{+8.1}_{-5.5}\times10^{-3}$\\
$128.00-256.00$&$\phn6.00-\phn8.00$&$9.9^{+9.3}_{-3.9}\times10^{-3}$&$9.7^{+5.5}_{-5.2}\times10^{-3}$\\
$128.00-256.00$&$\phn8.00-12.00$&$2.09^{+0.96}_{-0.80}\times10^{-2}$&$1.90^{+0.62}_{-0.65}\times10^{-2}$\\
$128.00-256.00$&$12.00-16.00$&$<7.9\times10^{-3}$&$<5.9\times10^{-3}$\\
\hline
$256.00-500.00$&$\phn0.50-\phn0.75$&$<7.9\times10^{-1}$&$<8.2\times10^{-1}$\\
$256.00-500.00$&$\phn0.75-\phn1.00$&$<6.0\times10^{-1}$&$<5.7\times10^{-1}$\\
$256.00-500.00$&$\phn1.00-\phn1.25$&$<3.5\times10^{-1}$&$<3.4\times10^{-1}$\\
$256.00-500.00$&$\phn1.25-\phn1.50$&$<2.3\times10^{-1}$&$<1.6\times10^{-1}$\\
$256.00-500.00$&$\phn1.50-\phn1.75$&$1.66^{+1.19}_{-0.96}\times10^{-1}$&$1.28^{+0.89}_{-0.75}\times10^{-1}$\\
$256.00-500.00$&$\phn1.75-\phn2.00$&$1.59^{+1.02}_{-0.83}\times10^{-1}$&$1.12^{+0.82}_{-0.66}\times10^{-1}$\\
$256.00-500.00$&$\phn2.00-\phn2.50$&$5.9^{+5.4}_{-3.6}\times10^{-2}$&$4.1^{+3.8}_{-2.7}\times10^{-2}$\\
$256.00-500.00$&$\phn2.50-\phn3.00$&$10.0^{+5.9}_{-3.8}\times10^{-2}$&$7.6^{+3.3}_{-3.9}\times10^{-2}$\\
$256.00-500.00$&$\phn3.00-\phn4.00$&$3.4^{+2.4}_{-2.0}\times10^{-2}$&$2.7^{+2.4}_{-1.7}\times10^{-2}$\\
$256.00-500.00$&$\phn4.00-\phn6.00$&$3.2^{+1.5}_{-1.5}\times10^{-2}$&$2.6^{+2.1}_{-1.5}\times10^{-2}$\\
$256.00-500.00$&$\phn6.00-\phn8.00$&$<2.2\times10^{-2}$&$<1.7\times10^{-2}$\\
$256.00-500.00$&$\phn8.00-12.00$&$2.1^{+1.5}_{-1.1}\times10^{-2}$&$1.62^{+1.07}_{-0.70}\times10^{-2}$\\
$256.00-500.00$&$12.00-16.00$&$<2.0\times10^{-2}$&$<1.5\times10^{-2}$\\
\enddata

\tablecomments{Estimated occurrence rates for DR25 KOI catalog planet candidates associated with FGK stars using two different vetting efficiency schemes.}
\end{deluxetable*}

\section{Discussion}
\label{secDiscussion}
Using Kepler DR25, Gaia DR2, SysSim and ABC model, we provide accurate estimates of the exoplanet occurrence rate around FGK main sequence stars as a function of planet size and orbital period.
Our results span a wide range of orbital periods and planet radii, including the $\eta_\oplus$ regime, without the need to extrapolate.  
Our Bayesian approach provides direct and meaningful constraints on the $\eta_\oplus$ regime, despite the limited number of such planet candidates (see \S\ref{secEtaEarth}).  
We find planet occurrence rates somewhat lower than some previous studies \citep[e.g., Figure 17 of ][]{BCM+2015}.  
Just as importantly, we derive statistically valid uncertainties in planet occurrence rates, while accounting for many effects that have often been neglected in previous studies.
The slightly lower rate and larger uncertainty may have significant implications for scientists planning future planet surveys or characterization missions.  

We performed extensive tests of our model and examined the relative importance of various upgrades to the model.
One of the most significant factors resulting in differences with the Q1-Q16 rates presented in \citet{HFR+2018} is the usage of Gaia DR2 stellar parameters.
In addition to improving the accuracy of planet radii, the Gaia data enabled the selection of a clean sample of FGK main sequence target stars for occurrence rate studies.
The number of small planets at long orbital periods was significantly reduced after focusing on a clean FGK target star sample and incorporating improved stellar radii.  
This is one factor that contributes to our inferred occurrence rates being lower than some previous studies.

We find that it is important to account for the vetting efficiency when characterizing the occurrence rate of small planets or planets with orbital periods greater than a month  (see \S\ref{secVet}).  
We provide an improved model for the combined detection efficiency (i.e., including both being identified by the Kepler planet search pipeline and labeled as a planet candidate by the robovetter) in Eqn. \ref{eqnDetAndVet}.  
We conclude with a discussion of the areas for further improvement in future occurrence rate studies (\S\ref{secFutureRates}) and implications for mission planning (\S\ref{secRecommendations}).

\begin{figure}
\centering
\includegraphics[scale=0.28]{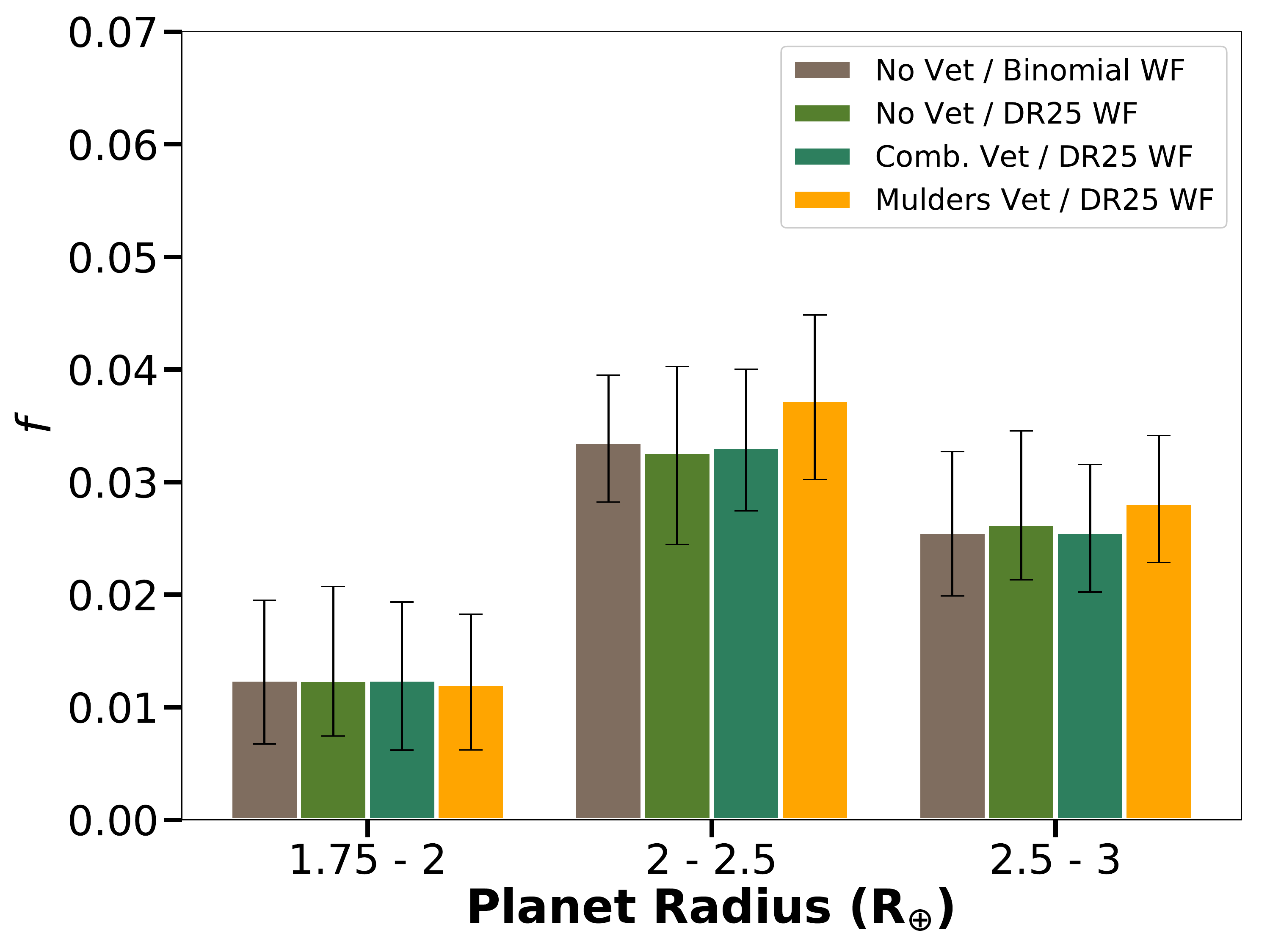}
\includegraphics[scale=0.28]{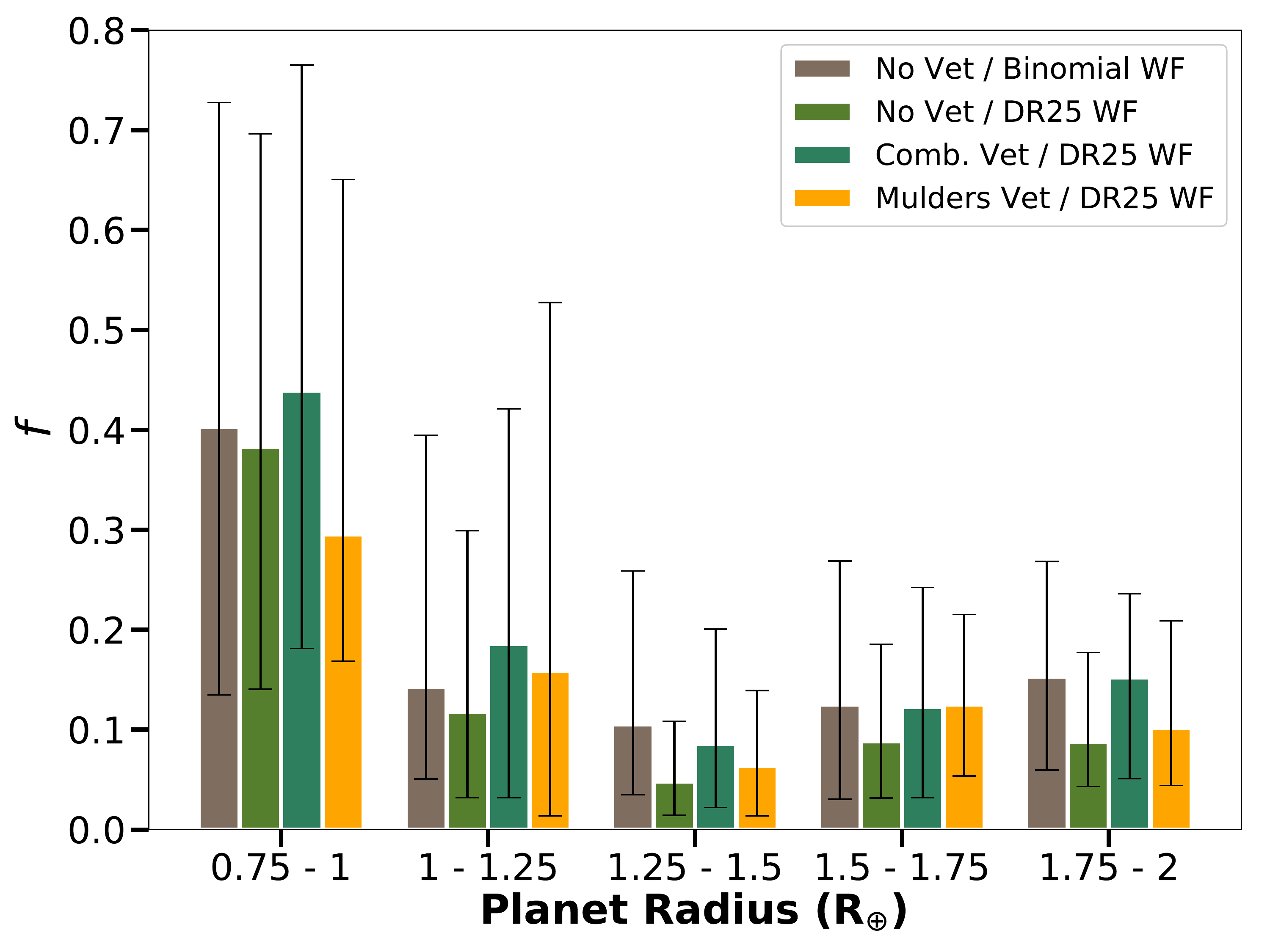}
\caption{Occurrence rate estimates over several radii bins over the range (Top) $P = 8-16$d and (Bottom) $P = 237-500$d with different vetting efficiencies and window functions using a uniform prior with individual bin parameterization.  
The upper panel is representative of the majority of the grid in planet size and orbital period, where the planet occurrence rates are well-constrained and the choice of window function and vetting efficiency model has a small impact on the inferred planet occurrence rate.  
However, both the window function and model for vetting efficiency are important for planets with longer orbital periods, as illustrated in the bottom panel.  
The DR25 catalog does not include any planet candidates (associated with our cleaned sample of Kepler targets) in the bins with $P = 237-500$d and  $R_p =$ $\{0.75-1$, $1-1.25$, $1.25-1.5\}$ R$_{\oplus}$.  
Therefore, the posterior distribution in these bins is broad and sensitive to the choice of prior.
Nevertheless, the lack of detections provides robust upper limits on the planet occurrence rates in these bins.  
Our baseline results are based on using the DR25 window function and the combined detection and vetting efficiency model, as it was directly fit to pixel-level transit injection simulations.  
}
\label{figVet}
\end{figure}

\subsection{Implications of Vetting Efficiency}
\label{secVet}
Previous studies of planet occurrence rates have assumed that all true planets detected by Kepler's Transiting Planet Search module are labeled as planet candidates in the Kepler planet catalog.
With Kepler's DR25 data release, vetting of threshold crossing events was performed by an automated process, the ``robovetter''.  
This opened the door to explicitly modeling the vetting efficiency, i.e. the probability with which the Kepler robovetter identifies a true exoplanet as a planet candidate.

In order to explore the importance of vetting efficiency, we calculated the occurrence rate using three different models:
\begin{itemize}
 \item Using the detection efficiency model in Eqn. \ref{eqnDetC2017} and setting the vetting efficiency to unity.  This is analogous to previous exoplanet occurrence rate studies.  
 \item Using the detection efficiency model in Eqn. \ref{eqnDetC2017} and adopting the vetting efficiency in Eqn.\ \ref{eqnVetMulders}, a function of orbital period and planet radius based on the model of \citet{MPA+2018}. 
 \item Using a single model for the combined detection and vetting efficiency in Eqn. \ref{eqnDetAndVet}.
\end{itemize}
The occurrence rates produced by these three different models are compared for two bins in different detection regimes  in Fig. \ref{figVet}.
When comparing results using a vetting efficiency of unity to results using the Mulders vetting efficiency model, we find that the Mulders model can significantly affect the inferred planet occurrence rate, particularly for small, long-period planets (bottom panel).  
As the occurrence rate of such planets is of particular interest, this motivated us to develop an improved model for the vetting efficiency.  
We make use of the pixel-level transit injections that were processed by both the Kepler planet search pipeline and the robovetter to empirically fit a model for the probability of a true transiting planet being both detected by the pipeline and labeled as a planet candidate.

Our results demonstrate the importance of both accounting for the vetting efficiency and of using an accurate vetting efficiency model.  
When comparing the results with our model for the combined detection and vetting efficiency, we find significant changes in the inferred planet occurrence rates relative to previous calculations that neglected vetting efficiency, only for bins with $P > 32$d or $R_{p} < 2 R_{\oplus}$.
For the remainder of the bins, the difference in the median of the ABC-posteriors is typically considerably less than the width of the ABC-posterior.

We compared the results using our combined detection and vetting efficiency to results using the Mulders model in two regimes:  short orbital periods with well-characterized occurrence rates with $P=8-16$d and $R_p=1.75-3 R_\oplus$ (using 3 bins), and long orbital periods with weaker constraints on the occurrence rates ($P=237-500$d and $R_p=0.75-2 R_\oplus$ (using 5 bins).  We find that in both of these cases, the inferred occurrence rates are consistent within measurement uncertainties.  The Mulders vetting model yields a mean rate only slightly higher (no more than $\sim~12\%$) in the first case.   In the long-period regime, the Mulders vetting model finds a lower rate.  While the fractional difference in the mean occurrence rates is up to $\sim~33\%$, this is still significantly smaller than the uncertainty due to the Kepler sample size.  We conclude that our inferences for planet occurrence rates are robust to the choice of vetting model.  

\subsection{Occurrence Rates of Planets near the Habitable Zone}
\label{secEtaEarth}
Given the widespread interest in the occurrence rate of small planets in or near the habitable zone ($\eta_\oplus$), we performed additional calculations for the occurrence rate of planets with radii $\{0.75-1$, $1-1.25$, $1.25-1.5$, $1.5-1.75$, $1.75-2\}$ R$_\oplus$ and orbital periods 237-500 days using different choices of window functions and vetting efficiencies.

\subsubsection{Importance of Individual Model Improvements}
Our results using an independent uniform prior for each bin's occurrence rate can be seen in Fig. \ref{figVet} (lower panel) where we present occurrence rate estimates assuming:
1) no loss of planets due to the vetting process and the simple binomial window function (comparable to the methodology of \citet{HFR+2018}), 
2) no loss of planets due to the vetting process and DR25 target-by-target window functions, 
3) combined detection and vetting efficiency and DR25 target-by-target window functions, and 
4) \citet{MPA+2018} vetting efficiency and DR25 target-by-target window functions.

The inclusion of a target-by-target set of window functions can result in diminishing the inferred occurrence rates in the $\eta_{\oplus}$ regime due to better accounting for detection  probability when there are few transits.  
While this effect is small compared to measurement uncertainties for the smallest planets (i.e., $<1.25 R_\oplus$), it can be significant for larger planets.

Accounting for the vetting efficiency can significantly affect the inferred occurrence rate compared to previous estimates that have assumed all planets pass the vetting process.  
The occurrence rates based on the combined detection and vetting efficiency model generally increases relative to assuming perfect vetting. 
This demonstrates the importance of accounting for the vetting process when computing planet occurrence rates.

For orbital periods beyond 128 days, Kepler observations place interesting upper limits on the occurrence rate of Earth-size planets, even if there are no detections in a given period-radius bin.  
Table \ref{tab:occ_rates} and Figure \ref{figRates} provide upper limits based on adopting independent uniform priors for each bin's occurrence rate.  
While appealing in its simplicity, adopting independent uniform priors for each bin is equivalent to assuming a total rate of planets in that period range that is peaked at large planet occurrence rates.
This can be understood by considering the sum of $n$ multiple independent random variables, each uniformly distributed between zero and $A$.  The probability distribution for the sum is an Irwin-Hall distribution which has mean of $nA/2$ and a standard deviation of $\sqrt{nA/12}$.  Assuming independent uniform distributions for the rate of planets in each of several radius bins is equivalent to assuming a prior distribution for the total rate that is more concentrated than assuming a uniform prior for the total rate.  The location of the prior mode is at half of the sum of the upper limits of the individual priors.  Thus, the shape of the posterior at low rates is sensitive the choice of prior, when there are no (or few) detections.   Even in the case of no detections, the posterior still drops off rapidly towards high rates, since the lack of detections strongly excludes high rates.  These effects can be seen clearly by comparing the dotted lines in panels f and l of Figure \ref{figPost}.

While the choice of prior did not have a significant effect for most of the bins in our period-radius grid, it does affect the rates derived in the $\eta_\oplus$ regime.  
Given the relatively weak observational constraints, the use of a peaked prior on the total rate of small planets in the habitable zone seems unwise.  

Therefore, for the results presented in \S\ref{secEtaEarthNumbers}, we performed additional simulations specifically focusing on this regime using a Dirichlet prior over radius bins with boundaries subset $R_p =$ \{0.5, 0.75, 1, 1.25, 1.5, 1.75, 2, 2.5\}.  This allows us to compute the posterior distribution for the rate of small, long-period planets, while assuming a uniform prior on the rate of planets of all sizes between 0.5 and 2.5 R$_\oplus$.  We do not include larger planets, so as to improve the computational efficiency of our ABC algorithm.
The rate of planets in each bin larger than 2.5 R$_\oplus$ is well constrained and the sum is less than $\simeq~40\%$.  In principle, one could attempt to use the rate of larger planets to motivate a narrower prior on the rate of planets between 0.5 and 2.5 R$_\oplus$.  Based on the posteriors computed, we confirm that including larger planets would have a negligible effect on the ABC posterior.  Therefore, we do not reduce the upper limit for the total rate in $P = 237-500d$ and $R_{p} = 0.5-2.5$ R$_\oplus$ planets, as any corrections would be modest and the reliability of the larger planet candidates may not be well known.

\subsubsection{$\eta_{\oplus}$ Rate Estimates}
\label{secEtaEarthNumbers}
\begin{deluxetable*}{lrrrr}
\tablewidth{0pt}
\tablecaption{HZ Occurrence Rate Estimates\tablenotemark{a}
\label{tab:HZrate}}
\tablehead{
&
\multicolumn{4}{c}{Prior}\\
\colhead{Radius}&
\multicolumn{2}{c}{Dirichlet}&
\multicolumn{2}{c}{Uniform}\\
\colhead{(R$_\oplus$)}&
\colhead{$f$}&
\multicolumn{1}{c}{$\Gamma$}&
\colhead{$f$}&
\colhead{$\Gamma$}
}
\startdata
$\phn0.75-\phn1.00$ & $<0.14$                   & $<0.66$ & $<0.70$ & $<3.3$\\
$\phn1.00-\phn1.25$ & $<0.10$                   & $<0.61$ & $<0.31$ & $<1.9$\\
$\phn1.25-\phn1.50$ & $<0.058$                  & $<0.43$ & $<0.14$ & $<1.0$\\
$\phn1.50-\phn1.75$ & $0.077^{+0.061}_{-0.040}$ & $0.67^{+0.53}_{-0.35}$ & $0.12^{+0.07}_{-0.06}$ & $1.06^{+0.63}_{-0.53}$\\
$\phn1.75-\phn2.00$ & $0.12^{+0.06}_{-0.05}$    & $1.2^{+0.6}_{-0.5}$    & $0.14^{+0.09}_{-0.07}$ & $1.39^{+0.90}_{-0.72}$\\
$\phn0.75-\phn1.50$ & $<0.27$                   & $<0.51$ & $<1.1$ & $<2.0$\\
$\phn1.00-\phn1.75$ & $0.17^{+0.09}_{-0.07}$    & $0.40^{+0.10}_{-0.07}$ & $0.35^{+0.21}_{-0.14}$ & $0.83^{+0.50}_{-0.34}$\\
\enddata
\tablecomments{Occurrence rate estimates over the HZ region for $P = 237-500$d and several radii bins with two prior choices.  Upper limits correspond to the 84.13th percentile. \tablenotetext{a}{The true rate of planets in this period range could be lower by as much as a factor of $\sim~2$, due to uncertainty in the reliability of planet candidates \citep[see \S\ref{secReliability} \& ][]{TCH+2018,BCM+2008}.  Upper limits are believed robust to $\sim~10\%$ (see \S\ref{secFutureRates}).} }
\end{deluxetable*}

\begin{figure*}[p]
\centering
\gridline{\fig{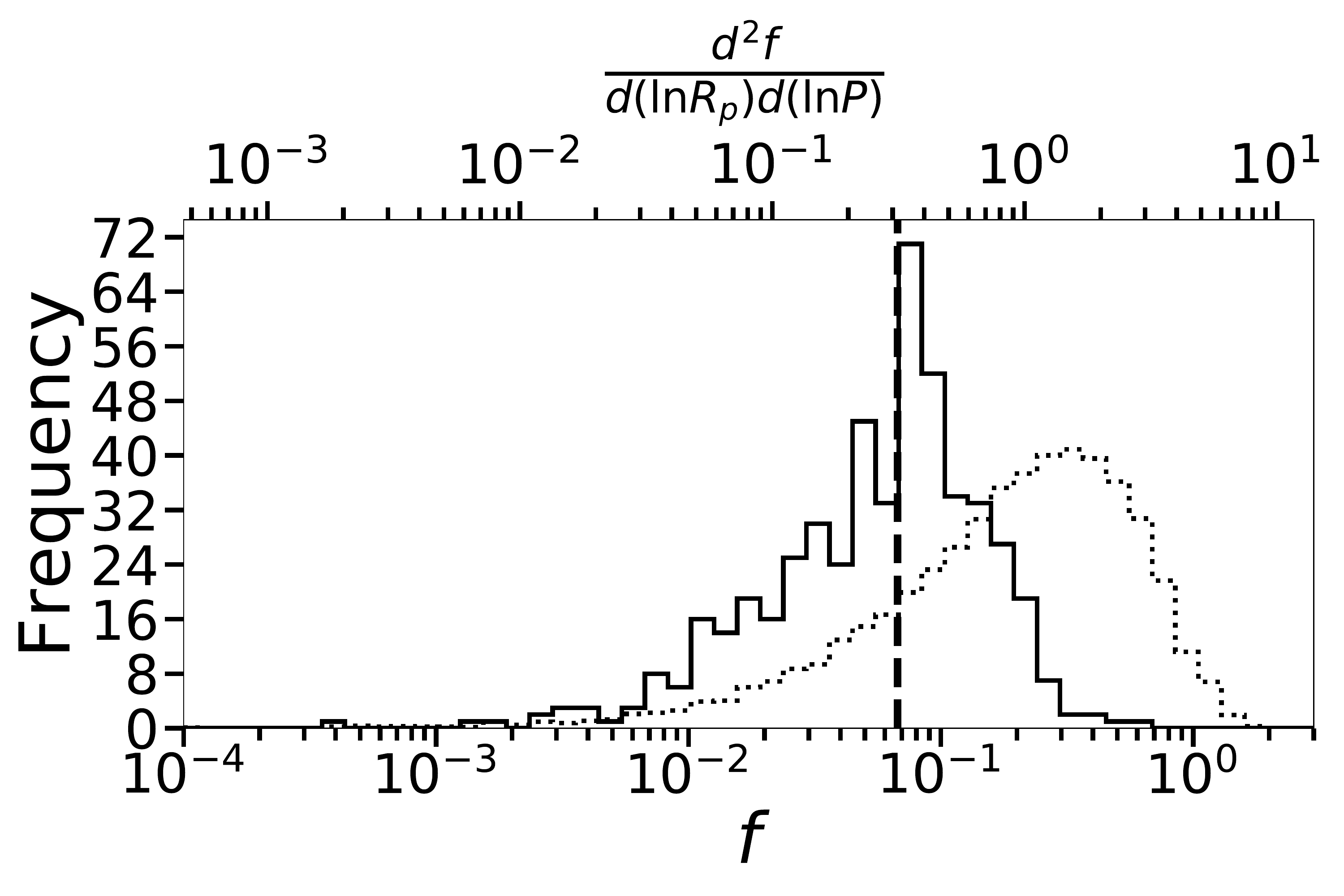}{0.33\textwidth}{(a) $R_p = 0.75-1$ R$_\oplus$, Dirichlet prior}
\fig{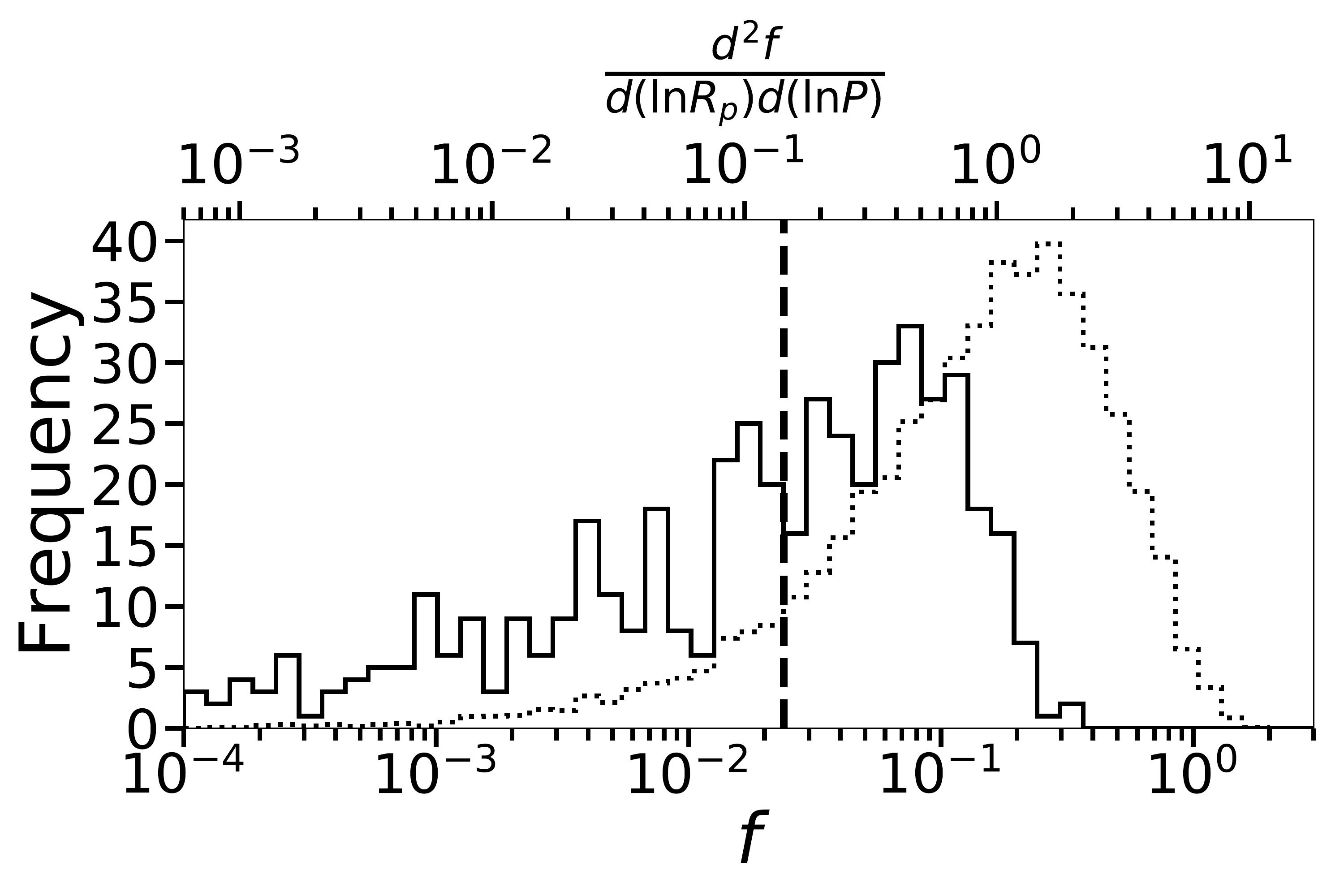}{0.33\textwidth}{(c) $R_p = 1-1.25$ R$_\oplus$, Dirichlet prior}
\fig{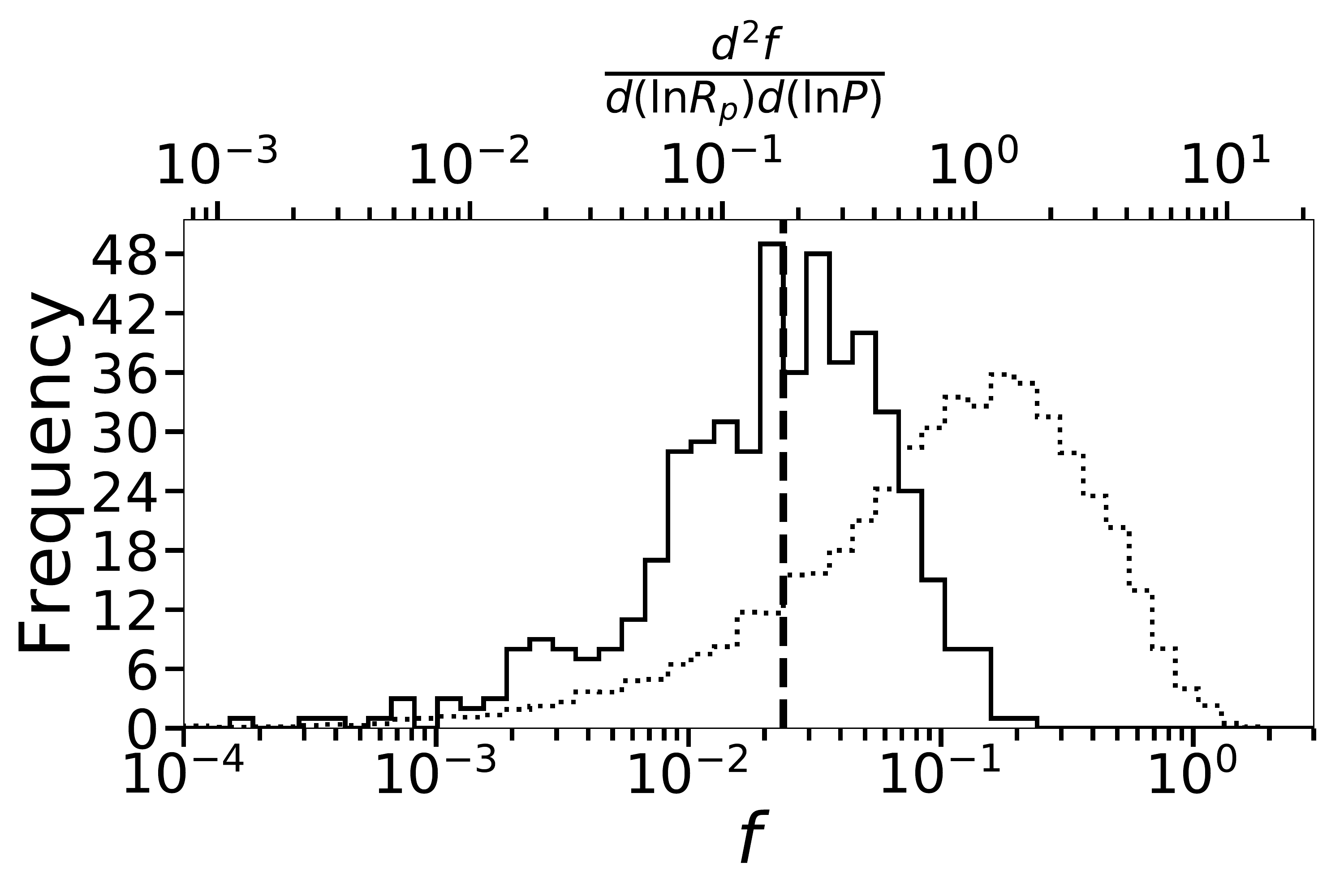}{0.33\textwidth}{(e) $R_p = 1.25-1.5$ R$_\oplus$, Dirichlet prior}}
\gridline{\fig{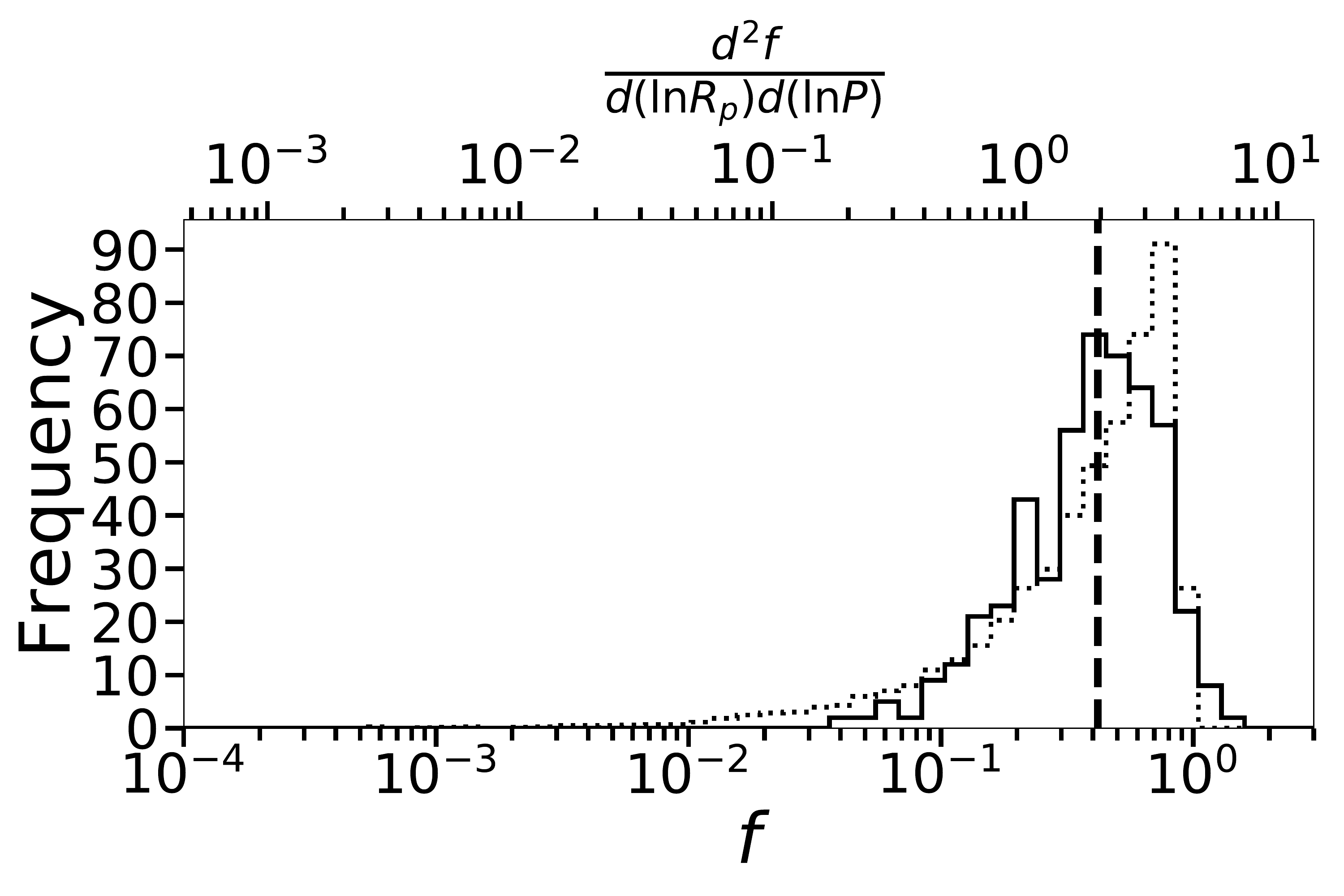}{0.33\textwidth}{(b) $R_p = 0.75-1$ R$_\oplus$, Uniform prior}
\fig{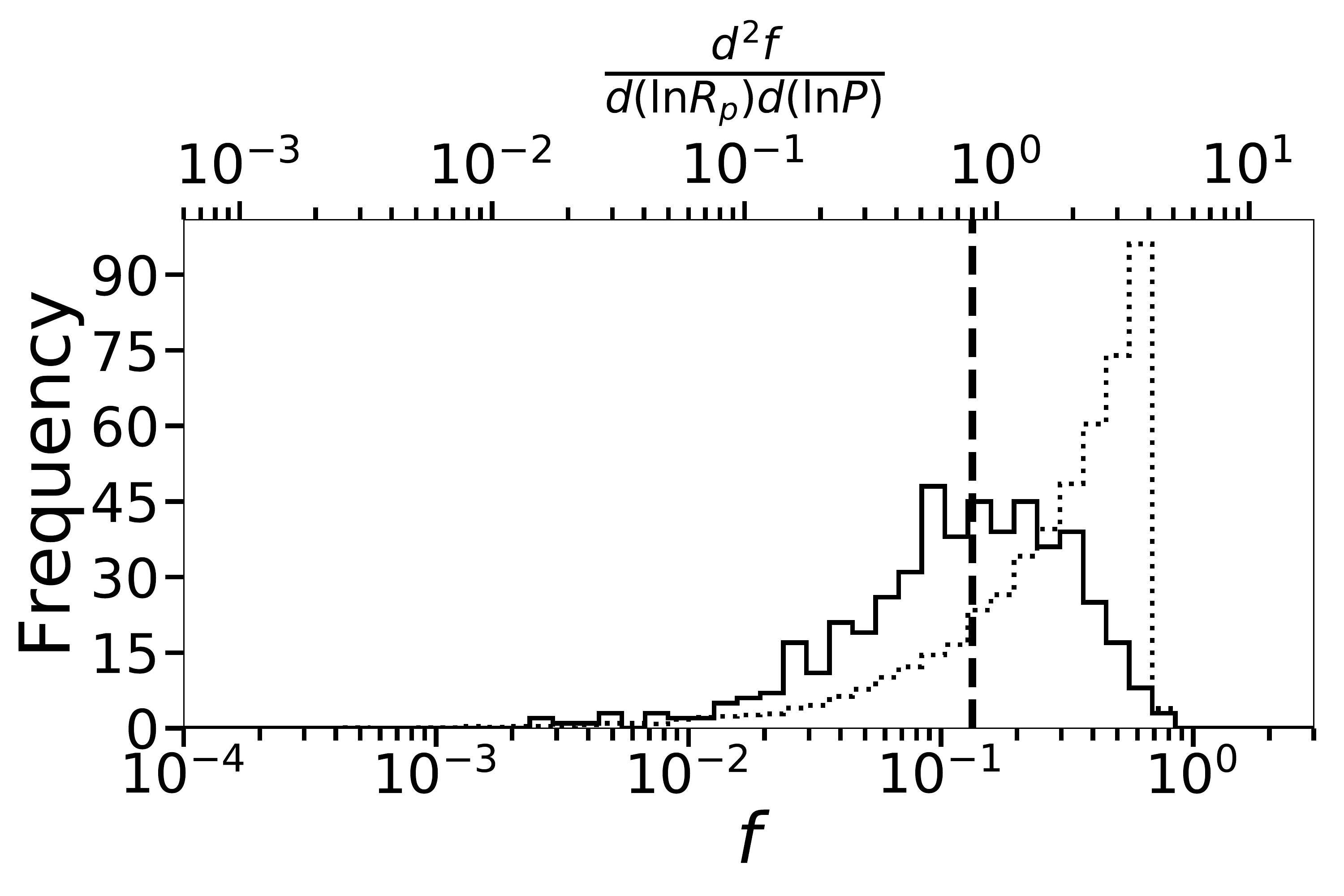}{0.33\textwidth}{(d) $R_p = 1-1.25$ R$_\oplus$, Uniform prior}
\fig{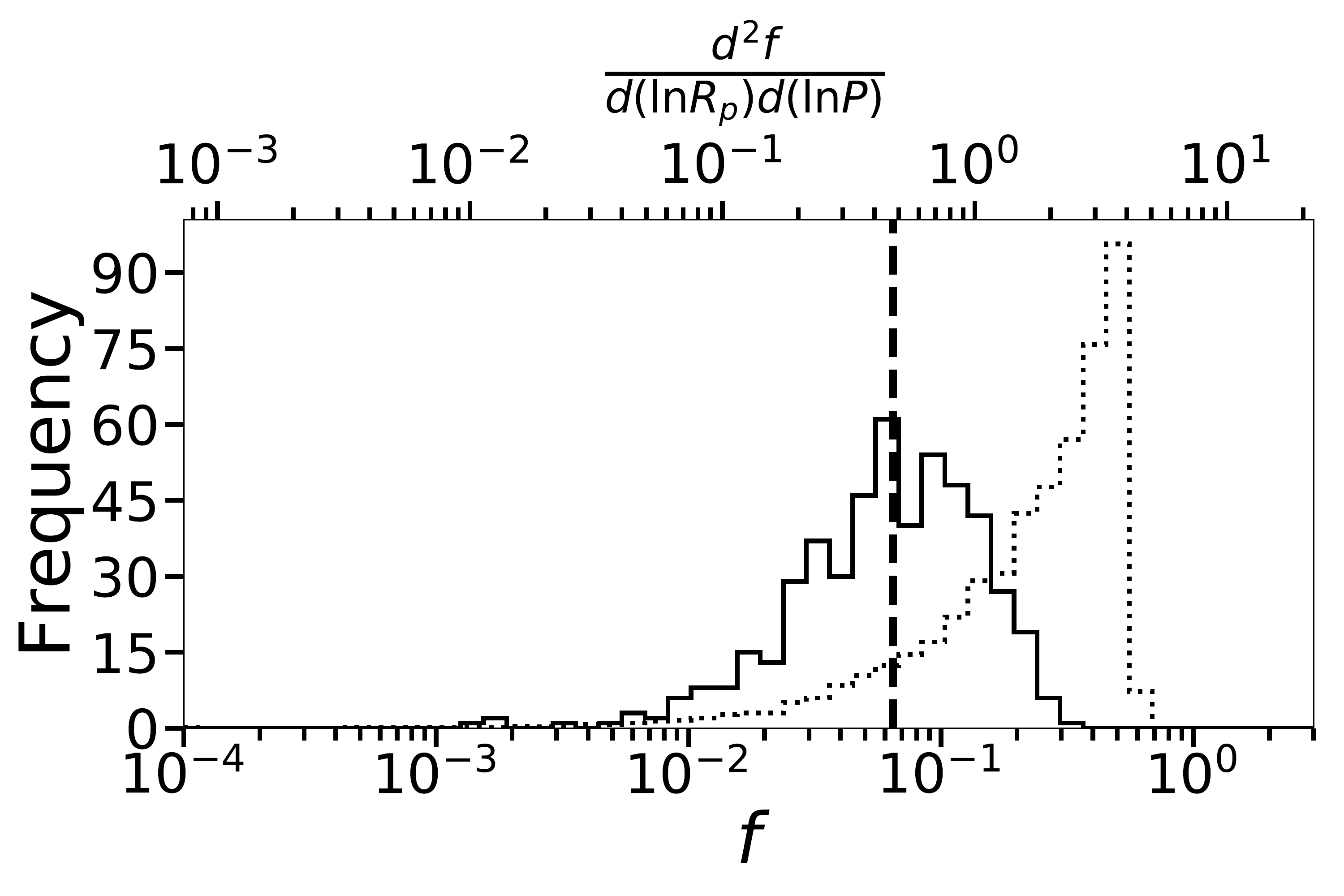}{0.33\textwidth}{(f) $R_p = 1.25-1.5$ R$_\oplus$, Uniform prior}}
\hrulefill
\gridline{\fig{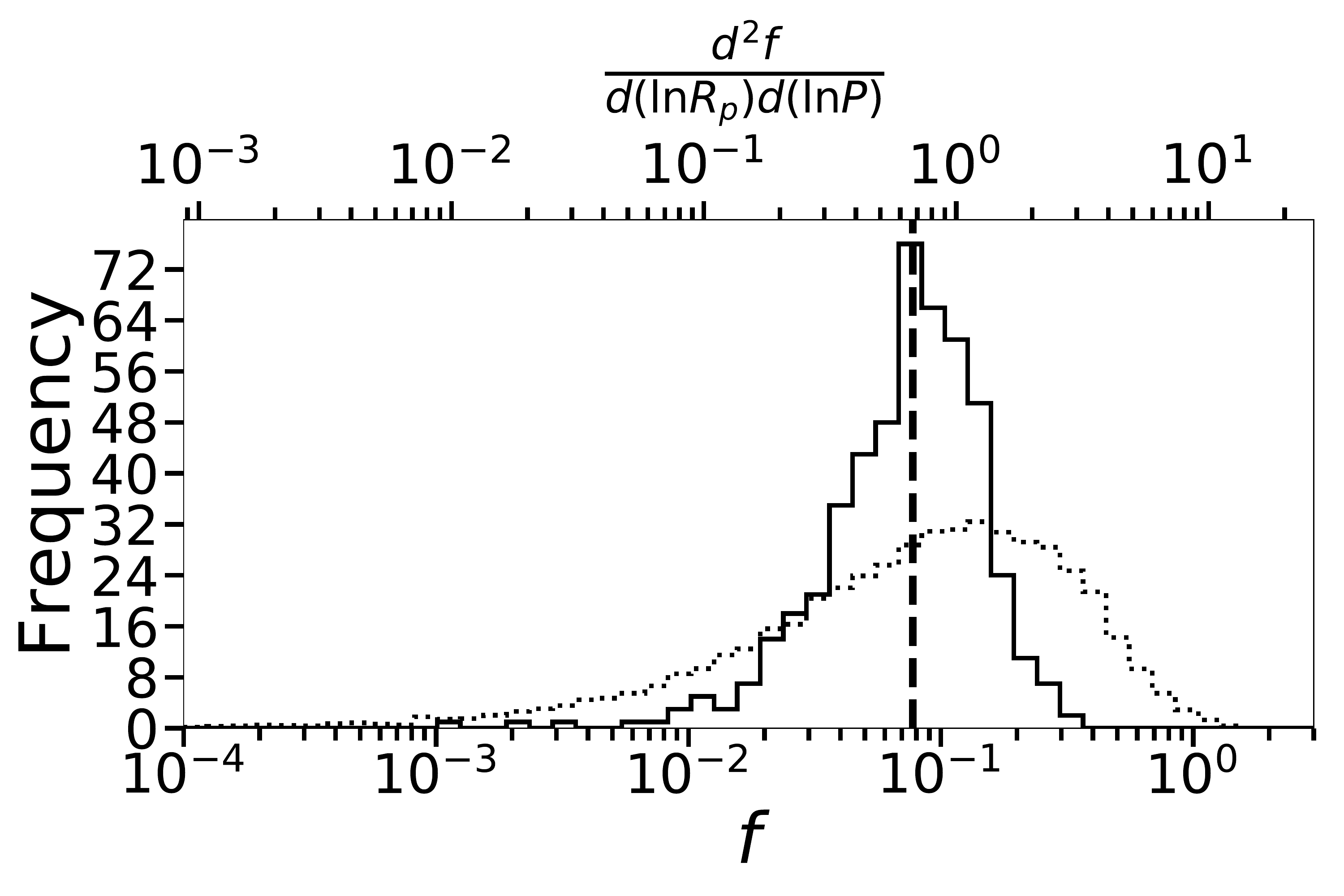}{0.33\textwidth}{(g) $R_p = 1.5-1.75$ R$_\oplus$, Dirichlet prior}
\fig{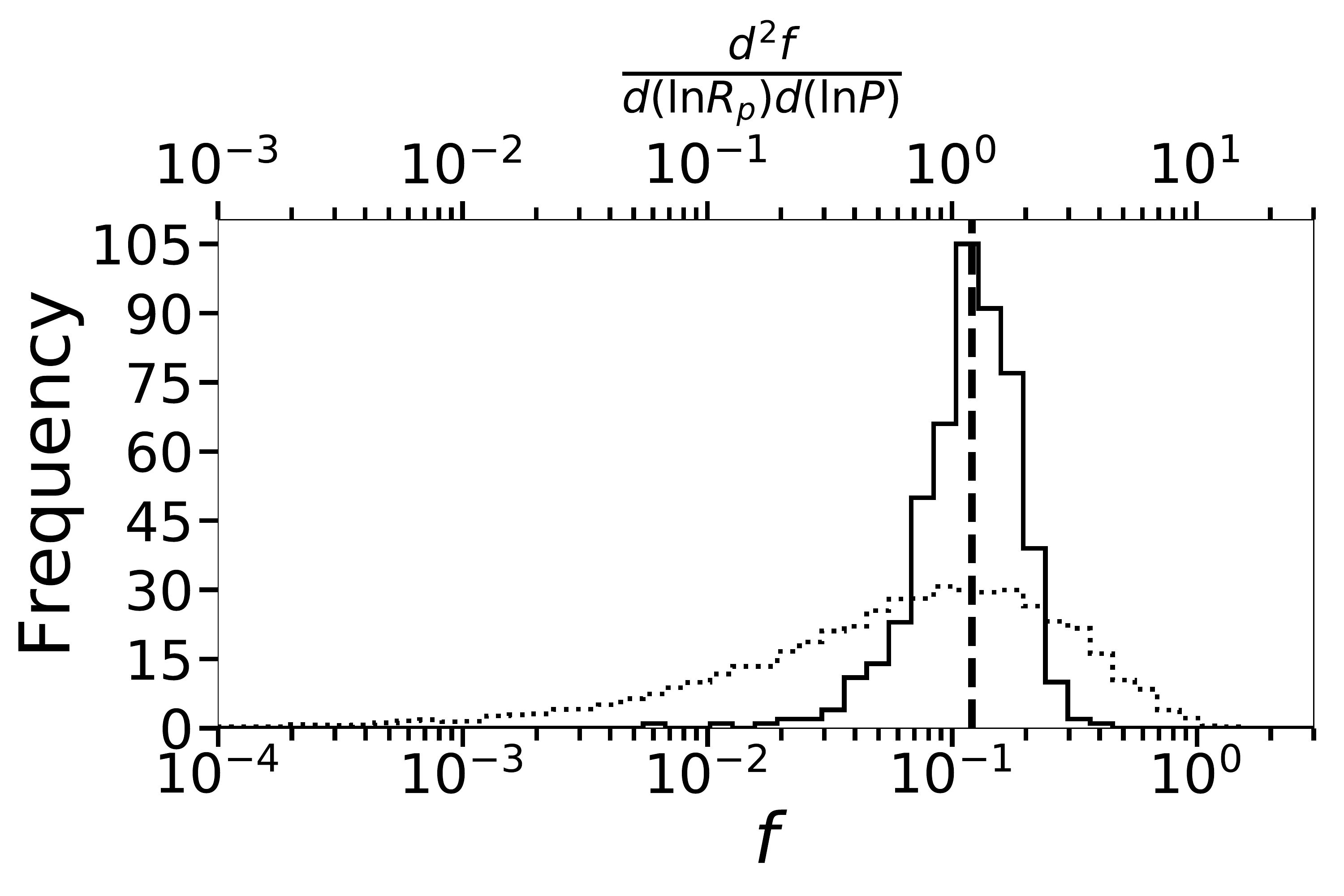}{0.33\textwidth}{(i) $R_p = 1.75-2$ R$_\oplus$, Dirichlet prior}
\fig{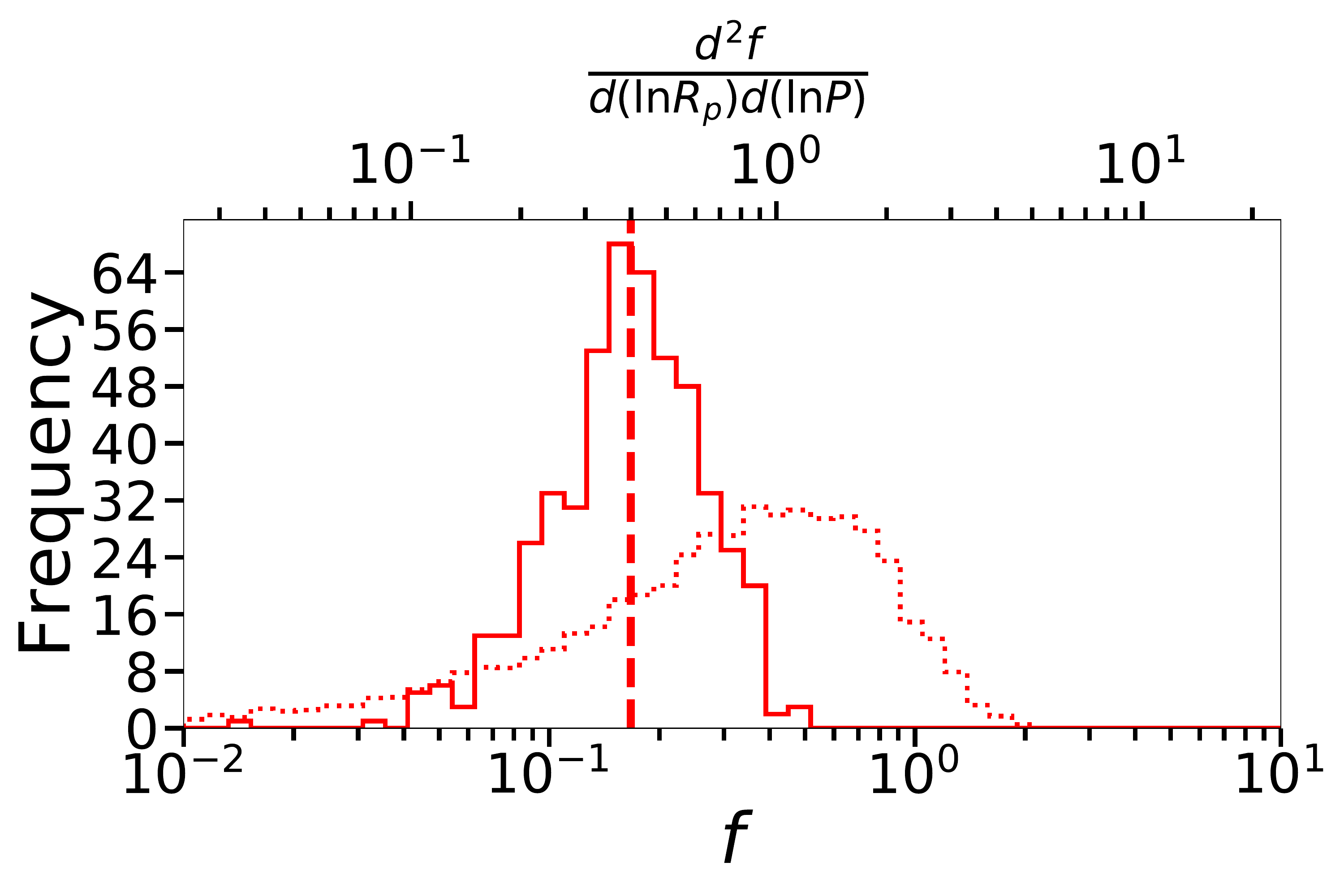}{0.33\textwidth}{(k) $R_p = 1-1.75$ R$_\oplus$, Dirichlet prior}}
\gridline{\fig{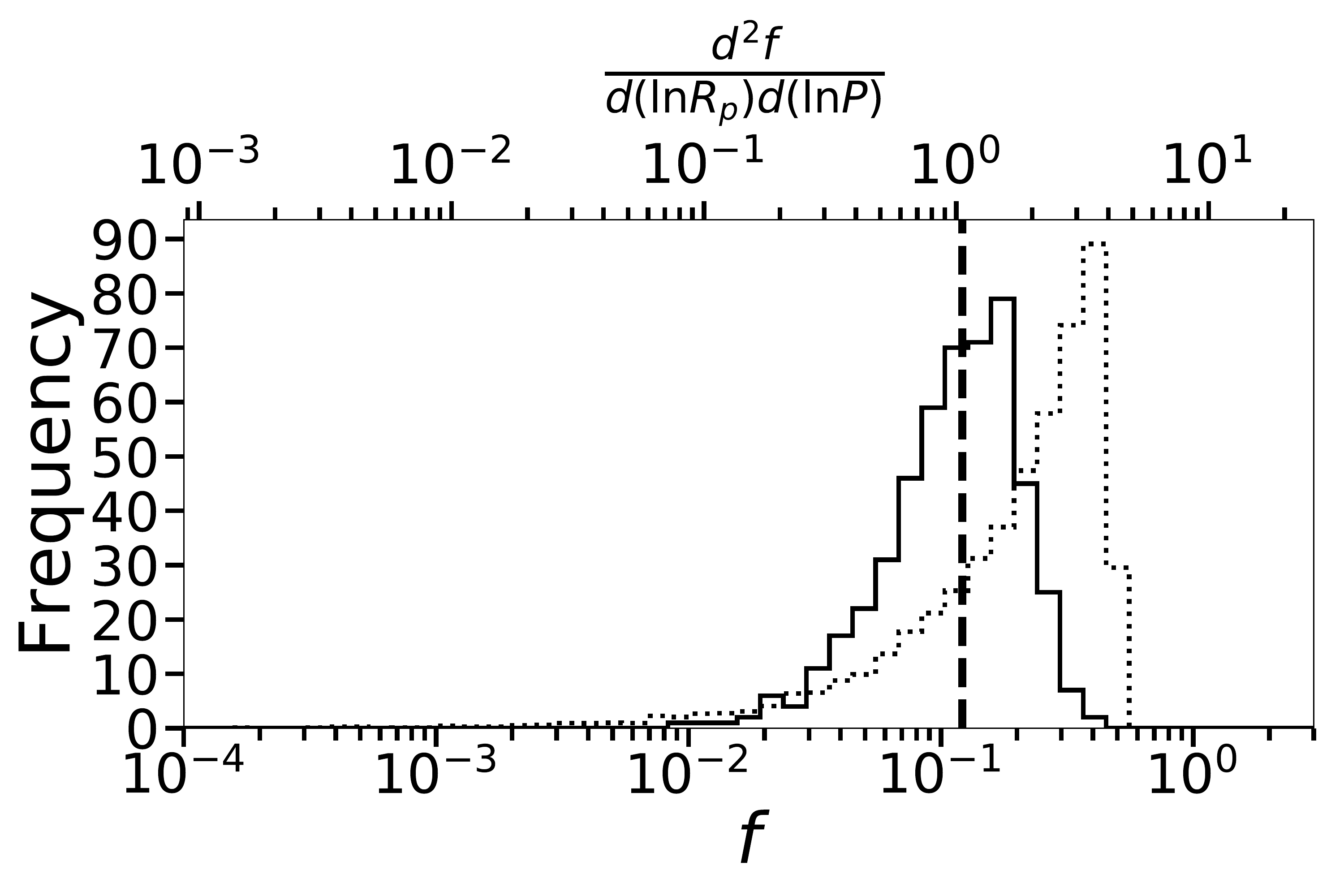}{0.33\textwidth}{(h) $R_p = 1.5-1.75$ R$_\oplus$, Uniform prior}
\fig{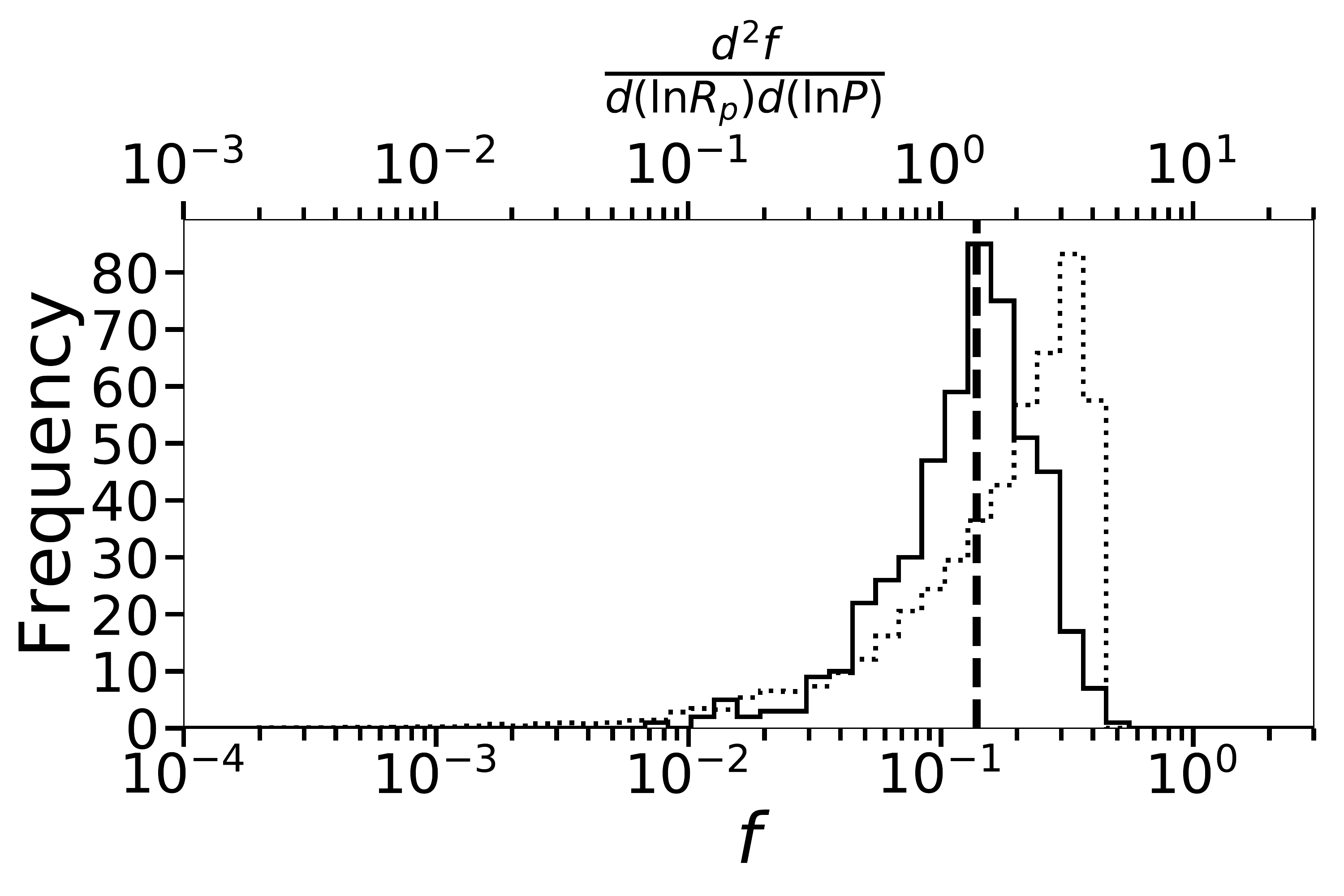}{0.33\textwidth}{(j) $R_p = 1.75-2$ R$_\oplus$, Uniform prior}
\fig{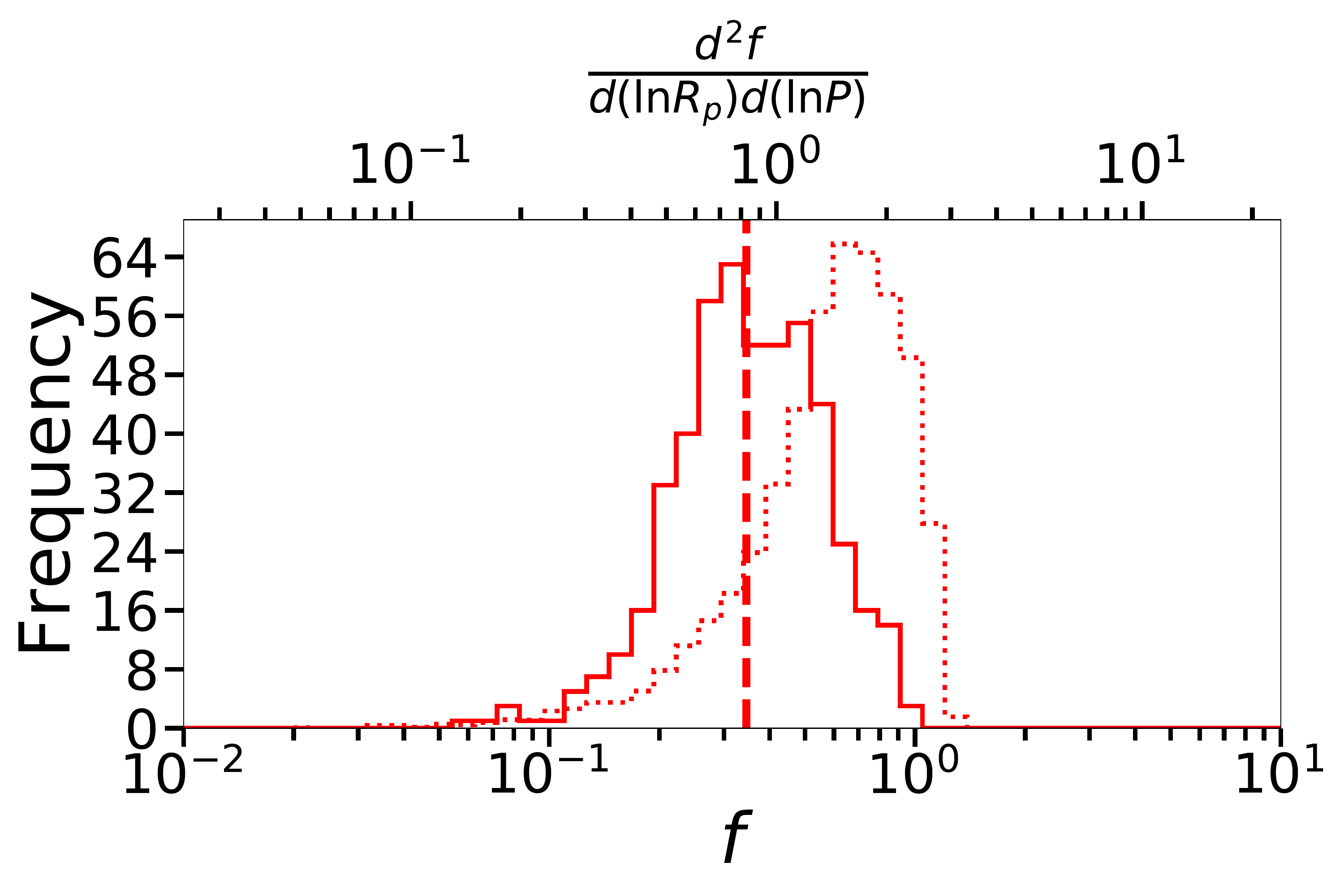}{0.33\textwidth}{(l) $R_p = 1-1.75$ R$_\oplus$, Uniform prior}}
\caption{ABC posterior samples for the planet occurrence for planets with orbital periods of  $P = 237-500$d around FGK stars using two prior choices.  The panels show results for several radius bins in the $\eta_{\oplus}$ regime using the combined detection and vetting efficiency with two choices of prior.  For comparison, a sample from the prior distribution is shown by the dotted line.  
The median (i.e. 50th percentile) is indicated by the vertical dashed line.  
For all size ranges shown, the inferred upper limits (e.g., 84.13th or 95th percentiles) are significantly less than the same percentile for the prior, even for radius ranges that do not include any detected planet candidate.  
Panels a - f show bins where Kepler detected no planets (for targets included in our cleaned sample) while panels g - j show bins with Kepler planet candidates.  
Panels k \& l (bottom right) show the integrated rate over 1-1.75 R$_\oplus$ (assuming the occurrence rate is piecewise constant over each of 1-1.25, 1.25-1.5, and 1.5-1.75 R$_\oplus$).
Kepler data provides a 95th percentile empirical upper limit of 0.34 (Dirichlet prior) and 0.7 (Uniform prior) on the rate of such planets.
}
\label{figPost}
\end{figure*}

Our ABC posterior distributions for $\eta_\oplus$ (or $\Gamma_\oplus$) summarize the current state of our knowledge about the rate of small planets in the habitable zone of FGK stars (see panels a, c, e, g, i and k of Figure \ref{figPost}).
We summarize our results by reporting upper limits (84.13th percentile) and median (50th percentile) occurrence rate of small planets over the period range $P = 237-500$d using our combined detection and vetting efficiency and two choices of prior.
Note that the our recommended rates in this regime are based on the Dirichlet prior for relative rates, and so these rates differ from the rates reported in Table \ref{tab:occ_rates} and Figure \ref{figRates} based on independent uniform priors for each bin.  
Comparing the results with the two priors can be useful as a test to establish how sensitive these occurrence rates are to the choice of prior.  
Summing over the three interior bins, the \{5, 15.87, 50, 84.13, 95\}th percentiles for the summed occurrence rate assuming the Dirichlet prior over the range $P = 237-500$d, $R_{p} =$ $0.75-1.5$ R$_{\oplus}$ are $f_{R,P} =$ \{$0.06$, $0.10$, $0.16$, $0.27$, $0.36$\}.  
The corresponding percentiles for the differential rates are $\Gamma_\oplus \equiv (d^2 f)/[d(\ln P)~d(\ln R_{p})] = $ \{ $0.12$, $0.19$, $0.32$, $0.51$, $0.69$\} for the same range.
We consider the above rates to represent the current best estimates.  

\label{secEtaEarthSensitivity}

For the sake of examining the sensitivity of the results to the choice of prior, we performed a similar analysis, adopting an independent uniform prior for each radius bin.  
The results can be seen in panels b, d, f, h, j and l of Figure \ref{figPost}. 
Using this prior, the \{5, 15.87, 50, 84.13, 95\}th percentiles for the summed occurrence rate over the range $P = 237-500$d, $R_{p} =$ $0.75-1.5$ R$_{\oplus}$ are $f_{R,P} =$ \{$0.27$, $0.40$, $0.67$, $1.1$, $1.3$\} while the differential rates are $\Gamma_\oplus = $ \{ $0.51$, $0.78$, $1.3$, $2.0$, $2.5$\}.
While we expect use use of independent uniform priors to result in an overestimate of summed occurrence rates, we include these values to quantify the sensitivity of these results to the choice of prior.  
While these two priors are quite different (e.g., compare dotted lines for panels k \& l), the ABC posteriors are qualitatively similar for large rates (where data provide strong constraints), but differ significantly for small rates (where the posterior shape is primarily based on information coming from the prior).  

\subsubsection{Comparison to previous studies}
\begin{figure}
\centering
\includegraphics[scale=0.28]{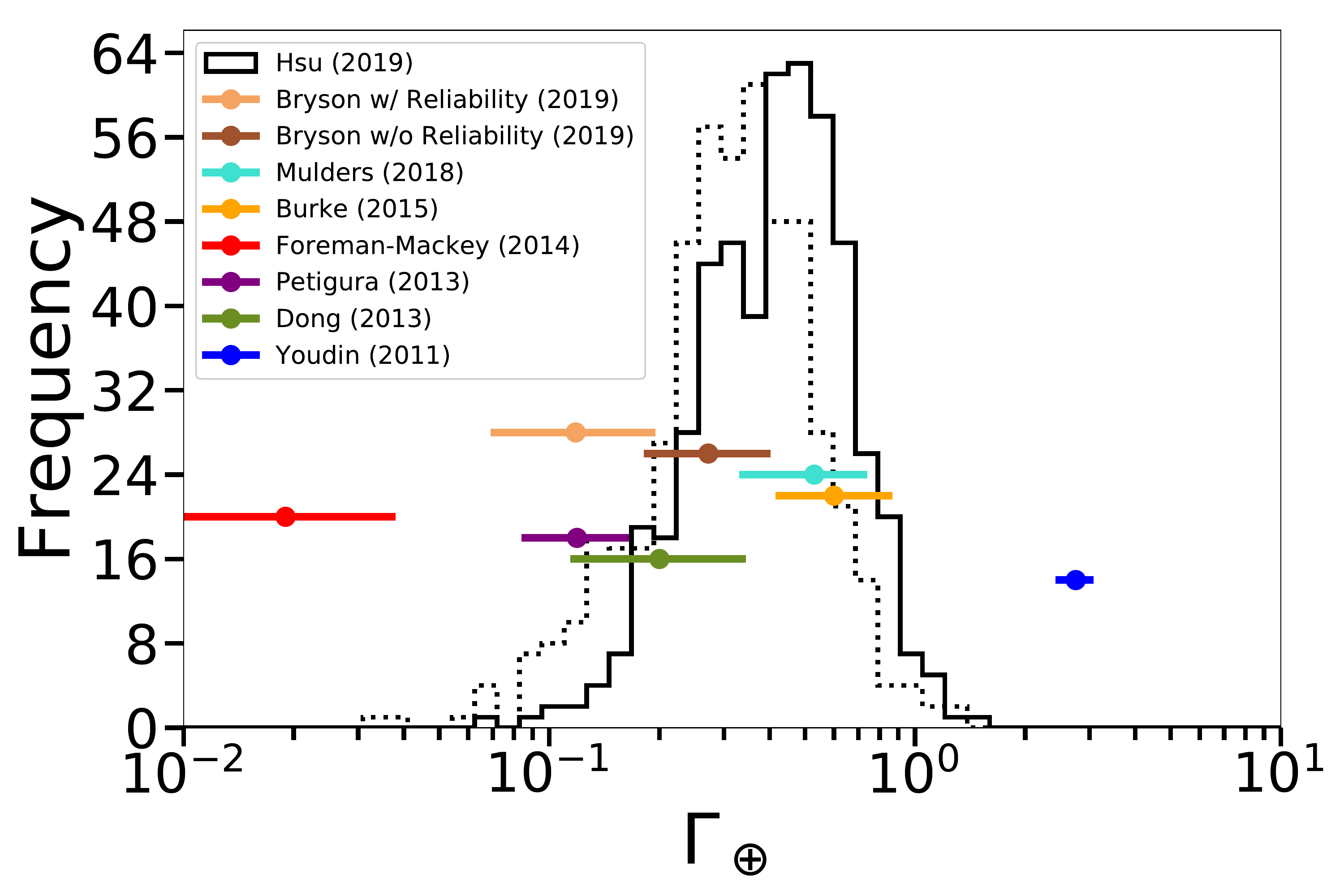}
\caption{Inferred HZ occurrence rate density ($\Gamma_\oplus$)
comparison between various studies. Two posteriors for $P = 237-500$d and different radius ranges are provided for this study: (Solid black) $R_p = 1-1.75$ $R_\oplus$ and (dotted black) $R_p = 0.75-1.5$ $R_\oplus$.  Median estimates with uncertainties are provided from \citet{Y2011, DZ2013, PHM2013, FHM2014, BCM+2015, MPA+2018, BCB+2019}.  The estimate from \citet{BCM+2015} is taken from the baseline extrapolated posterior of Figure 17. The y-axis values for the median estimates of other studies have no meaning except to separate the different estimates for easier readability.
}
\label{figRateComp}
\end{figure}

The median of our posterior for the occurrence rate of small planets in or near the habitable zone of sun-like stars (see Fig. \ref{figPost}, panel k) is similar to estimates based on extrapolation from recent studies and well within the range of possibilities suggested by sensitivity analyses \citep[e.g.,][]{BCM+2015,MPA+2018,BCB+2019}. Fig. \ref{figRateComp} shows the ABC posterior distribution for $\Gamma_{\oplus}$ from our study, along with the results of several previous studies.
In order to make a meaningful comparison, each studies results have been converted into a differential rate (using units of $R_\oplus$ and days, and natural log rather than base 10 or base 2).  
Other studies are summarized with a median and separate upper and lower uncertainties.  

The most discrepant estimates are readily understood as due to using a much earlier and very different catalog of planets \citep[i.e.,][]{Y2011, PHM2013, FHM2014, DZ2013}.  Only \citet{MPA+2018} and \citet{BCB+2019} used the Kepler DR25 planet catalog.  Results for $\Gamma_\oplus$ from this study are intermediate, i.e., between the estimates from these studies.  Equally important, we suggest that the uncertainty estimate from our study is more realistic than those of either \citet{MPA+2018,BCB+2019}.  While the \citet{BCM+2015} analysis used a Q1-16 planet catalog, they still found a similar rate to \citet{MPA+2018}.  

Both the \citet{BCM+2015} and \citet{MPA+2018} studies arrived at higher estimates for $\Gamma_\oplus$ than in this analysis, perhaps due to their use of a parametric model for the planet occurrence rate (i.e., power law or broken power-law in planet radius or planet-star radius ratio).  Since the rate of planets in the $\eta_\oplus$ regime is only weakly constrained by data, assuming such a parametric distribution, essentially results in an extrapolation in the $\eta_\oplus$ regime and a large proportion of the inferred rate is based on planets that are only rarely detectable in the Kepler sample.  

In contrast, a preliminary draft of \citet{BCB+2019} reports lower estimates for $\Gamma_\oplus$ than this study, regardless of whether they attempt to account for potential false alarms (i.e., ``No Reliability'' versus ``With Reliability'').  There is significant overlap between our posterior $\Gamma_\oplus$ and that of \citet{BCB+2019}.  We attribute the offset in the mean of our posteriors for $\Gamma_\oplus$ and their ``No Reliability'' estimate of $\Gamma_\oplus$ as likely due to some combination of their use of a parametric model, the different selection criteria for target stars, and/or difference in the stellar properties (based on an unpublished catalog, similar to \citet{BHG+2018}).    
There is a more significant different in our posterior for $\Gamma_\oplus$ and the ``With Reliability'' estimate of \citet{BCB+2019}.  As discussed in \S\ref{secReliability}, we caution that our posteriors for occurrence rates at $P>250$d and for $\Gamma_\oplus$ might overestimate the true rate by as much as a factor of $\sim~2$ due to potential false alarms at long orbital periods, but our upper limits should be robust, even at these long orbital periods.  Using the stellar sample and stellar properties from \citet{BCB+2019} results in $2-4$ planet candidates in the $\eta_\oplus$ regime.  Their analysis finds that these have low ``reliability'' and deweights these planets accordingly.  We note that none of those planet candidates directly contribute to our estimate of $\Gamma_\oplus$, as they were either removed or assigned a radius larger than $1.5R_\oplus$, given our set of stellar properties and target selection criteria.  According to \citet{BCB+2019}, larger planets tend to have greater reliability, and thus the impact of false alarms on our planet occurrence rates are likely mitigated.

We have demonstrated that there are substantial uncertainties in occurrence rates in and near the $\eta_\oplus$ regime, making extrapolation even more dangerous.  Therefore, we caution against overinterpreting apparent differences between this study and \citet{BCM+2015}, \citet{MPA+2018}, or \citet{BCB+2019}, as the differences in the inferred rates among these three studies are smaller than the uncertainties in $\Gamma_{\oplus}$  (which are primarily set by the effective number of stars searched).

Our upper limits are significantly more accurate and robust than previous upper limits, due to a combination of effects including the use of Kepler's final DR25 planet candidate catalog, updated and homogeneous stellar properties from Gaia, and use of DR25 data products such as the target-specific window functions. 
While \citet{BCM+2015}, \citet{MPA+2018}, and \citet{BCB+2019} report more precise values for $\Gamma_{\oplus}$, these are based on assuming a parametric model for the planet occurrence rate.  They infer model parameters based on the occurrence rate of planets at much shorter orbital periods, and assume that these planets provide information about the occurrence rate in the habitable zone.  For example, \citet{BCM+2015} measure a differential occurrence rate between $P=50-300$d and assume that it remains unchanged at larger orbital periods.  In contrast, this study reports a direct measurement of $\Gamma_{\oplus}$ between $P=237-500d$. 
While our upper limits should be robust, the lower limits are sensitive to the choice of prior and the unknown reliability of planet candidates in this regime \citep[see \S\ref{secReliability} \&][]{TCH+2018,BCB+2019}.  

\subsubsection{Extrapolations to Longer Periods}
\label{secEtaEarthExtrap}

While the aforementioned studies have offered estimates of $\eta_{\oplus}$ for FGK stars \citep[i.e.,][]{PHM2013,FHM2014,BCM+2015,MPA+2018}, we caution that such estimates involved extrapolating a model that is constrained based on planets with larger sizes and/or shorter orbital periods.  
On one hand, we do not find compelling evidence for rapidly changing occurrence rates as a function of orbital period.  On the other hand, we demonstrate that the uncertainties in planet occurrence rates in and near the eta-Earth regime are substantial, in some cases of the same order of magnitude as the median value.  Extrapolation based on these rates to larger orbital periods risks amplifying errors.
While it's possible that the occurrence rate of planets as a function of planet size and period varies sufficiently slowly to render such extrapolations accurate, there are both observational and theoretical reasons to be cautious about such extrapolations.  
Theoretically, the origin and evolution of planets with significant gas envelopes could quite different than that of rocky planets.  
Indeed, studies have suggested a local minimum in the occurrence rates as a function of planet size, perhaps due to photoevaporation or core-powered mass-loss \citep{OW2017,GS2018}.
Such a feature is not captured by parametric models which assume planet occurrence rates to be proportional to a power-law (or broken power-law) in planet radius.  
Additionally, there may be qualitative changes in the occurrence rate and properties of planets as orbital periods approach that of the ice line.  
From an observational perspective, radial velocity surveys find a substantial rise in the frequency of more massive planets starting at orbital periods close to one year.  
Similarly, we find the occurrence rate of larger planets increases substantially as orbital periods increase beyond $\sim~128$d.  
Therefore, we recommend that scientists and mission planners be cautious in adopting estimates of $\eta_\oplus$ that are primarily the result of extrapolating a parametric model into the $\eta_\oplus$ regime.

Instead, we recommend making use of the ABC-posterior distribution for $\eta_\oplus$ (or $\Gamma_\oplus$) which encapsulates the current direct observational constraints of the occurrence rate of small planets in the habitable zone of FGK stars.  
This represents a direct measurement based on Kepler data for the relevant planet sizes and orbital periods and not an extrapolation.  
Our constraints on $\eta_\oplus$ are only possible thanks to the Kepler project providing DR 25 data products for both actual Kepler data and simulated data with injected transits.   

For purposes of planning missions to directly image potentially Earth-like planets, mission concepts are computing expected mission yields based on the ranges $P = $ $338-788$d and $R_p = $ $0.8-1.4 R_{\oplus}$ for a solar twin.
Unfortunately, Kepler's mission duration and sensitivity do not place meaningful constraints on small planets at such long orbital periods.  
If one were to extrapolate our results using the Dirichlet prior, assuming the differential occurrence rate derived over the interval $P = 237-500$d and $R_{p} =$ $0.75-1.5 R_\oplus$, then one obtains an upper limit (95th percentile) of 0.33 planets per sun-like star with $P = $ $338-788$d and $R_p = $ $0.8-1.4 R_{\oplus}$.   
The inclusion of long-period planets with sizes less than 1 R$_\oplus$, results in such an extrapolation being sensitive to the choice of prior.  For example, adopting independent uniform priors for the rates of each radius bin results in the 95th percentile of the posterior increasing to 1.2 planets per sun-like star with $P = $ $338-788$d and $R_p = $ $0.8-1.4 R_{\oplus}$.  
This demonstrates that such estimates inevitably are sensitive to the choice of prior and have large uncertainties.  
As emphasized above, we recommend considerable caution when adopting occurrence rates that are based on an extrapolation from shorter orbital periods.

\subsection{Future Prospects for Occurrence Rate Studies}
\label{secFutureRates}
The model used in this study could be improved further to account for additional effects.  
We summarize the assumptions of our current model and our rationale for postponing consideration of these effects.

\label{secLimitations}

\label{secImprovingDetectionModel}
{\em Detection Model:}
We model the probability that a planet is detected as a function of the expected effective SNR and the number of transits observed.  In principle, the detection model could be further improved to incorporate additional information (e.g., transit duration, stellar properties, sky group).

\label{secFutureContam}
{\em Uncertainty in Contamination:}  
This study models the effects of contamination due to known stars, but does not account for the uncertainty in the contamination.  
If only a fraction of the flux observed for a given Kepler target comes from the planet-hosting star, then the transit depth is diluted by the contaminating flux.
While this has been considered in detail for individual systems \citep{HCH+2017,TeCH+2018}, 
it is unclear how much contamination affects planet occurrence rate studies.  
In this study, we account for the dilution of transit depths due to contamination from known stars when computing the transit depth and SNR.
We adopt the contamination provided for each target in the Kepler Input Catalog (KIC).  
Of course, there may be additional dilution from unknown sources, particularly bound multiple star systems.  
We mitigate the effects of unrecognized contamination from multiple star systems by focusing our analysis on a subset of the Kepler target stars which excludes targets with either astrometric data suggestive of binary companion or that appear to reside significantly above the main sequence.
While binary stars with more disparate luminosities will survive our target selection process, they have limited effect on the planet occurrence rate. 
If a planet transits a star significantly fainter than the primary star, then the detection probability is significantly reduced by contamination from the primary.  
Therefore, such planets will rarely be detected for targets in our cleaned target list. 
If a planet transits the primary star in a system with a significantly fainter secondary star, then the effect of dilution is modest. 
Therefore, we expect that the size of the effect of unrecognized dilution will have a modest effect on the occurrence rates inferred using our cleaned target list.

\label{secMultiDetectProb}
{\em Multiple Transiting Planets \& TCEs:}
For targets with multiple transiting planets, the Kepler pipeline searched for planet candidates in an iterative fashion.  After each ``threshold crossing event'' (TCE) was identified, the pipeline masked out a portion of the lightcurve during and immediately surrounding each putative transit before searching for additional planet candidates. The pipeline selected larger MES candidates first so that the presence of a larger and/or shorter-period planet causes the effective duty cycle to be reduced when searching for lower MES signals around the same star.  
Further complicating matters, the pipeline identified many TCEs that were later discarded as not being strong planet candidates.  
Nevertheless, all TCEs result in less data being available for detecting additional planets transiting that target \citep{ZCH2019}.
Fully accounting for these effects would be extremely complicated due to the complexity of the pipeline, and the available data products. 
Further, the measurement of planet occurrence rates would no longer decouple from the characterization of planetary system architectures.  
An initial investigation of this effect, suggested that the detectability of planets may decrease by $\simeq~5.5\%$ ($15.9\%$) for planets with orbital periods less (greater) than 200 days, if the system contains another transiting planet which the pipeline would detect first \citep{ZCH2019}.  
The DR25 pipeline run resulted in 34,032 TCEs, far greater than the $\simeq~8,000$ KOIs \citep{TJS+2016}.  
Since our combined detection and vetting efficiency is based on fitting to the results of pixel-level transit injections into light curves that already include these TCEs, our detection efficiency already accounts for this, but in an average sense, rather than using knowledge of specific TCEs for each star.  

The decrease in detectability due to additional planets only affects a small fraction of stars, since the ratio of planet candidates to target stars is $\simeq~0.03$ in our sample.  
If all planetary systems were strictly coplanar, then short-period planets would transit whenever the long-period planet transited.  
However, for orbital periods of 237-500 days, even very modest mutual inclinations ($\simeq~1^\circ$) cause the conditional transit probability to decrease substantially.  
Therefore, the impact of this effect on the overall planet occurrence rate will be diluted by at least half, even if this effect is  significant for analyzing the relative abundance of systems with single and multiple transiting planets.  
In summary, we estimate that any increase in the occurrence rate of long-period planets due to interactions of the Kepler pipeline and multiple planet systems is likely to be less than $\simeq~8\%$

\label{secPipelineTimeouts}
{\em Pipeline Timeouts:}
The actual Kepler DR25 transit search did not search every target for transit signals as small as the typical 7.1-$\sigma$ threshold. 
The DR25 stellar catalog reports the ``MES threshold'' and ``timeout'' indicators for when the search for transits was halted prematurely.
The DR25 stellar catalog data indicate that only 94\% of targets were searched to MES of 7.1 (the standard criterion for triggering a TCE) across all durations. When applying this only to stars in our clean FGK catalog, 96\% of targets were searched to full depth. However, of the 4\% that were not searched to MES of 7.1, the MES threshold was still below 8 in 80\% of the cases. 
In practice, the detection and vetting probability is so low for such cases that our assumption that the pipeline would have found detections between MES of 8 and 7.1 around these 3.2\% of stars would not affect our results significantly. The remaining 0.8\% of stars with somewhat higher MES thresholds are also not a significant source of systematic error.

\label{secReliability}
{\em Reliability:}
Our analysis assumes that all planet candidates (identified by DR25 pipeline and associated with target stars in our cleaned sample) were correctly identified as planet candidates, rather than being some sort of false positive (e.g., diluted eclipsing binary) or false alarm (e.g., astrophysical noise).  
Therefore, technically, the occurrence rates we infer should be regarded as the combined occurrence rate of true planets and any astrophysical false positives or instrumental false alarms that masquerade as planets within the given range of planet sizes and orbital periods.
Fortunately, the overall reliability of the Kepler planet candidate sample is quite high \citep{TCH+2018,BCB+2019}.
Nevertheless, validating any individual planet candidate at high reliability is difficult \citep{BMT+2019}.

\citet{BCB+2019} quantified the impact of vetting efficiency, false positives, and potential false alarms on planet occurrence rate estimates.  Their results indicate that it is important to consider vetting efficiency for a wide range of planet periods and the reliability of DR25 planet candidates is very high, except for orbital planets greater than 250 days, which likely suffer from some level of false alarms.  The results from our study include a detailed model for vetting efficiency, so we expect the occurrence rates we report in Figures \ref{figRates} \& \ref{figInteg} and Table \ref{tab:occ_rates} are accurate planet occurrence rates for all but the bins with the longest orbital period.  For the last bin ($P=256-500$d) and for habitable zone estimates (see Figures \ref{figPost} \& \ref{figRateComp} and Table \ref{tab:HZrate}), we caution that the occurrence rates we report could be interpreted as upper limits, due to the potential for a significant fraction of long-period planet candidates being false alarms.  Results of \citet{TCH+2018,BCB+2019} suggest that this is likely less than  a factor of $\sim~2$.

For bins with a small number of detected planets (e.g., small, long-period planets, but also planets with very short-periods and very large planets), a few planet candidates being a false positive or false alarm could cause our inferred rate to be larger than the true planet rate.  
While this could be a large fractional error, it would not contribute significantly to the total number of planets integrated over a broad range of orbital periods and planet sizes.  
Reliability concerns are particularly acute at long orbital periods (e.g., our 256-500d bin), as shown in \citet{TCH+2018,BCB+2019}.  
Complicated interactions of detector noise and the spacecraft orbital period likely contribute to some of the planet candidates at long orbital periods.  
We have attempted to mitigate reliability concerns by analyzing a cleaned set of target stars.
Nevertheless, we regard the apparent increase in the occurrence rate of R=$1.5-6$ R$_\oplus$ planets beyond 256d as highly suggestive, but worthy of a more detailed investigation that includes a detailed treatment of the reliability of long-period planet candidates.  
We note that reliability concerns will only lead to revisions downward in the inferred rate.  
Therefore, our upper limits on the occurrence rates of small planets are robust to reliability concerns.

For periods of 128-256d and 256-500d, upper limits for the sum of occurrence rate over all radii exceeds one planet per star.  
Previous studies have shown that the fraction of stars with planets is significantly less than the average number of planets per star, based largely on the observed distribution of targets with multiple transiting planets.  If the reliability of these long-period planet candidates is shown to be high, then future studies should also investigate the extent to which long-term orbital stability helps to inform the upper limit on the occurrence rate of long-period planets. A high occurrence rate of large-radii planets could dynamically restrict the occurrence rate of smaller radius planets. 

\subsubsection{Recommendations}
\label{secRecommendations}
In conclusion, we recommend that mission concepts aiming to characterize potentially rocky planets in or near the habitable zone of sun-like stars prepare compelling science programs that would be robust to the true rate of $0.75-1.5$ R$_\oplus$ planets with orbital periods in 237-500 days, $f_{R,P} \simeq$ $0.03-0.40$ or a differential rate of $(d^2 f)/[d(\ln P)~d(\ln R_{p})] \simeq$ $0.06-0.76$.  This larger range is based on starting from the 90\% credible interval derived from our calculations, but has been expanded to allow for the possibility of the reliability of such planets being as low as $\simeq~50\%$ or for the rate to increase by up to $\simeq~10\%$ due to issues discussed above.  

\acknowledgments
We thank the entire Kepler team for the many years of work that has proven so successful and was critical to this study.  
We thank an anonymous referee for useful suggestions for improving the analysis and manuscript.  
We thank Gijs Mulders for feedback and sharing updated values for the  Mulders et al.\ (2018) vetting model which allowed us to make more direct the comparisons of occurrence rates based on the different vetting models.
We thank Steve Bryson for discussions of the potential impact of false alarms and reliability.  
D.C.H, E.B.F., D.R., and K.A. acknowledge support from NASA Origins of Solar Systems grant \# NNX14AI76G and Exoplanet Research Program grant \# NNX15AE21.
E.B.F acknowledges support from NASA Kepler Participating Scientist Program Cycle II grant \# NNX14AN76G.
D.C.H, E.B.F. acknowledge support from the Penn State Eberly College of Science and Department of Astronomy \& Astrophysics, the Center for Exoplanets and Habitable Worlds and the Center for Astrostatistics.  
The citations in this paper have made use of NASA's Astrophysics Data System Bibliographic Services.  
This research has made use of the NASA Exoplanet Archive, which is operated by the California Institute of Technology, under contract with the National Aeronautics and Space Administration under the Exoplanet Exploration Program.
This work made use of the gaia-kepler.fun crossmatch database created by Megan Bedell.
We acknowledge the Institute for CyberScience (\url{http://ics.psu.edu/}) at The Pennsylvania State University, including the CyberLAMP cluster supported by NSF grant MRI-1626251, for providing advanced computing resources and services that have contributed to the research results reported in this paper.
This material was based upon work partially supported by the National Science Foundation under Grant DMS-1127914 to the Statistical and Applied Mathematical Sciences Institute (SAMSI). Any opinions, findings, and conclusions or recommendations expressed in this material are those of the author(s) and do not necessarily reflect the views of the National Science Foundation.
This study benefited from the 2013 SAMSI workshop on Modern Statistical and Computational Methods for Analysis of Kepler Data, the 2016/2017 Program on Statistical, Mathematical and Computational Methods for Astronomy, and their associated working groups.

\software{ABC \citep{ABC_Julia},
CORBITS \citep{BR2016},
ExoplanetsSysSim \citep{ExoplanetsSysSim2019},
ExoJulia SysSimData \citep{SysSimData},
Matplotlib \citep{Matploblib}
}

\bibliographystyle{aasjournal}
\bibliography{references}

\begin{thebibliography}{}
\expandafter\ifx\csname natexlab\endcsname\relax\def\natexlab#1{#1}\fi
\providecommand{\url}[1]{\href{#1}{#1}}
\providecommand{\dodoi}[1]{doi:~\href{http://doi.org/#1}{\nolinkurl{#1}}}
\providecommand{\doeprint}[1]{\href{http://ascl.net/#1}{\nolinkurl{http://ascl.net/#1}}}
\providecommand{\doarXiv}[1]{\href{https://arxiv.org/abs/#1}{\nolinkurl{https://arxiv.org/abs/#1}}}

\bibitem[{{Andrae} {et~al.}(2018){Andrae}, {Fouesneau}, {Creevey}, {Ordenovic},
  {Mary}, {Burlacu}, {Chaoul}, {Jean-Antoine-Piccolo}, {Kordopatis}, {Korn},
  {Lebreton}, {Panem}, {Pichon}, {Th{\'e}venin}, {Walmsley}, \&
  {Bailer-Jones}}]{AFC+2018}
{Andrae}, R., {Fouesneau}, M., {Creevey}, O., {et~al.} 2018, \aap, 616, A8,
  \dodoi{10.1051/0004-6361/201732516}

\bibitem[{{Beaumont} {et~al.}(2009){Beaumont}, {Cornuet}, {Marin}, \&
  {Robert}}]{BCM+2008}
{Beaumont}, M.~A., {Cornuet}, J.-M., {Marin}, J.-M., \& {Robert}, C.~P. 2009,
  Biometrika, 96, 983, \dodoi{10.1093/biomet/asp052}

\bibitem[{{Berger} {et~al.}(2018){Berger}, {Huber}, {Gaidos}, \& {van
  Saders}}]{BHG+2018}
{Berger}, T.~A., {Huber}, D., {Gaidos}, E., \& {van Saders}, J.~L. 2018, \apj,
  866, 99, \dodoi{10.3847/1538-4357/aada83}

\bibitem[{{Bezanson} {et~al.}(2017){Bezanson}, {Edelman}, {Karpinski}, \&
  {Shah}}]{Julia}
{Bezanson}, J., {Edelman}, A., {Karpinski}, S., \& {Shah}, V.~B. 2017, SIAM
  Review, 59, 65, \dodoi{10.1137/141000671}

\bibitem[{{Brakensiek} \& {Ragozzine}(2016)}]{BR2016}
{Brakensiek}, J., \& {Ragozzine}, D. 2016, \apj, 821, 47,
  \dodoi{10.3847/0004-637X/821/1/47}

\bibitem[{{Bryson} {et~al.}(2019){Bryson}, {Coughlin}, {Batalha}, {Berger},
  {Huber}, {Burke}, \& {Mullally}}]{BCB+2019}
{Bryson}, S., {Coughlin}, J., {Batalha}, N.~M., {et~al.} 2019, arXiv e-prints,
  arXiv:1906.03575.
\newblock \doarXiv{1906.03575}

\bibitem[{{Burke} \& {Catanzarite}(2017{\natexlab{a}})}]{BC2017Wf}
{Burke}, C.~J., \& {Catanzarite}, J. 2017{\natexlab{a}}, {Planet Detection
  Metrics: Window and One-Sigma Depth Functions for Data Release 25}, Tech.
  rep.

\bibitem[{{Burke} \& {Catanzarite}(2017{\natexlab{b}})}]{BC2017Flux}
---. 2017{\natexlab{b}}, {Planet Detection Metrics: Per-Target Flux-Level
  Transit Injection Tests of TPS for Data Release 25}, Tech. rep.

\bibitem[{{Burke} \& {Catanzarite}(2017{\natexlab{c}})}]{BC2017DetCon}
---. 2017{\natexlab{c}}, {Planet Detection Metrics: Per-Target Detection
  Contours for Data Release 25}, Tech. rep.

\bibitem[{{Burke} {et~al.}(2019){Burke}, {Mullally}, {Thompson}, {Coughlin}, \&
  {Rowe}}]{BMT+2019}
{Burke}, C.~J., {Mullally}, F., {Thompson}, S.~E., {Coughlin}, J.~L., \&
  {Rowe}, J.~F. 2019, \aj, 157, 143, \dodoi{10.3847/1538-3881/aafb79}

\bibitem[{{Burke} {et~al.}(2015){Burke}, {Christiansen}, {Mullally}, {Seader},
  {Huber}, {Rowe}, {Coughlin}, {Thompson}, {Catanzarite}, {Clarke}, {Morton},
  {Caldwell}, {Bryson}, {Haas}, {Batalha}, {Jenkins}, {Tenenbaum}, {Twicken},
  {Li}, {Quintana}, {Barclay}, {Henze}, {Borucki}, {Howell}, \&
  {Still}}]{BCM+2015}
{Burke}, C.~J., {Christiansen}, J.~L., {Mullally}, F., {et~al.} 2015, \apj,
  809, 8, \dodoi{10.1088/0004-637X/809/1/8}

\bibitem[{{Christiansen}(2017)}]{C2017Pixel}
{Christiansen}, J.~L. 2017, {Planet Detection Metrics: Pixel-Level Transit
  Injection Tests of Pipeline Detection Efficiency for Data Release 25}, Tech.
  rep.

\bibitem[{{Christiansen} {et~al.}(2015){Christiansen}, {Clarke}, {Burke},
  {Seader}, {Jenkins}, {Twicken}, {Catanzarite}, {Smith}, {Batalha}, {Haas},
  {Thompson}, {Campbell}, {Sabale}, \& {Kamal Uddin}}]{CCB+2015}
{Christiansen}, J.~L., {Clarke}, B.~D., {Burke}, C.~J., {et~al.} 2015, \apj,
  810, 95, \dodoi{10.1088/0004-637X/810/2/95}

\bibitem[{{Coughlin}(2017)}]{C2017Robo}
{Coughlin}, J.~L. 2017, {Planet Detection Metrics: Robovetter Completeness and
  Effectiveness for Data Release 25}, Tech. rep.

\bibitem[{{Dong} \& {Zhu}(2013)}]{DZ2013}
{Dong}, S., \& {Zhu}, Z. 2013, \apj, 778, 53,
  \dodoi{10.1088/0004-637X/778/1/53}

\bibitem[{{Evans}(2018)}]{E2018}
{Evans}, D.~F. 2018, Research Notes of the American Astronomical Society, 2,
  20, \dodoi{10.3847/2515-5172/aac173}

\bibitem[{Ford(2019)}]{SysSimData}
Ford, E. 2019, {ExoJulia/SysSimData: Initial release of Data Files for the
  Exoplanet System Simulator}, \dodoi{10.5281/zenodo.3255313}.
\newblock \url{https://doi.org/10.5281/zenodo.3255313}

\bibitem[{{Ford} {et~al.}(2018){Ford}, {He}, {Hsu}, \& {Ragozzine}}]{ABC_Julia}
{Ford}, E.~B., {He}, M., {Hsu}, D.~C., \& {Ragozzine}, D. 2018, {Approximate
  Bayesian Computing with Julia}, 1.0,  Zenodo, \dodoi{10.5281/zenodo.1198716}.
\newblock \url{https://doi.org/10.5281/zenodo.1198716}

\bibitem[{{Ford} {et~al.}(2019){Ford}, {He}, {Hsu}, \&
  {Ragozzine}}]{ExoplanetsSysSim2019}
---. 2019, {Planetary Systems Simulation \& Model of Kepler Mission for
  characterizing the Occurrence Rates of Exoplanets and Planetary
  Architectures}, 1.0,  Zenodo, \dodoi{10.5281/zenodo.3237792}.
\newblock \url{https://doi.org/10.5281/zenodo.3237792}

\bibitem[{{Foreman-Mackey} {et~al.}(2014){Foreman-Mackey}, {Hogg}, \&
  {Morton}}]{FHM2014}
{Foreman-Mackey}, D., {Hogg}, D.~W., \& {Morton}, T.~D. 2014, \apj, 795, 64,
  \dodoi{10.1088/0004-637X/795/1/64}

\bibitem[{{Fulton} {et~al.}(2017){Fulton}, {Petigura}, {Howard}, {Isaacson},
  {Marcy}, {Cargile}, {Hebb}, {Weiss}, {Johnson}, {Morton}, {Sinukoff},
  {Crossfield}, \& {Hirsch}}]{FPH+2017}
{Fulton}, B.~J., {Petigura}, E.~A., {Howard}, A.~W., {et~al.} 2017, \aj, 154,
  109, \dodoi{10.3847/1538-3881/aa80eb}

\bibitem[{{Gaia Collaboration} {et~al.}(2018){Gaia Collaboration}, {Brown},
  {Vallenari}, {Prusti}, {de Bruijne}, {Babusiaux}, {Bailer-Jones}, {Biermann},
  {Evans}, {Eyer}, {Jansen}, {Jordi}, {Klioner}, {Lammers}, {Lindegren},
  {Luri}, {Mignard}, {Panem}, {Pourbaix}, {Randich}, {Sartoretti}, {Siddiqui},
  {Soubiran}, {van Leeuwen}, {Walton}, {Arenou}, {Bastian}, {Cropper},
  {Drimmel}, {Katz}, {Lattanzi}, {Bakker}, {Cacciari}, {Casta{\~n}eda},
  {Chaoul}, {Cheek}, {De Angeli}, {Fabricius}, {Guerra}, {Holl}, {Masana},
  {Messineo}, {Mowlavi}, {Nienartowicz}, {Panuzzo}, {Portell}, {Riello},
  {Seabroke}, {Tanga}, {Th{\'e}venin}, {Gracia-Abril}, {Comoretto},
  {Garcia-Reinaldos}, {Teyssier}, {Altmann}, {Andrae}, {Audard},
  {Bellas-Velidis}, {Benson}, {Berthier}, {Blomme}, {Burgess}, {Busso},
  {Carry}, {Cellino}, {Clementini}, {Clotet}, {Creevey}, {Davidson}, {De
  Ridder}, {Delchambre}, {Dell'Oro}, {Ducourant}, {Fern{\'a}ndez-
  Hern{\'a}ndez}, {Fouesneau}, {Fr{\'e}mat}, {Galluccio}, {Garc{\'\i}a-Torres},
  {Gonz{\'a}lez-N{\'u}{\~n}ez}, {Gonz{\'a}lez-Vidal}, {Gosset}, {Guy},
  {Halbwachs}, {Hambly}, {Harrison}, {Hern{\'a}ndez}, {Hestroffer}, {Hodgkin},
  {Hutton}, {Jasniewicz}, {Jean-Antoine-Piccolo}, {Jordan}, {Korn},
  {Krone-Martins}, {Lanzafame}, {Lebzelter}, {L{\"o}ffler}, {Manteiga},
  {Marrese}, {Mart{\'\i}n-Fleitas}, {Moitinho}, {Mora}, {Muinonen}, {Osinde},
  {Pancino}, {Pauwels}, {Petit}, {Recio-Blanco}, {Richards}, {Rimoldini},
  {Robin}, {Sarro}, {Siopis}, {Smith}, {Sozzetti}, {S{\"u}veges}, {Torra}, {van
  Reeven}, {Abbas}, {Abreu Aramburu}, {Accart}, {Aerts}, {Altavilla},
  {{\'A}lvarez}, {Alvarez}, {Alves}, {Anderson}, {Andrei}, {Anglada Varela},
  {Antiche}, {Antoja}, {Arcay}, {Astraatmadja}, {Bach}, {Baker},
  {Balaguer-N{\'u}{\~n}ez}, {Balm}, {Barache}, {Barata}, {Barbato}, {Barblan},
  {Barklem}, {Barrado}, {Barros}, {Barstow}, {Bartholom{\'e} Mu{\~n}oz},
  {Bassilana}, {Becciani}, {Bellazzini}, {Berihuete}, {Bertone}, {Bianchi},
  {Bienaym{\'e}}, {Blanco-Cuaresma}, {Boch}, {Boeche}, {Bombrun}, {Borrachero},
  {Bossini}, {Bouquillon}, {Bourda}, {Bragaglia}, {Bramante}, {Breddels},
  {Bressan}, {Brouillet}, {Br{\"u}semeister}, {Brugaletta}, {Bucciarelli},
  {Burlacu}, {Busonero}, {Butkevich}, {Buzzi}, {Caffau}, {Cancelliere},
  {Cannizzaro}, {Cantat-Gaudin}, {Carballo}, {Carlucci}, {Carrasco},
  {Casamiquela}, {Castellani}, {Castro-Ginard}, {Charlot}, {Chemin},
  {Chiavassa}, {Cocozza}, {Costigan}, {Cowell}, {Crifo}, {Crosta}, {Crowley},
  {Cuypers}, {Dafonte}, {Damerdji}, {Dapergolas}, {David}, {David}, {de
  Laverny}, {De Luise}, {De March}, {de Martino}, {de Souza}, {de Torres},
  {Debosscher}, {del Pozo}, {Delbo}, {Delgado}, {Delgado}, {Di Matteo},
  {Diakite}, {Diener}, {Distefano}, {Dolding}, {Drazinos}, {Dur{\'a}n},
  {Edvardsson}, {Enke}, {Eriksson}, {Esquej}, {Eynard Bontemps}, {Fabre},
  {Fabrizio}, {Faigler}, {Falc{\~a}o}, {Farr{\`a}s Casas}, {Federici},
  {Fedorets}, {Fernique}, {Figueras}, {Filippi}, {Findeisen}, {Fonti},
  {Fraile}, {Fraser}, {Fr{\'e}zouls}, {Gai}, {Galleti}, {Garabato},
  {Garc{\'\i}a-Sedano}, {Garofalo}, {Garralda}, {Gavel}, {Gavras}, {Gerssen},
  {Geyer}, {Giacobbe}, {Gilmore}, {Girona}, {Giuffrida}, {Glass}, {Gomes},
  {Granvik}, {Gueguen}, {Guerrier}, {Guiraud}, {Guti{\'e}rrez-S{\'a}nchez},
  {Haigron}, {Hatzidimitriou}, {Hauser}, {Haywood}, {Heiter}, {Helmi}, {Heu},
  {Hilger}, {Hobbs}, {Hofmann}, {Holland}, {Huckle}, {Hypki}, {Icardi},
  {Jan{\ss}en}, {Jevardat de Fombelle}, {Jonker}, {Juh{\'a}sz}, {Julbe},
  {Karampelas}, {Kewley}, {Klar}, {Kochoska}, {Kohley}, {Kolenberg},
  {Kontizas}, {Kontizas}, {Koposov}, {Kordopatis}, {Kostrzewa-Rutkowska},
  {Koubsky}, {Lambert}, {Lanza}, {Lasne}, {Lavigne}, {Le Fustec}, {Le
  Poncin-Lafitte}, {Lebreton}, {Leccia}, {Leclerc}, {Lecoeur-Taibi},
  {Lenhardt}, {Leroux}, {Liao}, {Licata}, {Lindstr{\o}m}, {Lister}, {Livanou},
  {Lobel}, {L{\'o}pez}, {Managau}, {Mann}, {Mantelet}, {Marchal}, {Marchant},
  {Marconi}, {Marinoni}, {Marschalk{\'o}}, {Marshall}, {Martino}, {Marton},
  {Mary}, {Massari}, {Matijevi{\v{c}}}, {Mazeh}, {McMillan}, {Messina},
  {Michalik}, {Millar}, {Molina}, {Molinaro}, {Moln{\'a}r}, {Montegriffo},
  {Mor}, {Morbidelli}, {Morel}, {Morris}, {Mulone}, {Muraveva}, {Musella},
  {Nelemans}, {Nicastro}, {Noval}, {O'Mullane}, {Ord{\'e}novic},
  {Ord{\'o}{\~n}ez-Blanco}, {Osborne}, {Pagani}, {Pagano}, {Pailler},
  {Palacin}, {Palaversa}, {Panahi}, {Pawlak}, {Piersimoni}, {Pineau}, {Plachy},
  {Plum}, {Poggio}, {Poujoulet}, {Pr{\v{s}}a}, {Pulone}, {Racero}, {Ragaini},
  {Rambaux}, {Ramos-Lerate}, {Regibo}, {Reyl{\'e}}, {Riclet}, {Ripepi}, {Riva},
  {Rivard}, {Rixon}, {Roegiers}, {Roelens}, {Romero-G{\'o}mez}, {Rowell},
  {Royer}, {Ruiz-Dern}, {Sadowski}, {Sagrist{\`a} Sell{\'e}s}, {Sahlmann},
  {Salgado}, {Salguero}, {Sanna}, {Santana- Ros}, {Sarasso}, {Savietto},
  {Schultheis}, {Sciacca}, {Segol}, {Segovia}, {S{\'e}gransan}, {Shih},
  {Siltala}, {Silva}, {Smart}, {Smith}, {Solano}, {Solitro}, {Sordo}, {Soria
  Nieto}, {Souchay}, {Spagna}, {Spoto}, {Stampa}, {Steele},
  {Steidelm{\"u}ller}, {Stephenson}, {Stoev}, {Suess}, {Surdej}, {Szabados},
  {Szegedi-Elek}, {Tapiador}, {Taris}, {Tauran}, {Taylor}, {Teixeira},
  {Terrett}, {Teyssandier}, {Thuillot}, {Titarenko}, {Torra Clotet}, {Turon},
  {Ulla}, {Utrilla}, {Uzzi}, {Vaillant}, {Valentini}, {Valette}, {van Elteren},
  {Van Hemelryck}, {van Leeuwen}, {Vaschetto}, {Vecchiato}, {Veljanoski},
  {Viala}, {Vicente}, {Vogt}, {von Essen}, {Voss}, {Votruba}, {Voutsinas},
  {Walmsley}, {Weiler}, {Wertz}, {Wevers}, {Wyrzykowski}, {Yoldas},
  {{\v{Z}}erjal}, {Ziaeepour}, {Zorec}, {Zschocke}, {Zucker}, {Zurbach}, \&
  {Zwitter}}]{GBV+2018}
{Gaia Collaboration}, {Brown}, A.~G.~A., {Vallenari}, A., {et~al.} 2018, \aap,
  616, A1, \dodoi{10.1051/0004-6361/201833051}

\bibitem[{{Gupta} \& {Schlichting}(2019)}]{GS2018}
{Gupta}, A., \& {Schlichting}, H.~E. 2019, \mnras, 487, 24,
  \dodoi{10.1093/mnras/stz1230}

\bibitem[{{Hirsch} {et~al.}(2017){Hirsch}, {Ciardi}, {Howard}, {Everett},
  {Furlan}, {Saylors}, {Horch}, {Howell}, {Teske}, \& {Marcy}}]{HCH+2017}
{Hirsch}, L.~A., {Ciardi}, D.~R., {Howard}, A.~W., {et~al.} 2017, \aj, 153,
  117, \dodoi{10.3847/1538-3881/153/3/117}

\bibitem[{{Hoffman} \& {Rowe}(2017)}]{H2017MCMC}
{Hoffman}, Kelsey, L., \& {Rowe}, J.~F. 2017, {Uniform Modeling of KOIs: MCMC
  Notes for Data Release 25}, Tech. rep.

\bibitem[{{Hsu} {et~al.}(2018){Hsu}, {Ford}, {Ragozzine}, \&
  {Morehead}}]{HFR+2018}
{Hsu}, D.~C., {Ford}, E.~B., {Ragozzine}, D., \& {Morehead}, R.~C. 2018, \aj,
  155, 205, \dodoi{10.3847/1538-3881/aab9a8}

\bibitem[{{Hunter}(2007)}]{Matploblib}
{Hunter}, J.~D. 2007, Computing in Science and Engineering, 9, 90,
  \dodoi{10.1109/MCSE.2007.55}

\bibitem[{{Jenkins} {et~al.}(2015){Jenkins}, {Twicken}, {Batalha}, {Caldwell},
  {Cochran}, {Endl}, {Latham}, {Esquerdo}, {Seader}, {Bieryla}, {Petigura},
  {Ciardi}, {Marcy}, {Isaacson}, {Huber}, {Rowe}, {Torres}, {Bryson},
  {Buchhave}, {Ramirez}, {Wolfgang}, {Li}, {Campbell}, {Tenenbaum},
  {Sanderfer}, {Henze}, {Catanzarite}, {Gilliland}, \& {Borucki}}]{JTB+2015}
{Jenkins}, J.~M., {Twicken}, J.~D., {Batalha}, N.~M., {et~al.} 2015, \aj, 150,
  56, \dodoi{10.1088/0004-6256/150/2/56}

\bibitem[{{Kipping}(2010)}]{K2010}
{Kipping}, D.~M. 2010, \mnras, 407, 301,
  \dodoi{10.1111/j.1365-2966.2010.16894.x}

\bibitem[{Lance \& Williams(1967)}]{LW1967}
Lance, G.~N., \& Williams, W.~T. 1967, Australian Computer Journal, 1, 15

\bibitem[{{Lopez} \& {Fortney}(2014)}]{LF2014}
{Lopez}, E.~D., \& {Fortney}, J.~J. 2014, \apj, 792, 1,
  \dodoi{10.1088/0004-637X/792/1/1}

\bibitem[{{Mandel} \& {Agol}(2002)}]{MA2002}
{Mandel}, K., \& {Agol}, E. 2002, \apj, 580, L171, \dodoi{10.1086/345520}

\bibitem[{{Mulders} {et~al.}(2018){Mulders}, {Pascucci}, {Apai}, \&
  {Ciesla}}]{MPA+2018}
{Mulders}, G.~D., {Pascucci}, I., {Apai}, D., \& {Ciesla}, F.~J. 2018, \aj,
  156, 24, \dodoi{10.3847/1538-3881/aac5ea}

\bibitem[{{Mullally} {et~al.}(2018){Mullally}, {Thompson}, {Coughlin}, {Burke},
  \& {Rowe}}]{MTC+2018}
{Mullally}, F., {Thompson}, S.~E., {Coughlin}, J.~L., {Burke}, C.~J., \&
  {Rowe}, J.~F. 2018, \aj, 155, 210, \dodoi{10.3847/1538-3881/aabae3}

\bibitem[{{Owen} \& {Wu}(2017)}]{OW2017}
{Owen}, J.~E., \& {Wu}, Y. 2017, \apj, 847, 29,
  \dodoi{10.3847/1538-4357/aa890a}

\bibitem[{{Petigura} {et~al.}(2013){Petigura}, {Howard}, \& {Marcy}}]{PHM2013}
{Petigura}, E.~A., {Howard}, A.~W., \& {Marcy}, G.~W. 2013, Proceedings of the
  National Academy of Science, 110, 19273, \dodoi{10.1073/pnas.1319909110}

\bibitem[{{Price} \& {Rogers}(2014)}]{PR2014}
{Price}, E.~M., \& {Rogers}, L.~A. 2014, \apj, 794, 92,
  \dodoi{10.1088/0004-637X/794/1/92}

\bibitem[{{Shabram} {et~al.}(2019){Shabram}, {Batalha}, {Thompson}, {Hsu},
  {Ford}, {Christiansen}, {Huber}, {Berger}, {Catanzarite}, {Nelson}, {Bryson},
  {Belikov}, {Burke}, \& {Caldwell}}]{SBT+2019}
{Shabram}, M.~I., {Batalha}, N., {Thompson}, S.~E., {et~al.} 2019, arXiv
  e-prints, arXiv:1908.00203.
\newblock \doarXiv{1908.00203}

\bibitem[{{Teske} {et~al.}(2018){Teske}, {Ciardi}, {Howell}, {Hirsch}, \&
  {Johnson}}]{TeCH+2018}
{Teske}, J.~K., {Ciardi}, D.~R., {Howell}, S.~B., {Hirsch}, L.~A., \&
  {Johnson}, R.~A. 2018, \aj, 156, 292, \dodoi{10.3847/1538-3881/aaed2d}

\bibitem[{{Thompson} {et~al.}(2018){Thompson}, {Coughlin}, {Hoffman},
  {Mullally}, {Christiansen}, {Burke}, {Bryson}, {Batalha}, {Haas},
  {Catanzarite}, {Rowe}, {Barentsen}, {Caldwell}, {Clarke}, {Jenkins}, {Li},
  {Latham}, {Lissauer}, {Mathur}, {Morris}, {Seader}, {Smith}, {Klaus},
  {Twicken}, {Van Cleve}, {Wohler}, {Akeson}, {Ciardi}, {Cochran}, {Henze},
  {Howell}, {Huber}, {Pr{\v s}a}, {Ram{\'{\i}}rez}, {Morton}, {Barclay},
  {Campbell}, {Chaplin}, {Charbonneau}, {Christensen-Dalsgaard}, {Dotson},
  {Doyle}, {Dunham}, {Dupree}, {Ford}, {Geary}, {Girouard}, {Isaacson},
  {Kjeldsen}, {Quintana}, {Ragozzine}, {Shabram}, {Shporer}, {Silva Aguirre},
  {Steffen}, {Still}, {Tenenbaum}, {Welsh}, {Wolfgang}, {Zamudio}, {Koch}, \&
  {Borucki}}]{TCH+2018}
{Thompson}, S.~E., {Coughlin}, J.~L., {Hoffman}, K., {et~al.} 2018, \apjs, 235,
  38, \dodoi{10.3847/1538-4365/aab4f9}

\bibitem[{{Twicken} {et~al.}(2016){Twicken}, {Jenkins}, {Seader}, {Tenenbaum},
  {Smith}, {Brownston}, {Burke}, {Catanzarite}, {Clarke}, {Cote}, {Girouard},
  {Klaus}, {Li}, {McCauliff}, {Morris}, {Wohler}, {Campbell}, {Kamal Uddin},
  {Zamudio}, {Sabale}, {Bryson}, {Caldwell}, {Christiansen}, {Coughlin},
  {Haas}, {Henze}, {Sanderfer}, \& {Thompson}}]{TJS+2016}
{Twicken}, J.~D., {Jenkins}, J.~M., {Seader}, S.~E., {et~al.} 2016, \aj, 152,
  158, \dodoi{10.3847/0004-6256/152/6/158}

\bibitem[{{Van Eylen} {et~al.}(2018){Van Eylen}, {Agentoft}, {Lundkvist},
  {Kjeldsen}, {Owen}, {Fulton}, {Petigura}, \& {Snellen}}]{VAL+2017}
{Van Eylen}, V., {Agentoft}, C., {Lundkvist}, M.~S., {et~al.} 2018, \mnras,
  479, 4786, \dodoi{10.1093/mnras/sty1783}

\bibitem[{{Youdin}(2011)}]{Y2011}
{Youdin}, A.~N. 2011, \apj, 742, 38, \dodoi{10.1088/0004-637X/742/1/38}

\bibitem[{{Zink} {et~al.}(2019){Zink}, {Christiansen}, \& {Hansen}}]{ZCH2019}
{Zink}, J.~K., {Christiansen}, J.~L., \& {Hansen}, B. M.~S. 2019, \mnras, 483,
  4479, \dodoi{10.1093/mnras/sty3463}

\end{thebibliography}

\appendix
\section{ExoPAG Grid DR25 Occurrence Rates}
Below in Table \ref{tab:exopag_occ_rates} we present our DR25 occurrence rates for 
F, G and K stellar types as defined by the Study Analysis Group (SAG) 13 of the NASA Exoplanet Exploration Program Analysis Group (ExoPAG).  The combined detection and vetting efficiency is used.
\startlongtable
\begin{deluxetable*}{rrrrr}
\tablewidth{0pt}
\tablecaption{ExoPaG Occurrence Rates
\label{tab:exopag_occ_rates}}
\tablehead{
\colhead{Period}&
\colhead{Radius}&
\colhead{Type F}&
\colhead{Type G}&
\colhead{Type K}\\
\colhead{(days)}&
\colhead{($R_{\oplus}$)}&
&
&
}

\startdata
$\phn10.00-\phn20.00$&$\phn0.67-\phn1.00$&$1.32^{+0.98}_{-0.65}\times10^{-1}$&$3.9^{+2.1}_{-1.7}\times10^{-2}$&$1.27^{+0.59}_{-0.49}\times10^{-1}$\\
$\phn10.00-\phn20.00$&$\phn1.00-\phn1.50$&$1.23^{+0.37}_{-0.34}\times10^{-1}$&$7.1^{+1.7}_{-1.6}\times10^{-2}$&$7.2^{+1.9}_{-1.7}\times10^{-2}$\\
$\phn10.00-\phn20.00$&$\phn1.50-\phn2.25$&$3.4^{+2.0}_{-1.4}\times10^{-2}$&$5.14^{+1.16}_{-0.94}\times10^{-2}$&$7.3^{+1.5}_{-1.3}\times10^{-2}$\\
$\phn10.00-\phn20.00$&$\phn2.25-\phn3.38$&$2.4^{+1.1}_{-1.0}\times10^{-2}$&$6.14^{+0.99}_{-0.86}\times10^{-2}$&$8.1^{+1.2}_{-1.1}\times10^{-2}$\\
$\phn10.00-\phn20.00$&$\phn3.38-\phn5.06$&$<7.9\times10^{-3}$&$8.8^{+4.8}_{-4.0}\times10^{-3}$&$6.5^{+5.7}_{-3.7}\times10^{-3}$\\
$\phn10.00-\phn20.00$&$\phn5.06-\phn7.59$&$<5.2\times10^{-3}$&$3.1^{+2.2}_{-1.6}\times10^{-3}$&$5.7^{+2.3}_{-3.4}\times10^{-3}$\\
$\phn10.00-\phn20.00$&$\phn7.59-11.39$&$<3.7\times10^{-3}$&$1.46^{+1.43}_{-0.86}\times10^{-3}$&$4.0^{+3.9}_{-2.2}\times10^{-3}$\\
$\phn10.00-\phn20.00$&$11.39-17.09$&$<4.3\times10^{-3}$&$1.7^{+1.2}_{-1.0}\times10^{-3}$&$<2.1\times10^{-3}$\\
\hline
$\phn20.00-\phn40.00$&$\phn0.67-\phn1.00$&$2.3^{+1.9}_{-1.2}\times10^{-1}$&$6.7^{+5.9}_{-3.8}\times10^{-2}$&$1.51^{+0.91}_{-0.70}\times10^{-1}$\\
$\phn20.00-\phn40.00$&$\phn1.00-\phn1.50$&$6.3^{+3.8}_{-2.7}\times10^{-2}$&$4.7^{+2.1}_{-1.6}\times10^{-2}$&$7.3^{+3.1}_{-2.7}\times10^{-2}$\\
$\phn20.00-\phn40.00$&$\phn1.50-\phn2.25$&$6.3^{+2.7}_{-2.5}\times10^{-2}$&$4.3^{+1.1}_{-1.2}\times10^{-2}$&$8.3^{+1.9}_{-1.6}\times10^{-2}$\\
$\phn20.00-\phn40.00$&$\phn2.25-\phn3.38$&$6.1^{+2.3}_{-2.1}\times10^{-2}$&$9.7^{+1.3}_{-1.2}\times10^{-2}$&$6.7^{+1.4}_{-1.2}\times10^{-2}$\\
$\phn20.00-\phn40.00$&$\phn3.38-\phn5.06$&$1.15^{+1.59}_{-0.75}\times10^{-2}$&$9.3^{+6.2}_{-4.4}\times10^{-3}$&$1.30^{+0.91}_{-0.67}\times10^{-2}$\\
$\phn20.00-\phn40.00$&$\phn5.06-\phn7.59$&$1.21^{+1.03}_{-0.63}\times10^{-2}$&$2.6^{+2.1}_{-1.5}\times10^{-3}$&$6.2^{+7.6}_{-3.8}\times10^{-3}$\\
$\phn20.00-\phn40.00$&$\phn7.59-11.39$&$<1.6\times10^{-2}$&$<2.0\times10^{-3}$&$5.8^{+4.2}_{-2.9}\times10^{-3}$\\
$\phn20.00-\phn40.00$&$11.39-17.09$&$<1.6\times10^{-2}$&$3.6^{+2.9}_{-1.9}\times10^{-3}$&$<3.4\times10^{-3}$\\
\hline
$\phn40.00-\phn80.00$&$\phn0.67-\phn1.00$&$<4.1\times10^{-1}$&$<1.7\times10^{-1}$&$<1.7\times10^{-1}$\\
$\phn40.00-\phn80.00$&$\phn1.00-\phn1.50$&$<7.9\times10^{-2}$&$8.2^{+4.0}_{-3.0}\times10^{-2}$&$9.3^{+5.2}_{-4.3}\times10^{-2}$\\
$\phn40.00-\phn80.00$&$\phn1.50-\phn2.25$&$4.9^{+3.8}_{-2.3}\times10^{-2}$&$4.7^{+2.1}_{-1.6}\times10^{-2}$&$9.8^{+3.1}_{-2.3}\times10^{-2}$\\
$\phn40.00-\phn80.00$&$\phn2.25-\phn3.38$&$3.8^{+2.9}_{-1.6}\times10^{-2}$&$1.34^{+0.21}_{-0.19}\times10^{-1}$&$7.1^{+2.8}_{-1.9}\times10^{-2}$\\
$\phn40.00-\phn80.00$&$\phn3.38-\phn5.06$&$<1.8\times10^{-2}$&$5.9^{+6.6}_{-4.0}\times10^{-3}$&$2.6^{+1.5}_{-1.2}\times10^{-2}$\\
$\phn40.00-\phn80.00$&$\phn5.06-\phn7.59$&$<2.4\times10^{-2}$&$1.03^{+0.91}_{-0.42}\times10^{-2}$&$6.9^{+6.4}_{-4.2}\times10^{-3}$\\
$\phn40.00-\phn80.00$&$\phn7.59-11.39$&$<1.6\times10^{-2}$&$5.0^{+4.5}_{-3.3}\times10^{-3}$&$7.5^{+7.5}_{-3.7}\times10^{-3}$\\
$\phn40.00-\phn80.00$&$11.39-17.09$&$<2.6\times10^{-2}$&$5.7^{+4.4}_{-2.9}\times10^{-3}$&$8.3^{+7.0}_{-5.1}\times10^{-3}$\\
\hline
$\phn80.00-160.00$&$\phn0.67-\phn1.00$&$<10.0\times10^{-1}$&$<6.6\times10^{-1}$&$<1.0\times10^{0}$\\
$\phn80.00-160.00$&$\phn1.00-\phn1.50$&$<4.5\times10^{-1}$&$<8.7\times10^{-2}$&$1.67^{+1.03}_{-0.89}\times10^{-1}$\\
$\phn80.00-160.00$&$\phn1.50-\phn2.25$&$9.7^{+5.4}_{-6.2}\times10^{-2}$&$7.7^{+3.7}_{-2.8}\times10^{-2}$&$1.01^{+0.46}_{-0.29}\times10^{-1}$\\
$\phn80.00-160.00$&$\phn2.25-\phn3.38$&$6.8^{+5.1}_{-4.0}\times10^{-2}$&$1.14^{+0.29}_{-0.19}\times10^{-1}$&$1.22^{+0.34}_{-0.28}\times10^{-1}$\\
$\phn80.00-160.00$&$\phn3.38-\phn5.06$&$<2.5\times10^{-2}$&$2.8^{+1.4}_{-1.1}\times10^{-2}$&$1.5^{+1.4}_{-1.0}\times10^{-2}$\\
$\phn80.00-160.00$&$\phn5.06-\phn7.59$&$3.1^{+3.9}_{-1.8}\times10^{-2}$&$4.9^{+5.6}_{-3.1}\times10^{-3}$&$1.97^{+1.32}_{-0.94}\times10^{-2}$\\
$\phn80.00-160.00$&$\phn7.59-11.39$&$<3.7\times10^{-2}$&$2.13^{+0.89}_{-0.87}\times10^{-2}$&$<1.8\times10^{-2}$\\
$\phn80.00-160.00$&$11.39-17.09$&$<2.4\times10^{-2}$&$1.14^{+1.06}_{-0.57}\times10^{-2}$&$<6.5\times10^{-3}$\\
\hline
$160.00-320.00$&$\phn0.67-\phn1.00$&$<9.2\times10^{-1}$&$<9.7\times10^{-1}$&$<9.6\times10^{-1}$\\
$160.00-320.00$&$\phn1.00-\phn1.50$&$<6.5\times10^{-1}$&$<4.0\times10^{-1}$&$<2.8\times10^{-1}$\\
$160.00-320.00$&$\phn1.50-\phn2.25$&$<2.7\times10^{-1}$&$1.26^{+0.81}_{-0.49}\times10^{-1}$&$1.27^{+0.69}_{-0.61}\times10^{-1}$\\
$160.00-320.00$&$\phn2.25-\phn3.38$&$1.57^{+1.17}_{-0.64}\times10^{-1}$&$9.6^{+4.7}_{-2.9}\times10^{-2}$&$1.17^{+0.57}_{-0.39}\times10^{-1}$\\
$160.00-320.00$&$\phn3.38-\phn5.06$&$<6.4\times10^{-2}$&$2.0^{+1.7}_{-1.2}\times10^{-2}$&$4.5^{+2.8}_{-2.5}\times10^{-2}$\\
$160.00-320.00$&$\phn5.06-\phn7.59$&$<4.5\times10^{-2}$&$2.7^{+1.3}_{-1.4}\times10^{-2}$&$<2.4\times10^{-2}$\\
$160.00-320.00$&$\phn7.59-11.39$&$4.4^{+3.9}_{-2.6}\times10^{-2}$&$2.10^{+1.58}_{-0.96}\times10^{-2}$&$3.0^{+4.7}_{-1.7}\times10^{-2}$\\
$160.00-320.00$&$11.39-17.09$&$<3.9\times10^{-2}$&$<9.6\times10^{-3}$&$<3.2\times10^{-2}$\\
\hline
$320.00-640.00$&$\phn0.67-\phn1.00$&$<8.3\times10^{-1}$&$<9.0\times10^{-1}$&$<9.0\times10^{-1}$\\
$320.00-640.00$&$\phn1.00-\phn1.50$&$<7.8\times10^{-1}$&$<6.9\times10^{-1}$&$<7.6\times10^{-1}$\\
$320.00-640.00$&$\phn1.50-\phn2.25$&$6.7^{+3.5}_{-3.7}\times10^{-1}$&$6.3^{+2.6}_{-2.4}\times10^{-1}$&$3.9^{+2.8}_{-2.0}\times10^{-1}$\\
$320.00-640.00$&$\phn2.25-\phn3.38$&$3.8^{+3.1}_{-2.2}\times10^{-1}$&$1.22^{+1.01}_{-0.63}\times10^{-1}$&$2.64^{+1.45}_{-0.96}\times10^{-1}$\\
$320.00-640.00$&$\phn3.38-\phn5.06$&$<2.4\times10^{-1}$&$3.5^{+3.3}_{-2.5}\times10^{-2}$&$1.32^{+1.12}_{-0.73}\times10^{-1}$\\
$320.00-640.00$&$\phn5.06-\phn7.59$&$<1.0\times10^{-1}$&$3.0^{+3.3}_{-1.5}\times10^{-2}$&$<7.0\times10^{-2}$\\
$320.00-640.00$&$\phn7.59-11.39$&$<1.1\times10^{-1}$&$<2.7\times10^{-2}$&$<7.8\times10^{-2}$\\
$320.00-640.00$&$11.39-17.09$&$<1.1\times10^{-1}$&$<2.9\times10^{-2}$&$<5.5\times10^{-2}$\\
\enddata

\tablecomments{ExoPAG grid estimated occurrence rates for DR25 KOI catalog planet candidates associated with FGK stars using the combined detection and vetting efficiency.}
\end{deluxetable*}

\listofchanges

\end{document}